\documentclass[pss,fleqn]{w-art}
\usepackage{comment}
\usepackage{times}
\usepackage{w-thm}
\usepackage[]{graphicx}
\usepackage{subfigure}
\usepackage{epsfig}
\usepackage{amsfonts}

\usepackage{dcolumn}
\usepackage{rotfloat}

\newcommand{\vect}[1]{{\mathbf #1}}

\newcommand{\bd}{\begin{displaymath}}
\newcommand{\ed}{\end{displaymath}}
\newcommand{\bea}{\begin{eqnarray}}
\newcommand{\eea}{\end{eqnarray}}

\newcommand{\be}{\begin{equation}}
\newcommand{\ee}{\end{equation}}
\newcommand{\bgc}{\begin{center}}
\newcommand{\ec}{\end{center}}
\newcommand{\rr}{\vect r}
\newcommand{\RR}{\vect R}
\newcommand{\cyl}{{\bf MIKA/cyl2}}
\newcommand{\doppler}{{\bf MIKA/doppler}}
\newcommand{\rsdot}{{\bf MIKA/RS2dot}}

\newcommand{\mika}{{\bf MIKA}}

\newcommand{\ie}{i.e.}

\newcommand{\phidual}{\tilde{\varphi}}
\newcommand{\psidual}{\tilde{\psi}}
\newcommand{\zetadual}{\tilde{\zeta}}
\newcommand{\hdual}{\tilde{h}}
\newcommand{\gdual}{\tilde{g}}

\newcommand{\mx}[1]{{#1}}

\newcommand{\orbital}[1]{{#1}}

\newcommand{\qmop}[1]{\hat{#1}}
\newcommand{\hamiltonianop}{\hat{H}}

\newcommand{\smin}{\mathrm{min}}

\newcommand{\etal}{et al.}

\newcommand{\figopt}{h}

\begin{document}

\bibliographystyle{thesis}

\DOIsuffix{theDOIsuffix}
\Volume{XX} \Issue{1} \Copyrightissue{01} \Month{01} \Year{2006}

\renewcommand{\copyrightyear}{2006}

\pagespan{1}{}
\Receiveddate{\sf zzz} \Reviseddate{\sf zzz} \Accepteddate{\sf
zzz} \Dateposted{\sf zzz}



\title[Three real-space discretization techniques in electronic structure calculations]
{Three real-space discretization techniques in electronic structure calculations}


\author[T. Torsti]{T. Torsti\footnote{Corresponding
     author: e-mail: {\sf T.E.Torsti@rug.nl}, Phone: +31\,50\,363\,4373, Fax:
     +31\,50\,363\,4441}\inst{1,2,3}}
\address[\inst{1}]{CSC -- Scientific Computing Ltd., P.O.Box 405, 02101 Espoo, Finland}
\address[\inst{2}]{Theoretical Chemistry, Rijksuniversiteit Groningen,
Nijenborgh 4, 9747AG Groningen, The Netherlands}
\author[T. Eirola]{T. Eirola\inst{4}}
\author[J. Enkovaara]{J. Enkovaara\inst{1}}
\author[T. Hakala]{T. Hakala\inst{3}}
\author[P. Havu]{P. Havu\inst{3}}
\author[V. Havu]{V. Havu\inst{4}}
\author[T. H\"oyn\"al\"anmaa]{T. H\"oyn\"al\"anmaa\inst{5}}
\author[J. Ignatius]{J. Ignatius\inst{1}}
\author[M. Lyly]{M. Lyly\inst{1}}
\author[I. Makkonen]{I. Makkonen\inst{3}}
\author[T. T. Rantala]{T. T. Rantala\inst{5}}
\author[J. Ruokolainen]{J. Ruokolainen\inst{1}}
\author[K. Ruotsalainen]{K. Ruotsalainen\inst{6}}
\author[E. R\"asanen]{E. R\"as\"anen\inst{3,7}}
\address[\inst{3}]{Laboratory of Physics, Helsinki University of Technology -- TKK, P.O.Box 1100, FI-02015 TKK, Finland}
\address[\inst{4}]{Institute of Mathematics, Helsinki University of Technology -- TKK, P.O. Box 1100, FI-02015 TKK, Finland}
\address[\inst{5}]{Institute of Physics, Tampere University of Technology, P.O. Box 692, FI-33101 Tampere, Finland} 
\address[\inst{6}]{Mathematics Division, Faculty of Technology, University of Oulu, P.O.Box 4500, FI-90014 Finland}
\address[\inst{7}]{Institut f\"ur Theoretische Physik, Johannes Kepler Universit\"at, A-4040 Linz, Austria}
\author[H. Saarikoski]{H. Saarikoski\inst{3}}
\author[M. J. Puska]{M. J. Puska\inst{3}}
\begin{abstract}
A characteristic feature of the state-of-the-art of real-space methods
in electronic structure calculations is the diversity of the
techniques used in the discretization of the relevant partial
differential equations.  In this context, the main approaches include
finite-difference methods, various types of finite-elements and
wavelets.  This paper reports on the results of several code
development projects that approach problems related to the electronic
structure using these three different discretization methods.  We
review the ideas behind these methods, give examples of their
applications, and discuss their similarities and differences.

\end{abstract}
\maketitle                   

\renewcommand{\leftmark}
{T. Torsti et al.: Three real-space discretization techniques in electronic structure calculations}

\pagebreak

\tableofcontents

\pagebreak




\section{Introduction}

In this paper, numerical methods for the solution of the Kohn-Sham
equations \cite{kohn65PR} of the density-functional theory (DFT)
\cite{hohenberg64PR} are discussed\footnote{As the emphasis is on the
numerical methods, it has to be pointed out that the discussion is in
fact more general and is equally well applicable to e.g. Hartree-Fock
(HF) equations or other formulations of computational quantum
mechanics.  One of our examples is in fact a HF calculation of atomic
orbitals (see Sec.~\ref{sec:waveletcalc}).}.  We are slowly moving our
emphasis from the ground state DFT towards its time-dependent (TD)
extension TDDFT \cite{runge84PRL}. Much of the discussion of the
present paper is relevant in that case as well.

In solid-state physics and quantum chemistry the standard
discretization methods  are today the plane-wave methodology
and the linear combination (LCAO) of atomic orbitals.  It has been
recognized that despite their merits, they both have certain
shortcomings which motivate the development of new methods.
The plane-wave basis gives accurate results with one single
convergence parameter, the cutoff energy. On the other hand, it gives
a uniform resolution across the entire calculation volume and
application of local refinements at the core regions of atoms do not
very naturally fit into this formalism\footnote{However, the method of
adaptive coordinates, now popular in finite-difference (FD)
\cite{waghmareCPC01,castro-private} and finite element (FE)
\cite{tsuchidaJPSJ98} methods, was originally introduced in the
plane-wave context \cite{gygi92EL,gygi93PRB}.}.  All-electron
calculations are inpractical, but pseudopotentials and the projector
augmented wave-method (PAW) \cite{blochl94PRB,blochl03} have been
developed to circumvent this problem. The domain decomposition method
for massive parallelization requires a description in terms of local
quantities.  In the large matrix inversion step in the Green's
function approach (Sec.~\ref{greensec}) one benefits from the sparsity
of the matrix, which is a consequence of the use of a local basis set.
Also in other contexts, such as the linear scaling method with
localized support functions\footnote{The terminology for these
localized functions varies in the litterature: Hernandez et
al. \cite{hernandez97PRB} used the term of support function, whereas
Skylaris et al. \cite{skylaris02PRB} call them nonorthogonal
generalized Wannier functions (NGWF's). The approach of Fattebert et
al. \cite{fattebert00PRB} suggest the term optimized localized orbital
(OLO).}, the use of a local basis set or description on a grid seems
to be a very natural choice.  The LCAO basis is a local basis and
typically tailored to the system so that big systems can be calculated
with a small number of basis functions.  On the other hand, with the
atom-centered basis functions it may be difficult to systematically
increase the size of the basis set towards the so-called basis set
limit. Extra care needs to be taken to choose special basis sets for
the description of excited states in TDDFT.  The description of
dynamical phenomena beyond linear response such as the ionization of
atoms or molecules under strong laser pulses may be problematic.

We present an introduction to three systematic
real-space methods: the finite-difference (FD) method, the finite-element (FE)
method and the wavelet method. The last two use local basis sets and
the FD method is based on a discretization of the
differential operator (or sometimes of the entire 
differential equation) that involves only local information.  Thus
these discretization methods are well suited for models where locality
plays a crucial role. They also are systematic in the sense that they
have relatively simple convergence parameters. They all are also
generic, which means that they are not tailor-made for a specific
problem. Besides of the similarities, all the methods have their own
characteristic properties, which make them better under some
circumstances and worse under others.

The ease of implementation of a particular method naturally depends on
the background of the developer, as well as on the availability of
software libraries that can be reused during the process.  One would
think that the finite difference method is the easiest approach,
because it does not depend on basis functions. Handling with basis
functions typically needs more work, at least in the first
place. However, the comparison is not so clear. In the case of the
FE-method, there exist  general purpose open-source program
packages \cite{netgen,elmerweb}, which include tools for the mesh
generation, construction of the matrices and for solving the resulting
linear systems of equations and eigenvalue problems so that the
programmer does not need to start everything from scratch. An example
of the utilization of such packages is given in
Sec.~\ref{elmersec}. Furthermore, the state-of-the-art pseudopotential
FD-method in fact involves nowadays also the use of an interpolation
basis, which is required in the double-grid treatment of the
pseudopotential operator \cite{ono99PRL}.

The generality of a numerical method is also connected with the
numerical error control. In the ultimate fully adaptive and systematic
method the user can predefine a level of numerical accuracy for the
physical observables.  The details of the calculation can then be
adjusted such that the desired accuracy is reached.  The requirement
of systematic convergence is easily available in the case of all
systematic methods with uniform accuracy as the accuracy is controlled
by a single parameter in such methods. For non-uniform basis sets the
convergence parameters vary in space and naturally are more
complicated.

Even if the convergence analysis with a non-uniform discretization is
more complicated, spatial adaptivity clearly is a desirable feature.
Many molecular systems include much empty space, where it is not
useful to invest so much for accuracy.  On the other hand, close to
the nuclei a greater resolution is needed, even when the Projector
Augmented Wave (PAW)-method \cite{blochl94PRB,blochl03} or
pseudopotentials are applied.  The system may also consist of a number
of different atoms, each of which require a different resolution in
its core region -- when uniform accuracy is demanded, the most
difficult atom defines the global resolution.  Although it may often
be wise to rely on pseudopotential or frozed core methods in practical
calculations, it is a desirable feature of the numerical method to
deliver also all-electron results when requested. Obviously, this is
unfeasible for methods with uniform spatial resolution.

When the computer resources increase the good parallelizability of the
method becomes important.  The local nature of each of the three
discretization methods allows for the utilization of domain
decomposition methods.  Another important point in the context of
large systems is the compatibility of the method to the linear scaling
formalisms. Popular varieties of these $O(N)$ methods involve
localized support functions which are most naturally expanded in terms
of localized basis functions
\cite{siesta}, or presented on a grid which spans only a part of the
large system \cite{fattebert00PRB}. However, it is also  possible to use a 
plane-wave description within these localized boxes
\cite{skylaris02PRB}.

Discretization of the differential equations is only a part of the
numerical work, solution of large linear systems of equations and
eigenvalue problems is not a trivial task. For large systems the
solution of the eigenvalue problem becomes the dominant part of the
calculation in the traditional approach, where the global orbitals
from the KS-equations are directly solved, scaling as the cube of the
number of atoms.  Each of the three discretization methods lead to
sparse matrix problems.  Many methods in linear algebra are generic in
the sense that they do not depend on the discretization, but others
may do so.  For example, the multi-grid methods require a hierarchy of
discretizations of different resolutions and restriction and extension
operators between them which are trivial to construct in the FD and
wavelet approaches, but slightly more challenging in the FE-context.

{\it Ab initio} molecular dynamics, either in its Car-Parrinello
\cite{car85PRL} or Born-Oppenheimer \cite{barnett93PRB} variety, is a
very important area of applications in our field. Every general
purpose program package has to be capable of performing such
calculations, and when a discretization technique is introduced,
it is important to address the question of its feasibility. All three
discretization methods discussed in this paper are in principle
compatible with {\it ab initio} molecular dynamics. With FE-methods in
adaptive coordinates, pioneering calculations have been recently
presented by Tsuchida \cite{tsuchida04JCP}. As we discuss in
Sec.~\ref{fepropaganda}, such calculations can also be performed with
a uniform FE-mesh or with the general unstructured tetrahedral mesh,
an approach to local refinements which we recommend instead of the
method of adaptive coordinates.  Also the finite-difference method has
been shown feasible in the context of Car-Parrinello type molecular
dynamics \cite{schmid04JCC}.

We have opted for a structure that follows the idea of separation
between the definition of the model of the physical system, typically
as a set of coupled integrodifferential equations, discretization of
the continuous equations, and solution of the discrete equations.  In
this paper we address the last two of these topics.  The detailed
structure of the paper is as follows: In Sec.~2. we introduce the
discretization methods. In Sec.~3. we discuss methods of linear
algebra for linear systems of equations and for eigenvalue problems.
In Sec.~4. we introduce briefly the three lines of work in which the
authors of this paper have been involved in, and the six code
development projects that are associated with these.  In Sec.~5. we
present some calculation examples generated by these projects.  In
Sec.~6. we discuss our future development plans and open questions.
In Sec.~7. we discuss the similarities and differences of the
discretization methods, and finally in Sec. \ref{sec:conclusions} we
summarize.

\section{Discretization methods} \label{sec:discreet}
In most of the calculations discussed in this paper the main
computational task is to solve numerically a single-particle
Schr\"odinger equation of the form
\begin{equation}
\label{eq:schrode}
\hat{H} \psi_i = \varepsilon_i \psi_i, \quad i=1,\ldots,N
\end{equation}
where $\hat{H}$ is a Hamiltonian operator. $\psi_i$ and
$\varepsilon_i$ are the single-particle orbitals and eigenvalues,
respectively.
In the case of electron transport problems (Sec.~\ref{greensec}), one
solves instead for the retarded Green's functions $G^r$ from the
following equation, which has to be solved repeatedly, for multiple
values of positions $\rr'$ and electron energies $\omega$:
\begin{equation} \label{greenRdisc}
\Big{(}\omega - \hat{H}  \Big{)} G^r({\bf
r},{\bf r}';\omega) = \delta({\bf r}-{\bf r}').
\end{equation}
In addition, the Poisson equation
\begin{equation}
\nabla^2 V_C(\rr) = -4\pi\rho(\rr)
\end{equation}
must be solved to obtain from the total charge density $\rho(\rr)$ the
electrostatic potential $V_C(\rr)$ which enters the Hamiltonian
$\hat{H}$ in Eq.~\ref{eq:schrode} or Eq.~\ref{greenRdisc}.  The
Hamiltonian $\hat{H}$ in the above equations may originate from the
density-functional theory or Hartree-Fock theory. In both cases, the
Hamiltonian depends on the solutions of the equations, which makes the
problem nonlinear.

For clarity, we define an example Hamiltonian $\hat{H}$ from
density-functional theory in the presence of Kleinman-Bylander (KB)
\cite{kleinmanPRL82} form of pseudopotentials by its action on a test
function $\eta(\rr)$:
\begin{equation}
\label{exampleham}
\hat{H} \eta(\rr) = -\frac12 \nabla^2 \eta(\rr) + 
V_{\rm eff}[n](\rr) \eta(\rr) + \sum_{alm}
c_n \xi^0_{alm}(\rr- \RR_a) \int \xi^0_{alm}(\rr'-\RR_a) \eta(\rr') d\rr'.
\end{equation}
The three terms of the Hamiltonian are referred to as the kinetic
energy operator, effective potential energy operator\footnote{The
notation emphasizes the functional dependence of the effective
potential on the density, which on the other hand is determined from
the eigenfunctions of $\hat{H}$. The detemination of the
self-consistent density is therefore a nonlinear problem,
which is further discussed in the preamble of Sec.~\ref{sec:solve}.}
$V_{\rm eff}[n](\rr)$ and the pseudopotential operator defined by the
help of atom centered (nucleus at ${\bf R_a}$) angular momentum
$(l,m)$ dependent projectors $\xi^0_{alm}(\rr-{\bf R_a})$ and the
pseudopotential operator\footnote{For convenience of notation, we
hereafter enumerate these projectors with a single index (such an
enumeration obviously exists) $n=n(a,l,m)$, and use projectors with
shifted origin $\xi_{n(a,l,m)}(\rr) = \xi^0_{alm}(\rr-\RR_a)$.}.

The numerical solution consists of a discrete presentation of the
infinite-dimensional problem and then solution of the resulting
discrete problem (this solution step is the topic of Sec.~\ref{sec:solve}).  
In this section we present three discretization
methods, all of which are often referred to as systematic real-space
methods. In Sec.~\ref{sec:fd} we introduce the finite-difference method.
The other two methods are both variational methods, thus we discuss
them first on a common footing in  Sec.~\ref{sec:galerkin}.  Thereafter
we discuss the finite-element method  in Sec.~\ref{sec:fem} and the 
wavelet approach in Sec.~\ref{sec:wavelet}.
Other systematic real-space methods exist as well,
for example, the discrete variable representation (DVR) method
\cite{liu03PRB}  and the Lagrange mesh method
\cite{baye02PRE}. For historical perspective, we mention
in this context also the highly accurate methods for diatomic
molecules by Pyykk\"o et al. \cite{kobus96CPC} 
and Becke et al.\cite{becke86PRA}, as well
as the method for polyatomic molecules by Becke et al. \cite{becke88JCP}
and for solids by Springer  \cite{springer98PRB}.

\subsection{Finite difference methods} \label{sec:fd}

The finite difference method is a popular numerical approach because
of its conceptual simplicity. No basis functions are involved, which
makes the implementation of a computer program very easy.  A uniform
grid is utilized, just as the Fourier grid which appears in fast
Fourier transform routines. The method is thus ideal, if one also
wants to reserve the opportunity to evaluate some operators in
$k$-space, as e.g. in the algorithm described in
Sec.~\ref{diffusionsec}. It is also straightforward to obtain the
hierarchy of coarser grids required in multilevel (multigrid) methods,
which are often the best available preconditioning methods for
iterative solvers in linear algebra.

The central idea of the simplest version of finite difference (FD)
method is a discrete representation of the partial differential
operators. In the case of our example Hamiltonian
(Eq.~\ref{exampleham}) the emphasis is thus on the kinetic energy
operator.  In some FD-approaches the entire differential equation is
discretized and not only the differential operators
\cite{briggs95PRB,collatz60,heiskanen98hodie}.  For clarity of
presentation, we discuss here only the simpler variety, which is also
more widely used in electronic structure calculations.

In the discretization procedure the function under scrutiny is sampled
in an evenly spaced point grid in three dimensions. This provides the
necessary mapping from the function $\eta(\rr)$ in the
infinite-dimensional Hilbert space of the model to a vector ${\bf e}$
in a finite-dimensional space. The second derivative at a grid point
is approximated by a weighted sum of the values of the function in
neighboring grid points
\begin{equation}
\label{fdkinetic}
\frac{\partial^2}{\partial x^2} \eta(x_i,y_j,z_k) \approx \sum_{n=-N}^N  c_n
{\bf e}_{(i+n)jk}, 
\end{equation}
where $(x_i,y_j,z_k)$ = $(ih_x,jh_y,kh_z)$ with the grid spacings
$h_{x,y,z}$. The coefficients $c_n$ are derived by requiring this
formula to be accurate for the $(2N)$th order polynomial fitted to the
values at the $2N+1$ sampling points above. The values of these
coefficients up to $N=6$ can be found from literature, for example,
from Ref. \cite{chelikowsky94PRB}.  Note that the polynomial is
different when the recipe is applied at neighboring points. Thus the
FD-scheme does not implicitly define an interpolating polynomial as an
element of the original Hilbert space.  Nevertheless this treatment of
the kinetic energy operator is quite accurate for smooth functions.

The projector functions $\xi_n$ in the pseudopotential operator are
also represented by their values $x_{nijk}$ at the grid points in the
simplest implementation of the FD-approach. The integrals occurring in
the pseudopotential operator are approximated by the trapetsoid rule
\begin{equation}
\label{bbfdpseudo}
  (\hat{V}^{\rm nl} \eta) (\rr_{ijk}) = \sum_n c_n \xi_n (\rr_{ijk})\int \xi_n(\rr') \eta(\rr') d\rr' \approx
\sum_n c_n x_{nijk} \sum_{i'j'k'} x_{ni'j'k'}
{\bf e}_{i'j'k'}.
\end{equation}

It has been recently widely recognized, that this sampling of the
pseudopotential projectors often requires very fine grids to be
sufficiently accurate\footnote{Even more problematic formulas occur
for the forces, as they involve the derivatives of the projectors
$\xi_l$ that vary more rapidly than the projectors themselves.}.  When
too coarse grids are employed, one encounters problems such as the
{\it egg box effect} \cite{nogueiraAPDFT}.  This is the spurious
dependence of the total energy as a function of the atom position when
the atom moves from one grid point to another, and is mainly a
consequence of the poor sampling of pseudopotential projectors.

In order to improve the discretization accuracy of the pseudopotential
operator, Ono and Hirose (OH) \cite{ono99PRL} proposed to use
polynomial interpolation to obtain values on a finer grid from the
discrete values ${\bf e}$, and use trapetsoid rule on this finer
grid\footnote{Note that the interpolating polynomial is necessarily
different from the polynomial used in obtaining the second derivative
in Eq.\ref{fdkinetic}, although a polynomial of the same order can be
used. The interpolation procedure involves the association of a
piecewise polynomial basis function of product form to each grid
point.}.  After a straightforward manipulation they find that the
discrete version of the pseudopotential operator can be evaluated on
the coarse grid as
\begin{equation}
  (\hat{V}^{\rm nl} \eta) (\rr_{ijk}) \approx
\sum_l c_l \tilde{x}_{lijk} \sum_{i'j'k'} \tilde{x}_{li'j'k'}
{\bf e}_{i'j'k'}.
\end{equation}
Here the discrete values for the projectors are given by the fine grid
trapetsoid rule approximation of
\begin{equation}
\tilde{x}_{lijk} = \int \xi_l(\rr) \phi_{ijk}(\rr) d \rr.
\label{tildex}
\end{equation}
Above the $\phi_{ijk}(\rr)$ is a piecewise polynomial basis function
from Lagrange interpolation, with value of unity at the point $(i,j,k)$ and zero
in other points. It is a cartesian product of three piecewise polynomials in
one dimension.  Although the matrix elements of the discrete pseudopotential
operator need not be stored, it is instructive to write the formula
\begin{equation}
V^{\rm nl}_{ijk,i'j'k'} =  \sum_l c_l \tilde{x}_{lijk} \tilde{x}_{li'j'k'} = \int \phi_{ijk} (\rr)
 \hat{V}^{\rm nl} \phi_{i'j'k'}(\rr) d \rr,
\end{equation}
where the required integrals are approximated by a trapetsoid rule on
a fine uniform grid.

The local potential energy operator is diagonal -- sampled by its values
at the grid points -- in all practical implementations
of the FD-approach. Therefore, in practical implementations
where the OH-scheme is used, it is beneficial to exploit the freedom of choice
within  the pseudopotential approach by picking a very smooth local part for the
pseudopotential.


\subsection{Variational methods with basis function discretization} \label{sec:galerkin}

Variational methods are those discretization methods which involve 
a set of basis
functions. In this paper we present two methods which use them -- the
finite-element method and the wavelet method.
The main idea in variational methods for solving partial differential
equations is to multiply the equation by a test function and then
integrate the identity over the computational domain. For instance,
this leads us to the eigenvalue problem
\begin{equation}
a_W (\psi, \eta) = \int \eta(\rr) \hat{H} \psi (\rr)  d\rr  =
\varepsilon \int \psi (\rr) \eta (\rr) d\rr =  \varepsilon (\psi,
\eta) 
\end{equation}
where $\eta$ is our test function, and $a_W (\cdot, \cdot)$ is the
bilinear form for our wavelet method, and the last equality sign
defines the inner product $(\cdot,\cdot)$. In the finite-element method
implementations in this paper, an
integration by parts of the laplacian is perfomed to obtain a
symmetric bilinear form, and to treat the boundary terms in a natural
way. Thus, with our example Hamiltonian the bilinear form for the
finite-element method would read
\begin{equation}
\begin{aligned}
a_{FE}(\psi, \eta) = & \int \left( \frac12 \nabla \psi(\rr) \cdot \nabla \eta(\rr) + V(\rr) \psi(\rr) \eta(\rr) + \sum_i^{N_{proj}}
c_i \xi_i(\rr) \int \xi_i(\rr') \psi(\rr') d\rr' \eta(\rr) \right)
d\rr  \\ 
	& + \hbox{boundary terms}.
\end{aligned}
\end{equation}
and the variational eigenproblem would be
\begin{equation}
a_{FE}(\psi, \eta) = \varepsilon (\psi, \eta)
\end{equation}
for all test functions $\eta$.

The discretization in variational methods is obtained simply by
choosing finite-dimensional spaces $V_h$ and $T_h$ to approximate the
functions $\psi$ and $\eta$ in the above eigenproblems. Then one
obtains with trial functions $\{\psi_j\}_{j=1}^n$, $\psi_j \in V_h$
and with test functions $\{\eta_j\}_{j=1}^n$, $\eta_j \in T_h$  the
Petrov-Galerkin condition
\begin{equation}
a_{W / FE} (\psi_h, \eta_k) = \varepsilon_h (\psi_h, \eta_k), \quad k=1,\ldots,n
\end{equation}
where
\begin{equation}
\psi_h = \sum_{j=1}^n c_j \psi_j.
\end{equation}
This leads to a finite-dimensional generalized eigenvalue problem
\begin{equation}
H c = \varepsilon_h S c
\end{equation}
where $H$ is the discretized hamiltonian, $H_{ij} = a_{W / FE}
(\psi_j, \eta_i)$, $S$ is the overlap matrix, $S_{ij} = (\psi_j,
\eta_i)$, and $c$ is the vector of the unknown coefficients $c_j$.

In variational methods it important to consider the choice of the
finite-dimensional subspaces $V_h$ and $T_h$. This is the point where
the finite-element method and the wavelet method we consider diverge
from each other. In particular, in the finite-element method we follow
the usual convention and select $V_h=T_h$ whereas our implementation
of the wavelet method uses different spaces for the trial and for the
test functions.


\subsection{Finite-element method \label{sec:fem}}

The finite-element method (FEM) \cite{braess,daBible,szabobook} is widely used in many different
fields, for example, in structural mechanics, fluid dynamics,
electromagnetics and heat transfer calculations. The popularity of the
FEM comes from it's flexibility.  It allows different geometries and
boundary conditions to be implemented in a straightforward way.
Possibility to use refinements and higher order polynomials in the basis reduce the
number of the basis functions in comparison for example to the number
of grid points in the finite difference approach. Because of the
popularity of the method there is a lot of theoretical work and useful
tools available. A recent review of the state of the art for FEM in
electronic structure calculations can be found in
Ref. \cite{pask04MSMSE}.

In the FEM the calculation domain $\Omega$ is divided into small
regions called elements. 
In
three-dimensional calculations tetrahedral, hexahedral and
pyramid-shaped elements are used. 
 Pyramids are needed to combine the meshes consisting of
hexahedra and tetrahedra together.  Hexahedral (box-shaped) elements
are good if the calculation volume is needed to be filled
uniformly. Instead tetrahedra have the advantage that their size can
easily vary inside the domain region for example to increase the
accuracy in the core regions of the atoms.
Generating high-quality meshes is a nontrivial task, and until
recently reliable, free and user friendly three-dimensional mesh
generators have not been easily available. 
The situation is changing rapidly however, as high
quality open-source mesh generators 
have become available \cite{easymesh,netgen}.

The basis functions are constructed conforming to the mesh of the
elements so that they are non-zero only in a few neighboring
elements. This ensures the local nature of the basis functions and the
resulting matrices become sparse. The utilization of local basis
functions also enables the parallelization of the problem based on
domain decomposition methods.  Small elements imply many basis
functions which yields a good numerical accuracy. This is how we can
increase the accuracy in the regions where the solution changes
fast. This is particularly useful in many atomistic calculations
involving hard norm-conserving pseudopotentials because it is easy to
increase the accuracy near the core regions of the atoms. Many systems
also contain a lot of empty space where large elements can be used.

\subsubsection{Finite-element $p$-basis \label{psection}}

There are some options on how to choose a good finite-element basis
\cite{braess}. The simplest choice is to use the linear elements so
that the basis function is unity in one of the nodes and declines
linearly to zero towards the boundaries of the element. The linear
basis is easiest to implement, but in order to achieve a better
convergence high-order elements are used. For a smooth solution there
is a remarkable difference in the accuracy between the linear and
higher-order element calculations with the same number of basis
functions.

The  $p$-elements are hierarchical in the sense that the higher-order
basis set includes also the lower-order basis sets
\cite{p_elementit}. A basis set has four types of functions, node-
edge-, face-, and element-based functions. The node-based functions,
which are linear, are nonzero only in the volume of the elements which
have a common node. Similarly an edge based function is nonzero in the
elements which have a common edge. The element-based functions have
their support only inside one element.

\begin{figure}[htb!]
\begin{center}
\epsfig{file=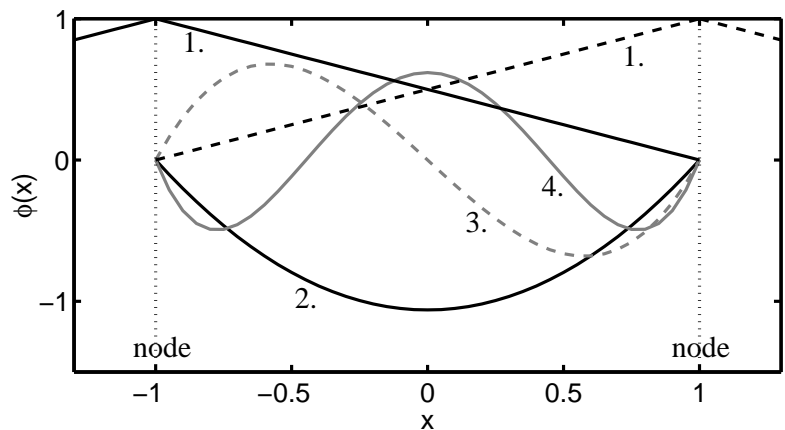,width=0.7\textwidth}
\end{center}
\caption{\label{kanta1D}{One-dimensional basis functions in the
    reference element [-1,1] up to the fourth order. Inside a element
    there ar two node-functions and three element-functions}}
\end{figure}

The basis function set of the $p$-elements is derived using the
Legendre polynomials.  This ensures that their derivatives are more
orthogonal to each other than in the traditional case of nodal basis
functions.  The orthogonality makes the solutions numerically stable
even when using polynomials of high order. Otherwise the conditioning
of linear systems can be a problem. The one-dimensional basis
functions in the reference element are shown in Figure~\ref{kanta1D}.

The high order element includes more basis functions than the low
order one.  Where one tetrahedral linear element includes four linear
basis functions, the fourth order one has 35. This means that the size
of the elements is bigger and the overlap between basis functions is
larger than in a linear mesh.  Therefore when the number of the basis
functions is reduced, but at the same time the filling of the
coefficient matrix is increased. The filling of the matrices increases
the CPU time and memory requirements when solving the equations, while
the reduction of the number of basis function works in the opposite
direction. However, a smooth solution converges faster for a larger
filling. Therefore the critical factor for a good convergence with
high-order polynomials is the smoothness of the solution. Naturally,
for each problem there is an optimal polynomial order. Because the
hierarchical nature of $p$-elements it is possible to change the order
of the elements inside one calculation mesh. This results in the
so-called $hp$ methods, where the polynomial order $p$ is variable in
space as well as the size $h$ of the elements. This is useful if the
nature of the solution has rapid changes in some part of the
calculation area, but otherwise the solution is smooth. This is the
case for example in the DFT calculations in the atomic core region.

In adaptive methods, an interesting question is how to choose whether
to increase the order or decrease the element size in a given part of
the mesh where resolution needs to be enhanced.

In the case of all-electron calculations, where the effective
potential has a singlarity of the type $1/r$, when the same mesh is
utilized for the interpolation of the effective potential and the
orbitals, the standard results of approximation theory
\cite{szabobook} suggest a refinement strategy where small elements
with low order $p$ are used in the vicinity of the nuclei. In other
parts of the space, the orbitals as well as the effective potential
are smooth, which suggests larger and higher order elements.

Multilevel methods can be also of the ``multi-$p$''-variety, where the
same mesh is used at each resolution level, but the order of the
elements varies.
 
\subsection{Wavelets}

The variational approximation scheme using the wavelet basis as the
trial and test functions will combine the best aspects both of the FEM
approach and Fourier approximation. From the FEM world we retain the
locality of the approximation. 
For a wide class
of operators their matrix representation \(H\) in the wavelet
basis is nearly diagonal.
This fact
together with the refinement equation makes it possible to carry out
the matrix-vector computations in FFT speed. It can be shown that
the wavelet based approximation schemes for operator equations have
the optimal computational complexity in the following sense: For a
given approximation tolerance \(\epsilon>0\) the wavelet
approximation, when computed using iterative and adaptive techniques,
requires the least number of arithmetical operations and storage locations
of all $N$-term approximation schemes with the same tolerance
\cite{stevenson2003}.

Furthermore, the wavelet transform yields powerful numerical schemes
including very efficient adaptive algorithms. So the wavelet
approximation is naturally adaptive. In addition to that the wavelet
based approximation method combined with the multilevel techniques
(i.e. the multiresolution analysis) are very good preconditioners for
linear systems \cite{dahlke2002}. It can be shown for a very general
class of operators that the scaled stiffness matrix has uniformly
bounded condition numbers \cite{dahmen1997}.

The rate of approximation depends on the smoothness of the function to
be approximated and the order of the wavelets.  Assuming that the
solution \(\psi\) of the eigenvalue problem (\ref{eq:schrode}) is
sufficiently smooth it can be shown (see \cite{lindemann05thesis})
that
\begin{equation}
\|\psi-\psi_h\|_{L^p(\mathbb{R}^d)} \leq  c(\psi)
n^{-\frac{s}{d}}.
\end{equation}
Here \(n\) denotes the number of basis functions
on the highest resolution level of the wavelet approximation, \(s\) is
the approximation order of the wavelets and the constant
\(c(\psi)\) is independent on \(n\). This approximation
property is exactly the same as for the polynomial finite
element spaces.

The application of wavelet analysis in \(\mathbb{R}^d\) has been
considered inpractical because the number of scaling functions
increases which affects the size of the wavelet basis. This holds for
the tensor product wavelets, or separable wavelets, which will be
obtained from the dyadic wavelet basis in one dimensions. However, by
using the non-separable wavelet basis we will retain the same
situation as in one-dimensional case. We have only one scaling
function. These wavelet basis will be obtained by using the isotropic
scaling matrix \(M\) with the determinant \(\det(M)=2\), i.e. the
matrix has integer coefficients. The non-separable wavelet basis have
one more advantage over conventional separable wavelet
functions. Because of their isotropy they are useful for rotationally
invariant systems.

\subsubsection{Interpolating wavelet basis set}

A biorthogonal wavelet family of degree \(m\) is characterized by a
mother scaling function \(\varphi\), a mother dual scaling function
\(\phidual\), and four finite filters \(h_j\), \(g_j\), \(\hdual_j\),
and \(\gdual_j\). We follow the convention in \cite{g98} where the
nonzero elements of the filters lie in the range \(j = -m, \ldots,
m\).  The mother wavelet \(\psi\) and the mother dual wavelet
\(\psidual\) are determined by the mother scaling function, the mother
dual scaling function, and the four filters.

Interpolating wavelets \cite{chui-li1996,donoho1992,g98} are one
biorthogonal wavelet family \cite{d92,g98}. Interpolating wavelets
enable simple calculation of matrix elements and expansion of
functions in a basis function set because of the special form of the
dual scaling functions and dual wavelets. Actually the interpolating
dual scaling functions and interpolating dual wavelets are not
functions but distributions.

For interpolating wavelet family of degree \(m\), the mother scaling
function \(\varphi\) is constructed by recursively applying polynomial
interpolation of degree \(m\) to data \(s_i = \delta_{i,0}\) and the
mother dual scaling function is
\begin{equation}
  \label{eq:interp-dual-phi}
  \phidual(x) = \delta(x) ,
\end{equation}
where \(\delta\) is the Dirac delta function.
For interpolating wavelets the coefficients \(h_j\) satisfy \cite{g98}
\begin{equation}
  \label{eq:interp-h}
  h_j = \varphi(j/2), \;\; j = -m,\ldots,m .
\end{equation}

For biorthogonal wavelets the matrix elements of operators are not
computed as ordinary inner products but they are computed as integrals
involving the basis functions and so called dual basis functions. The
matrix elements of an operator \(\qmop{A}\) in a basis set consisting
of biorthogonal wavelets are defined by
\begin{equation}
  \label{eq:sof-elements}
  A_{ij} = \int_{-\infty}^{\infty}
  \zetadual_i(x)\;\qmop{A}\;\zeta_j(x) dx ,
\end{equation}
where \(\zeta_j, \; j = 1,\ldots,N \) are the basis functions,
\(\zetadual_i, \; i = 1,\ldots,N \) are the dual basis functions, and
\(N\) is the size of the finite basis function set.

\section{Linear algebra} \label{sec:solve}

The numerical task of solving the discretized versions of the
Kohn-Sham equations, Hartree-Fock equations or equations for the
Green's functions consists of repeated solutions of linear systems of
equations and/or eigenvalue problems, whereas the real-time
propagation methods of TDDFT involve exponentiation of large matrices.
In other words, problems of linear algebra \cite{golub89MC}.  All
three discretization techniques discussed in Sec.~\ref{sec:discreet}
lead to sparse matrix problems. Typically these matrices are so large
that it is important to store only their nonzero elements, using
e.g. the CRS (Compressed Row Storage) format \cite{dongarraCRS} or
define the matrices only by a subroutine which operates upon a
vector\footnote{One often, e.g. in the method of
Secs.~\ref{diffusionsec} and \ref{sec:Linz} as well as in the standard
plane-wave approach, defines the ``matrix'' for the kinetic energy
operator as a sequence of an FFT transform, multiplication by
$-\frac12 k^2$, and an inverse FFT transform. In that case, of course,
the underlying discretization method is none of the three of
Sec.~\ref{sec:discreet}.}.  Notably, in the FD-method with a uniform
grid the nondiagonal elements of the discretized Laplacian are the
same on each row and need not be stored.

Because of the large size of the matrices, straightforward utilization
of standard linear algebra packages for dense matrices, such as {\bf
Lapack} \cite{lapack}, is not an option\footnote{There exists,
however, at least one promising starting point for the equivalent
standard library for sparse (and ``matrix-free'', see below) matrix
problems, called {\bf Sparskit} \cite{sparskit}. Utilization and
further development (if necessary) of such libraries is important in
our opinion -- the recycling of as many tools as possible is one of
the main current trends in real-space electronic structure
calculations as well as more generally in computational science. }.
Availability of a good selection of efficient tools for sparse matrix
problems is thus a necessary prerequisite for large scale real-space
electronic structure calculations.

It was emphasized in the notation of Eq.~\ref{exampleham}, that a part
of the Hamiltonian $\hat{H}$, the effective potential $V_{\rm
eff}[n](\rr)$, is a functional of the electron density, which on the
other hand needs to be determined from the sum of the squared moduli
of the occupied orbitals, which are the eigenfunctions of
$\hat{H}$. We are thus presented with a nonlinear eigenvalue
problem\footnote{Strictly speaking, an eigenvalue problem with fixed
potential is also a  nonlinear problem, as the eigenvalue and
eigenfunction need to be determined simultaneously. Nevertheless, such
problems are generally regarded as subfields of linear algebra.}.
There exists a general multigrid formulation called Full Approximation
Storage (FAS) \cite{brandt77MOC} which is in principle directly
applicable to such nonlinear problems.  Results from work on the
development of nonlinear FAS eigenvalue solvers have been reported by
Beck et al.\footnote{However, this work has thus far been mainly 
focused on applying the FAS algorithm of Ref.~\cite{Brandt83SIAM} to 
the eigenvalue problem with fixed potential, and 
updating the potential in an outer loop.} 
\cite{wang00JCP,wijesekara03JTCC} and Costiner and Ta'asan \cite{costinerPRE95b}.  
Apart from that, one
can try to minimize directly the corresponding nonquadratic total
energy functional using the preconditioned steepest descent (SD)
\cite{briggs95PRB,briggs96PRB} or conjugate-gradient (CG)
\cite{payne92RMP} method, in which the electron density and hence the
effective potential can be updated after each update of the
approximate orbitals.  However, an iterative diagonalization scheme,
in which the Hamiltonian is repeatedly diagonalized for a fixed
$V_{\rm eff}$, and the self-consistent $V_{\rm eff}$ is found using
some ``mixing scheme'' in an outer loop, can be at least equally efficient
\cite{kresse96PRB,kresseCMS96}.  We have opted for the latter approach
in all of our example calculations presented in
Sec.~\ref{sec:calculations}. Some of the systems considered require
special care in the choice of the mixing scheme in order to find a
convergent iteration\footnote{It is expected, that similar convergence
problems are present in the direct approaches as well.}.
Specifically, we want to mention here the GRPulay method
\cite{bowler00CPL} and the Newton-Raphson (or response function)
method of Refs.~\cite{newton,Collective} which we have found useful
in Sec.~\ref{greensec} and in  Secs.~\ref{rsdotsec} and \ref{axialsec}, respectively.  
For more information on our recent work related to self-consistency iterations see
Refs.~\cite{torsti04PSIK,torsti05nano}.


Apart from the nonlinear self-consistency problem which we solve in an
outer loop, we are left with only standard problems of linear algebra.
The problems relevant for our work can be divided in three classes:
systems of linear equations (Sec.~\ref{sec:linsystems}), eigenvalue
problems (Sec.~\ref{sec:eigen}) and matrix exponentiation, which is
needed in time propagation schemes \cite{Castro04}, as briefly
discussed in Sec.~\ref{jatkoprojsec}.

\subsection{Systems of linear equations} \label{sec:linsystems}

Large linear systems of equations occur in real-space electronic structure
calculations in many contexts. In the example calculations
of Sec.~\ref{sec:calculations} the discretized Poisson equation, the
inversion step in the shift and invert mode of the Lanczos method 
(Sec.~\ref{sec:lanczos}), the equations of full-response (Eq.~\ref{eq:cylrespo}) and collective
approximation (\cite{Collective}) formulations of the response function method, 
and the matrix inversion for obtaining the Green's function from Eq.~\ref{greenR}
(after discretization with the finite-element method) occur. Usually it is 
most convenient to solve such systems with iterative methods \cite{vorstbook}. However, for
matrix inversion, where the same equation has to be solved many times with different
right-hand sides, direct methods are more efficient. 

\subsubsection{Iterative solvers} \label{sec:iterative}

For symmetric positive definite matrices, the conjugate gradient (CG) method is
often the standard choice. For the dense ``matrix-free'' problems\footnote{We refer to 
matrices whose action on a vector can be easily implemented as a subroutine, but the storing
of whose matrix elements in impractical due to their large size, as matrix free.}
occuring in the application  of the response function method to two-dimensional quantum
dot problems (Sec.~\ref{rsdotsec}) the CG method was found efficient even in the 
absence of any preconditioner. 

Often, however, it is important to accelerate the convergence of the CG method
through preconditioning. Preconditioning the equation $Ax = b$ corresponds to
multiplying the equation by an approximate inverse $B \approx A^{-1}$, resulting
in $x \approx BAx = Bb$. In our applications of the finite-element
method to all-electron calculations of molecules (Sec.~\ref{elmersec}) 
the CG method with preconditioners based on incomplete LU factorization (ILU) 
\cite{ilu86} and the multigrid method \cite{briggs00MTSE,hackbush85MGMA,wesseling92AITMM}
was applied to the linear system of equations
occuring in the inversion step of the  shift and invert mode of the Lanczos
method for the eigenvalue problem (see Sec.~\ref{sec:lanczos}).

A multigrid method 
where the Gauss-Seidel method is used as 
a smoother, is used as a solver for the linear system of equations that results
from the FD-discretization of the Poisson equation within the MIKA-package 
\cite{torsti04PSIK,torsti05nano}.



In the axial symmetry, the matrix for the Laplacian within
the FD-method is not symmetric. However, we have found that our
MG-method with the Gauss-Seidel method as a smoother converges.
On the other hand, the CG-method for the full-response equation
\ref{eq:cylrespo} is not applicable. The best method we could 
find for this problem was the  GMRES method
with no preconditioning \cite{vorstbook}. For the corresponding 
equation of the collective approximation \cite{Collective} we found
no useful iterative scheme. This problem obviously 
calls for further work. A good starting point is probably to 
take a look at the algorithms available in the {\bf Sparskit}
package \cite{sparskit}.


\subsubsection{Direct solvers} \label{sec:direct}

When solving for the inverse of a matrix as is the case in the
calculations employing the Green's function (see Sec.~\ref{greensec})
it is not usually feasible to use iterative methods since the
iteration must usually be performed separately for each column of the
inverse. Instead, direct methods for sparse matrices provide an
attractive alternative since part of the computational work needs to
performed only once per inverse. Modern direct methods for sparse
matrices are based on the frontal factorisation algorithm of the
sparse matrix \cite{duff,duff-reid}. The algorithm starts with
symbolic factorisation of the matrix, i.e. heuristically finding a
permutation that is intended to minimise the fill-in in the numerical
factors. Next, a sparse Cholesky or LU-factorisation is computed using
block operations with dense BLAS kernels. Finally the problem is
solved with backward and forward substitutions. If the entire inverse
is desired only the final substitution steps must be performed for
each column of the inverse whereas the matrix factors remain
unaltered. Several implementations of the frontal method are available
as software libraries, e.g. \cite{hsl,umfpack,watson}.

\subsection{Eigenproblem solvers} \label{sec:eigen}

Mostly these eigenproblems are Hermitian, so that Lanczos based
iterations are the starting point, if one looks at the problem from
the point of view of a numerical analyst \cite{saad92NMLEP}. For
physicists the starting point is often the preconditioned conjugate
gradient method (PCG)\footnote{Interestingly, the performance of PCG
and Lanczos methods has been compared in the context of plane-wave
methods in Ref.~\cite{wang94CMS}, where it was found that for the
diagonalization step in a fixed potential $V_{\rm eff}$ of a 900-atom
Si-cluster, the Lanczos method is about an order of magnitude faster
than the PCG method.}
\cite{payne92RMP}.  Also nonhermitian
eigenvalue problems sometimes occur. This happens e.g.  when
generalized finite-difference discretizations are used (e.g.  the
Mehrstellen discretization of Ref.~\cite{briggs95PRB,briggs96PRB}), if
the method of adaptive coordinates is applied without extra care to
guarantee symmetric matrices \cite{modine97PRB,castro-private}, and
also when the FD-method is applied in axial symmetry
\cite{tuomas03thesis}.

In this section we describe those eigenproblem solvers that are used
in the example calculations of Sec.~\ref{sec:calculations}, as well
as the residual minimization method (Sec.~\ref{sec:rmmdiis}), 
which is at the core of the 
{\bf GridPaw}-code (Sec.~\ref{sec:gridpaw}), 
on which the main line of development within the
MIKA-project is currently based on. 

This is not an exhaustive list of eigenproblem solvers. 
In finite-difference based electronic structure calculations also the
method of steepest descent with multigrid preconditioning on global
grids \cite{briggs95PRB,briggs96PRB} as well as in an almost linear
scaling implementation based on localized orbitals (support functions
presented on a grid)
\cite{fattebert00PRB} is used.  The highly efficient method utilized
in {\bf Parsec} is based on a taylor-made parallel generalized
Davidson method \cite{stathopoulos00CSE}.  Recently, a new
preconditioned, Krylov-space technique has been introduced in the
mathematical litterature by A. Knyazev \cite{knyazevSIAM,knyazevETNA}
-- this method is claimed to be more efficient than its precursors,
and certainly deserves to be properly tested in challenging
applications. Relatively large\footnote{The dimension of these 
matrices is small in comparison to the Hamiltonian matrices occuring in typical
FD-based calculations, but large in comparison to those 
occuring in typical calculations with atom-centered basis functions} 
and dense eigenvalue problems
of a different structure  occur when computing excitation energies
and oscillator strengths from TDDFT from the Casida equation
\cite{casida95RADFM,jamorski96JCP}. 
Here, the Davidson method \cite{davidsonJCP75,davidsonCP93} 
is often used \cite{gisbergenCPC99,stratmanJChP98} for finding
a few of the lowest excitation energies.  According to Ref.~\cite{burdickCPC03}, however,
it is also feasible to utilize the {\bf Lapack} \cite{lapack} 
or {\bf ScaLapack} \cite{scalapack}
routines for all eigenvalues of dense matrices
in this case. 

\subsubsection{Lanczos and Arnoldi methods} \label{sec:lanczos}

The Arnoldi Package {\bf Arpack} \cite{arpack} contains an implementation of the Lanczos
method as well as its generalization to nonsymmetric problems, the Arnoldi
method. It is used through a reverse communication user interface, i.e. the 
user has to provide his or her own subroutines for matrix-vector products and
solvers for linear systems of equations, and {\bf Arpack} calls these user defined 
subroutines when needed. Thus the efficiency of {\bf Arpack} actually depends heavily
on the efficiency of these user defined subroutines. 

We have used {\bf Arpack} in the so-called shift and invert mode in the 
finite-element example calculations presented in Sec.~\ref{elmersec}.
The computationally expensive step in these calculations was the 
solution of a linear system of equations, for which the CG-method was 
applied as discussed above in Sec.~\ref{sec:iterative}. 




\subsubsection{Rayleigh-quotient multigrid} \label{rqmgsec}
The Rayleigh-quotient multigrid (RQMG) method  \cite{Heiskanen} is originally 
developed for the generalized eigenproblem
\begin{equation}
H  {\bf u} = \epsilon B {\bf u}
\end{equation}
that arises from a FD-discretization of an equation of the form of 
Eq. \ref{eq:schrode}. Unless generalized finite-difference methods
(such as those of Refs. \cite{briggs95PRB,collatz60,heiskanen98hodie}) 
are used, $B=I$ and $H$ is Hermitian 
(in fact real and symmetric).
In the general case, it is assumed that $B^{-1}H$ is Hermitian\footnote{This is indeed the case for
the Mehrstellen discretization introduced in Refs. \cite{briggs95PRB,briggs96PRB}, the authors
are not aware if this holds for the more  general high-order compact (HOC) discretizations of 
Ref~\cite{heiskanen98hodie}.}, 
thus the eigenvectors are
orthonormal\footnote{We have also considered a generalization
which requires an overlap matrix $S$ here: ${\bf u}_i^HS{\bf u}_j = \delta_{ij}$. Such a generalization would be needed within the generalized double 
grid scheme  \cite{torsti04PSIK,torsti05nano}, where the discretization matrices $H,B,S$ are derived through  "Galerkin transfer" from the FD-matrices $H,B,I$ on an auxiliary  finer level. 
In a  finite element implementation of RQMG, the 
matrices $B$ and $S$ are the same, and both $H$ and $S$ are 
also Hermitian.} in $\mathbb{R}^N$: ${\bf u}_i^H{\bf u}_j = \delta_{ij}$. 
Given the eigenvectors $\{ {\bf u}_i; 1 \leq i\leq k\}$, ${\bf u}_{k+1}$ is
obtained by finding the minimum of the functional
\begin{equation}
\label{rqmgneq}
F[{\bf u}_{k+1}] = \frac{{\bf u}_{k+1}^HH{\bf u}_{k+1}}  {{\bf u}_{k+1}^HB{\bf u}_{k+1}}  +
\sum_{i=1}^k q_i\frac{\left|{\bf u}_i^H{\bf u}_{k+1}\right|^2} {{\bf u}_i^H{\bf u}_i {{\bf u}_{k+1}^H{\bf u}_{k+1}}}.
\end{equation}
In the minimization process, a hierarchy of grids is utilized, and on each grid, one degree of
freedom is varied at a time and the minimum of the functional is found along the corresponding line in the 
space of {\bf u}'s. Varying one degree of freedom on a coarse grid corresponds to varying the 
{\bf u} values at multiple grid points on the fine grid at the same time. More details on the 
RQMG method can be found from Refs. \cite{Heiskanen,torsti03IJQC,tuomas03thesis,torsti04PSIK,torsti05nano}. Our implementation of the RQMG method was motivated by the method due to Mandel
and McCormick \cite{mandel89JCC}, which was designed for the solution of the eigenvector
with lowest eigenvalue from a generalized eigenproblem derived from a finite-element discretization. 

\subsubsection{Residual minimization method} \label{sec:rmmdiis}

Variants of the residual minimization method of Wood and Zunger \cite{wood85JPA}
are used as eigensolvers in the plane-wave code {\bf VASP} \cite{kresse96PRB,kresseCMS96}
as well as in the FD-PAW package {\bf GridPaw} \cite{mortensenPRB}. The basic idea is
to update the orbitals at each iteration by taking a step along the preconditioned
residual $P{\bf R}_n$:
\begin{equation}
{\bf u}'_n = {\bf u}_n + \lambda P {\bf R}_n.
\end{equation}
The residual is defined as 
\begin{equation}
{\bf R}_n = (H - \varepsilon_n S) {\bf u}_n,
\end{equation}
where $\varepsilon_n$ is the present estimate 
of the eigenvalue computed as the Rayleigh quotient, and the preconditioner $P$ 
is discussed below.
$\lambda$ is chosen such that the residual ${\bf R}'_n$ at the new
guess ${\bf u}'_n$ 
\begin{equation}
{\bf R}'_n = (H - \varepsilon_n S) {\bf u}'_n = 
{\bf R}_n + \lambda(H - \varepsilon_n S)P{\bf R}_n
\end{equation}
is minimized. This amounts to finding the minimum of a second order
polynomial in $\lambda$. In the implementation of Ref.~\cite{mortensenPRB},
an additional step is then performed by taking also a step of length $\lambda$
in  the direction of $P{\bf R}'_n$
\begin{equation}
{\bf u}_n \leftarrow {\bf u}_n + \lambda P {\bf R}_n + \lambda P {\bf R}'_n.
\end{equation}
In the implementation of Refs. \cite{kresse96PRB,kresseCMS96} the residual is
minimized in a multidimensional space spanned by the present guess ${\bf u}_n$
and several preconditioned residuals from previous iterations - thus the name
RMM-DIIS (Residual metric minimization with 
direct inversion in the iterative subspace). 
The general idea of minimizing a (preconditioned) residual in a multidimensional space 
spanned by previous residuals is  the same as in the well-known
Pulay (DIIS) mixing schemes \cite{pulay80CPL,pulay82JCC,kresse96PRB,kresseCMS96,bowler00CPL}.

The optimal preconditioned residual would be obtained by solving
$\tilde{\bf R}_n= P {\bf R}_n$ from $(H - \varepsilon S)\tilde{\bf R}_n = {\bf R}_n$.
In Ref.~\cite{mortensenPRB} one solves instead approximately the simpler equation
$-\frac12\nabla^2\tilde{\bf R}_n = {\bf R}_n$ with the aid of a single multigrid
V-cycle.

\subsubsection{Diffusion algorithm}   \label{diffusionsec}

In connection with the program package developed in
the group of Prof. Eckhard Krotscheck (see Sec.~\ref{sec:Linz})
the lowest $n$ solutions of the single-electron 
Schr\"odinger equation are solved by applying the evolution operator,
$\mathcal{T}(\epsilon)\equiv\text{e}^{-\epsilon H}$,
repeatedly to a set of states $\{\psi_j,\, 1\leq j\leq n\}$, and
orthogonalizing the states after every step. Instead of the commonly
used second-order factorization in combination with the Gram-Schmidt
orthogonalization, the fourth-order factorization for the
evolution operator~\cite{imstep} is used. It is given by
\begin{equation}
\mathcal{T}^{(4)}\equiv\text{e}^{-\frac{1}{6}\epsilon V}\text{e}^{-\frac{1}{2}\epsilon T} \text{e}^{-\frac{2}{3}\epsilon\tilde{V}}\text{e}^{-\frac{1}{2}\epsilon T}
\text{e}^{-\frac{1}{6}\epsilon V}=
\text{e}^{-\epsilon[H+\mathcal{O}(\epsilon^4)]};  \qquad \tilde{V} = V + \frac{1}{48} \epsilon^2 [V,[T,V]],
\end{equation}
where one operates with the potential energy operators $V,\tilde{V}$ in real-space, and 
with the kinetic energy operator $T$ in the $k$ space. 
Switching between spaces is expedited via fast fourier transforms (FFT).
Instead of Gramm-Schmidt orthogonalization, the Hamiltonian is diagonalized 
in the subspace of present approximate solutions and a new set of orthonormal
states is thereby obtained. This step, often referred to as a subspace rotation, 
is in fact an application of the Petrov-Galerkin method of Sec~\ref{sec:galerkin} 
with the present approximate solutions as basis functions.

Depending on the physical system this method can be faster by up
to a factor of 100 in comparison to the second-order factorization.

\section{Six projects} \label{sec:efforts}

This paper collects together material from essentially three separate
lines of work.  One of them is the MIKA project (see
Sec.~\ref{sec:mika}), which is in fact an umbrella for a number of
activities related to real-space methods.  The first priority in the
MIKA-project is on the expansion of these activities to include also
time-dependent problems.  The MIKA-project has already spawned
important international collaborations with two other projects (see
Secs.~\ref{sec:Linz} and \ref{sec:gridpaw}).
Another line of work is defined by its aim to promote the applications
of the finite-element methods in electronic structure calculations.
Actual work related to the finite-element approach has been performed
with two codes - one of them is the well established general purpose {\bf Elmer} 
package (see Sec.~\ref{sec:elmer})  developed at CSC, and the other one has
been recently developed for transport calculations (see Sec.~\ref{sec:paula}).
The third and thus far entirely independent line is centered around 
the wavelet approach (see Sec.~\ref{sec:wavelet}). 

\subsection{MIKA - a project and a program package \label{sec:mika}}

The finite-difference program package MIKA has been described in
Refs. \cite{torsti03IJQC,torsti04PSIK,torsti05nano}. Example
calculations using its components are described in those references as
well as in Secs. \ref{rsdotsec}, \ref{axialsec} and \ref{dopplersec}
of this article. The common denominator of these programs is the
utilization of the RQMG-method (Sec.~\ref{rqmgsec},
Ref. \cite{Heiskanen}) as an eigensolver.  The \mika-package can be
downloaded freely from the internet at {\bf
http://www.csc.fi/physics/mika/}.

The MIKA-project, on the other hand should not be understood
as too tightly bound to the {\bf MIKA}-package. Instead, it is 
a project for the development of real-space methods in general,
and currently the main emphasis is on the expansion of the activities
to dynamic phenomena described by the time-dependent density-functional
theory (TDDFT) \cite{runge84PRL}. We have chosen the {\bf GridPaw}-code
(Sec.~\ref{sec:gridpaw}) as the basis for this development, although
we, at the same time, monitor the development of the alternative 
FE-approach based on the {\bf Elmer}-package (Sec.~\ref{sec:elmer}).


\subsection{Diffusion-algorithm and response-function package} \label{sec:Linz}

The real-space program package developed at the University of 
Linz~\cite{newton,imstep,Collective,michael,magalg} for solving 
the KS equations can be thought as an 
alternative to the approach used in MIKA. Instead of using
multigrid methods, the eigenvalue problem is solved in this program by 
using a diffusion algorithm, namely a highly efficient fourth-order 
factorization of the evolution operator (see Sec.~\ref{diffusionsec}). 
Secondly, the number of self-consistent iterations is significally 
reduced by applying a response-function method.
Within the MIKA package, the response-function algorithm has  been implemented 
into \cyl \ (see Sec.~\ref{axialsec}).
The  research applications  of the two programs to quantum dot studies
have been rather similar, and the latest research results summarized
in Sec.~\ref{rsdotsec} have largely originated from collaborations 
between the two projects.

\subsection{GridPaw} \label{sec:gridpaw}

Also the  {\bf GridPaw}  code developed in 
the  Center for Atomic-scale Materials Physics (CAMP) of the Technical
University of Denmark is
a finite-difference program package for the Kohn-Sham
equations, in which strategies 
different from those in the {\bf MIKA} -package have been 
chosen \cite{mortensenPRB}.
Instead of norm-conserving pseudopotentials, the Projector
Augmented Wave (PAW) approach \cite{blochl94PRB,blochl03} 
is implemented. A very important
technical detail in this implementation is the 
utilization of the double-grid method \cite{ono99PRL}, discussed in Sec.~\ref{sec:fd}.
Instead of the RQMG method,
the residual minimization method with multigrid preconditioning is applied to solve
the eigenvalue problem as described in Sec.~\ref{sec:rmmdiis}. 
Within the second phase of the MIKA-project, a
close collaboration with the GridPaw project is essential, as 
described in Sec.~\ref{jatkoprojsec}.

\subsection{Elmer} \label{sec:elmer}

Elmer is a general purpose finite element solver and collection of
library subroutines for solving systems of partial differential equations
by the finite element method.

Given a finite element partition of an $n$-dimensional computational domain,
the program automatically constructs graphs for system matrices in the
compressed row storage format, optimizes bandwidth, builds up the quadtree
search structures and, if required, the restriction and extension
operators for projecting functions between different levels of finite
element spaces in multigrid analysis.

The program provides interpolation basis functions and their gradients for simplices
and $n$-cubes for any given polynomial degree up to ten, and automatically
determines the optimal quadrature for numerically evaluating their inner
products. The local element matrices representing the disretized differential
operators are finally assembled in global structures using the predetermined
graphs. The global system is solved either with standard multifrontal
decomposition methods or Krylov-type iteration schemes. The iterative
solvers may be preconditioned by standard incomplete decompositions,
or multigrid methods. For eigenvalue problems the program utilizes the
Arnoldi-method as implemented in the  {\bf Arpack} -package \cite{arpack} 
in the shift-and-invert mode, with its own linear
system solvers applied to the computationally expensive invertion-step
of the algorithm.

Basically, the program is well suited for solving problems that can be
posed in variational form. The classical Kohn-Sham scheme of DFT,
for example, fits into this framework extremely well. So far, we have
performed some preliminary tests for computing the electronic
charge density for simple molecules like CO and C$_{60}$ (see Sec.~\ref{elmersec}).
The results of these all-electron calculations are promising, but there is some work
that could be done to enhance the overall performance of the code. The
major problems are not in the finite element discretization itself,
but in the performance of the eigenvalue solver, and the relaxation
techniques needed to avoid the charge sloshing phenomenon of the 
Kohn-Sham scheme. Here one can directly utilize experience gathered
using more traditional discretization schemes (see Sec.~\ref{sec:solve}).

For more information, see {\bf http://www.csc.fi/elmer/}.

\subsection{Finite-element method for electron transport}  \label{sec:paula}

The program for modeling transport properties of the nanostructures is
written in the Laboratory of Physics within a co-operation with the
Institute of Mathematics both in the Helsinki University of
Technology. The project is in detail presented in the thesis
\cite{Paulan_vaikkari}.

A nanostructure between leads is a typical electron transport
problem. This kind of systems are problematic to the DFT with
eigenfunction methods, because the system is of infinite size without
periodicity. We have solved this problem using the Green's function
methods with the open boundary conditions. The code has three
versions: one-, two-, and three-dimensional, so that different types
of nanostructures can be modeled.
The numerical implementation is done using the finite-element method
with $p$-elements up to the fourth order. We use mesh generator {\bf
Easymesh} \cite{easymesh} in the two-dimensional calculations, and
{\bf Netgen} \cite{netgen} in the three-dimensional ones.

\subsection{Wavelet package} \label{sec:wavelet}

The interpolating wavelet package for the computation of the
atomic orbitals has been written in the Institute of Physics
in the Tampere University of Technology in co-operation with
the Mathematics Division of the Faculty of Technology from the
University of Oulu. The aim of the effort is to develop the
wavelet method for the quantum mechanical applications.
Instead of using the orthogonal Daubechies wavelets the
biorthogonal interpolating wavelets have been utilized in the
implementation of the code. Up to now the atomic orbitals have
been computed for some light many-electron atoms (ions). In
the implementation we have used the nonstandard operator
form, since it provides an efficient method to carry out the
multi-resolution analysis in the electronic structure
calculations.

\section{Calculation examples} \label{sec:calculations}
In this section we present example calculations from each of the three
lines of work. The MIKA-project emphasizes the finite-difference
method, and thus far has also been centered on the {\bf MIKA}-package
\cite{torsti04PSIK,torsti05nano}.
Applications of this package to quantum dots, surface nanostructures
and positron calculations are presented in Sections \ref{rsdotsec},
\ref{axialsec} and \ref{dopplersec}, respectively. Note, however, that
many of the results reported in Sec.~\ref{rsdotsec} have been
calculated with the response function package described in
Sec.~\ref{sec:Linz}. The introduction of the finite-element method to
electronic structure problems materializes in Sec.~\ref{elmersec},
where all-electron calculations for CO and C$_{60}$ are presented, and
in Sec~\ref{greensec}, where transport calculations based on the
nonequilibrium Green's functions with norm-conserving pseudopotentials
are presented. Preliminary results of our implementation of the wavelet
approach to electronic structure calculations are presented in
Sec.~\ref{sec:waveletcalc}.



\subsection{Recent real-space calculations on two-dimensional quantum dots} \label{rsdotsec}



In this Section we present a brief review on our recent 
computational results for two-dimensional quantum dots 
(QD's)~\cite{quantumdots}. 
We consider QD's fabricated at the interfaces of
semiconductor heterostructures (e.g. GaAs/AlGaAs), where the
two-dimensional electron gas (2DEG) is restricted.
The many-electron Hamiltonian for a QD in the presence of a magnetic
field is written in SI units as
\begin{equation}
H = \frac{1}{2m^*}\sum^N_{i=1}\left[-i\hbar\nabla_i+e\mathbf{A}
({\mathbf r}_i)\right]^2
+\sum^N_{i<j}\frac{e^2}{4\pi\epsilon_0\epsilon|{\mathbf r}_i-{\mathbf r}_j|}
+ \sum^N_{i=1}\left[V_{\rm ext}({\mathbf r}_i)+g^*\mu_BBs_{z,i}\right],
\label{qdhami}
\end{equation}
where the vector potential is chosen in the symmetric gauge
to define the magnetic field perpendicular to the dot plane.
We use the effective-mass approximation with the
material parameters for GaAs, i.e., the effective mass $m^*$=0.067
$m_e$ and the dielectric constant $\epsilon=12.4-13$.
The external confining potential is determined by 
$V_{\rm ext}({\mathbf r})$ and the
last term is the Zeeman energy.

In the calculations we apply the SDFT in
the self-consistent KS formulation.
In high magnetic fields we have also employed
the computationally more demanding
current-spin-density-functional theory (CSDFT),
which does not, however, represent a considerable
qualitative improvement over the SDFT.
A detailed comparison between these two methods
for a six-electron quantum dot can be found in
Ref.~\cite{lsda}.

In the numerical process of solving the KS equations
we have used both the response-function package
(see Sec.~\ref{sec:Linz}) as well as the 2D version of the 
MIKA package (\rsdot). 
Recent QD applications studied using these methods
can be found in Secs.~\ref{statistics} and ~\ref{magnet},
respectively. 
Within the both methods, the effective 
single-electron Schr\"odinger equation is solved
on a two-dimensional point grid.
In practical calculations, the number of grid points
is set between 128 and 196 in one direction. This 
gives less than $\sim 1$ nm for a typical grid spacing, which
is sufficient for describing electrons in GaAs.
Within \rsdot, obtaining full convergence typically takes 
$100\dots 500$ self-consistency iterations. By
using the response-function methods this number can
be typically reduced by a factor of ten.

\subsubsection{Statistics of quantum-dot ensembles} \label{statistics}
 
We have applied the response-function algorithm presented in 
the previous section to study the statistical properties of quantum 
dots (in zero magnetic fields) 
affected by external impurities~\cite{statistics1,statistics2}. 
In the Hamiltonian given in 
Eq.~\ref{qdhami} the external potential $V_{\rm ext}({\mathbf r})$
consists of a parabolic confinement $V_{\rm c}(r)=m^*\omega_0^2 r^2/2$
with and the impurity potential
\begin{equation}
V_{\rm imp}({\mathbf r})=\sum^{N_{\rm imp}}_{k=1}\frac{-e}
{4\pi\epsilon\epsilon_0\sqrt{({\mathbf r}-{\mathbf R}_k)^2+d^2_k}},
\end{equation}
where $N_{\rm imp}$ is the number of impurities
and ${\mathbf R}_k$ and $d_k$ are their (random)
lateral and vertical positions in the ranges of
$0\leq R_k\leq 100\,{\rm nm}$ and $0\leq d_k\leq 10\,{\rm nm}$, respectively.
For each fixed $N_{\rm imp}=5\ldots 50$ we apply 1000 random 
impurity configurations.

In a noninteracting system the addition
energy for a certain electron number $N$ is equal to the
eigenlevel spacing, i.e., $\Delta_0(N)=\epsilon_{N/2+1}-\epsilon_{N/2}$,
where the divisor of two follows from the spin
degeneracy. We have shown that in this case the resulting 
addition-energy distribution is a
combination of Poisson and Wigner-Dyson distributions,
corresponding to regular and chaotic systems, 
respectively~\cite{statistics1}.

In the interacting many-electron system the addition
energy is given as the second energy difference,
$\Delta(N)=E(N-1)-2E(N)+E(N+1)$. The ground-state
energies are chosen from the spin states with
lowest energies. We calculated all the relevant
spin configurations for different electron numbers $N$,
so that taking the impurity configurations into account,
a total number of $\sim 10^5$ self-consistent SDFT 
calculations were performed. We note that this would not
have been manageable (in a reasonable time) using
conventional mixing schemes in solving the KS equations.

There is an essential difference in the
addition-energy behavior of noninteracting and interacting
systems as a function of $N_{\rm imp}$.
Namely, when the impurity number is increased the 
distribution of the interacting system becomes 
symmetrical with Gaussian tails.
This result is qualitatively similar to what has been found in  
experiments for large QD's~\cite{sivan,patel}, as well as with previous
DFT calculations for disordered QD's~\cite{hirose2,jiang}.
The result demonstrates
the necessity for comprehensive treatment of the
{\em e-e} interactions beyond the random matrix theory 
combined with the constant-interaction model. In our system, 
the approximate impurity density required for the symmetrization
of the addition-energy distribution is $\sim 10^{-3}\,{\rm nm}^{-2}$,
which is of the same order of magnitude as the one used by
Hirose {\em et al.}~\cite{hirose2} in their calculations of
disordered QD's.

\begin{figure}
\begin{center}
\epsfig{file=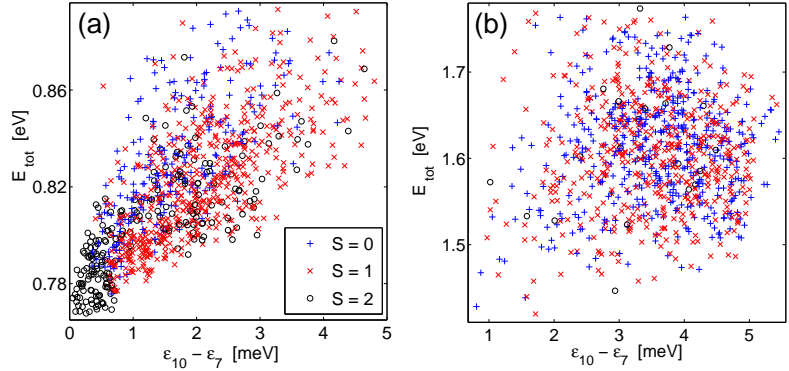,width=0.7\textwidth}
\end{center}
\caption{\label{corre}{(Color online) Total energy for different
spin states in a 16-electron quantum dot
vs. the sum of the noninteracting level
splittings on the fourth energy
shell. The number of impurities are
$N_{\rm imp}=5$ (a) and $N_{\rm imp}=20$ (b).
Reprinted with permission from E. R\"as\"anen and M. Aichinger, Phys. Rev. B 72, 045352 (2005).
Copyright (2005) by the American Physical Society. }}
\end{figure}

We have also analyzed the dependence of the spin states
on the impurity density and on the corresponding noninteracting
level statistics~\cite{statistics2}. Figure~\ref{corre}
shows the ground-state spins ($S=0,1,2$) for a 16-electron
QD with four impurity numbers $N_{\rm imp}=5$ (a) and $20$ (b).
The results are shown in a plane spanned by the sum
of the level splittings on the fourth shell,
$\epsilon_{10}-\epsilon_7$, and the total energy $E_{\rm tot}$.
The fraction of the $S=2$ states strongly decreases as
$N_{\rm imp}$ is increased. However, the relation of the
$S=0$ and $S=1$ states gradually saturates toward one.
The saturation effect is analyzed in Ref.~\cite{statistics2} in
detail. In the case of a small impurity density, the
resulting spin states show strong statistical dependence
on the level splittings: when the level spacing is small,
partial spin polarization is probable due to Hund's
rule. This applies especially for $S=2$ states which
are clustered in Fig.~\ref{corre}(a). However, when the impurity
density is increased the spin-state correlation
totally disappears. This indicates that in strongly
distorted systems the spin of the many-electron ground state can not be
predicted from the single-electron spectrum due to
the complicity of the e-e interactions.

\subsubsection{Quantum Hall regime} \label{magnet}

The real-space method based on \rsdot has proven to be a reliable and
efficient density-functional approach to study finite 2D electron
systems in magnetic
fields~\cite{henricluster,henristability,henrienergy,esagiant,henrimagnet}.
The original inspiration for our work is related with the well-known
quantum Hall effect in 2DEG~\cite{chakraborty}, as well as in the
electron transport~\cite{oosterkamp} and
magnetization~\cite{markkumusta} measurements, which both have
revealed a rich variety of transitions in the energetics of finite
electron systems as a function of the magnetic field.

\begin{figure}
\begin{center}
\epsfig{file=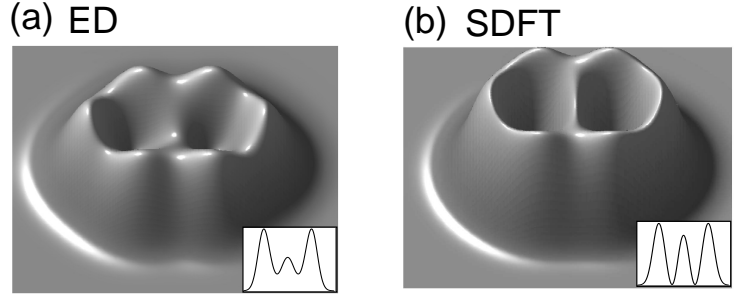,width=0.65\textwidth}
\end{center}
\caption{\label{ellipse}{Electron densities (gray scale) 
for two-vortex solutions in
elliptical quantum dots (${\rm eccentricity}=1.2$) calculated
using the ED (a) and SDFT within \rsdot (b).
Reprinted with permission from H. Saarikoski et al., 
Phys. Rev. B 71, 035421 (2005). Copyright (2005) by the American Physical Society. }}
\end{figure}

We have recently shown using both the SDFT and the exact
diagonalization (ED), that in QD's representing {\it finite-size}
quantum Hall droplets at strong magnetic fields the off-electron zeros
may localize~\cite{henricluster}.  In a parabolic confinement,
vortices form stable, clustered configurations with successive
transitions between them, as the magnetic field (and thus angular
momentum) is increased~\cite{henricluster,manninen}.
The mean-field SDFT densities and the conditional wave 
functions of the ED results shows similar vortex 
structures. Furthermore, QD's defined by a non-circular, e.g., elliptic
or rectangular confining potential, exhibit a good
agreement between the total densities obtained with SDFT
and ED calculations, respectively~\cite{henristability}.
Figure~\ref{ellipse} 
shows the electron density 
obtained using the ED (a) and SDFT (b) for two-vortex solutions 
in elliptical QD's. In the SDFT results the vortices are 
strongly localized with the electron density close to zero 
in the core, which reflects the fact that the mean-field 
method is not able to include all the correlations of the exact 
many-body state. We have found similar differences between the 
ED and SDFT for the so-called giant-vortex states, 
which are formed if a quartic term (flatness) is included 
to the otherwise parabolic potential~\cite{esagiant}.

The state transitions in the quantum Hall regime are also 
visible as oscillations in the QD energetics, e.g., in the 
chemical potentials $\mu(N)=E(N)-E(N-1)$ (Ref.~\cite{henrienergy}).
At magnetic fields below the total spin polarization 
we have found finite-size counterparts of the integer and
half-integer quantum Hall states, as well as
a developing pattern of de Haas--van Alphen oscillations 
in the magnetization
$M(N)=-\partial E(N)/\partial B$ (Ref.~\cite{henrimagnet}). 
These results are qualitatively consistent with the experimental
magnetization data of large dot arrays~\cite{markkumusta}.
At higher magnetic fields the signatures of above 
discussed vortex formation inside the QD are clearly visible in the 
magnetization oscillations, and the agreement between the SDFT and QMC results 
is good. These oscillations should be observable in accurate magnetization
measurements of QD devices.

 \subsection{Surface nanostructures studied with \cyl}  \label{axialsec}

 \subsubsection{Calculations in axial symmetry}

Many interesting nanostructures, such as adatoms on the surfaces,
circular quantum dots and quantum corrals can be modelled in axial
symmetry. Numerically the problem is reduced from 3D to 2D which makes
the calculations drastically easier, and allows modelling of much
larger systems.

In an axially symmetric potential the Schr\"odinger equation written
in the cylindrical coordinates is separable. The Kohn-Sham orbitals are
given as a product of the angular and (\emph{r,z}) -dependent parts,
i.e.,

\begin{equation}
\label{eq:cylwave}
\psi_{mkn}(\mathbf{r})=e^{im\phi}U_{mkn}(r,z).
\end{equation}
The formalism is nice from the computational point of view as the
eigenstates with different $m$ and \textbf{$\mathbf{k}$} are automatically
orthogonal. This allows the problem to be splitted into various subproblems
and we can take the full advantage of massively parallel computing
environment.

It is noteworthy that we can make use of this special form of
wave functions also in the response-function scheme.
The density correction is determined from Eq.~(2.5) of Ref.~\cite{newton}, which reads
after writing out the dielectric function,

\begin{equation}
\Delta\rho^{\left(k\right)}(\mathbf{r})-\delta\rho^{(k)}(\mathbf{r})=2\sum_{p,h}\frac{\psi_{p}(\mathbf{r})\psi_{h}(\mathbf{r})}{\epsilon_{p}-\epsilon_{h}}\int d^{3}r'd^{3}r''\psi_{p}(\mathbf{r}')\psi_{h}(\mathbf{r}')V_{p-h}(\mathbf{r}',\mathbf{r}'')\delta\rho^{(k)}(\mathbf{r}'').\label{eq:cylrespo}\end{equation}
Inserting (\ref{eq:cylwave}) and rewriting the integrals in the cylindrical
coordinates we see that the integral over $r'$ happily vanishes unless
the wave functions share the same angular momentum quantum number. This
drastically reduces the number of terms in the above summation which
is welcome when one tries to solve the linear equation using an iterative
method such as conjugate gradient or GMRES where efficient calculation
of the matrix operator is essential. Furthermore, {}``particle-hole
interaction'' $V_{p-h}$ can also be crudely approximated by the
Coulomb part of the effective potential (remember that it does not
affect the final result, only the speed of the iteration process), i.e.,
$V_{p-h}(\mathbf{r},\mathbf{r}')\approx\frac{\delta V_{c}(\mathbf{r})}{\delta\rho(\mathbf{r}')}$.
Thus the integration over double primed coordinate gives just the electrostatic
potential {}``caused'' by the charge distribution $\delta\rho^{(k)}(\mathbf{r}')$,

\begin{equation}
\int d³r'\frac{\delta V_{c}(\mathbf{r})}{\delta\rho(\mathbf{r}')}\delta\rho^{(k)}(\mathbf{r}')=\int d³r'\frac{\delta\rho^{(k)}(\mathbf{r}')}{\left|\mathbf{r}-\mathbf{r}'\right|}=V(\mathbf{r}),\end{equation}
so that the time consuming second integration can be circumvented
by simply solving the corresponding Poisson equation,
$\nabla^2V(\mathbf{r})=-4\pi\delta\rho(\mathbf{r}),$ that can be done 
efficiently in the axial symmetry. With the response-function scheme
we have been able to solve self-consistently electron systems consisting
of over 10000 Kohn-Sham orbitals in approximately 100 cpu hours with
IBMpower4 processors.

\subsubsection{Cu(111) surface and structures}

The Cu(111) surface is an example of a crystallographic cut of the
material that places the Fermi energy of electrons propagating normal
to the surface inside the bulk bandgap. Conduction electrons on the
surface have energy close to the Fermi energy and are sandwiched on
the surface, unable to escape into the vacuum because of the potential
barrier, and forbidden to enter the bulk because of the band gap.
They can however move parallel to the surface and thereby form a kind
of two-dimensional electron gas. Adatoms deposited on the surface
can affect the surface electron distribution, which offers interesting
possibilities to study the properties of confined electrons, their
interaction with adsorbates and many-body physics in general.

We have used the \cyl \  software to study axial symmetric surface structures
such as single adatoms and circular quantum corrals. The Cu substrate
in our calculations is described by a 1D model potential; the
Cu(111) planes, perpendicular to our symmetry axis (z), are considered
to have uniform density, while in the z-direction we have an oscillating
periodic potential that mimics the periodic structure. This particular
model was proposed by Chulkov et al. \cite{chulkov-ps}. In addition
to the work function and the bandgap, the model potential produces
correctly the Shockley surface states and the image-potential states.
Previously, this model has been successfully used e.g. in studies
of dielectric response-functions and lifetimes of excited states \cite{chulk-selfe-im},
and electron confinement in a metallic slab on solid surfaces within 1D
self-consistent DFT calculations \cite{edu-slabs}. Once the surface
is constructed, suitable pieces of jellium can be added to mimic various
kinds of adsorbates.

\begin{figure}			  
\begin{tabular}{cc}
\includegraphics[
  height=0.25\paperwidth,
  keepaspectratio]{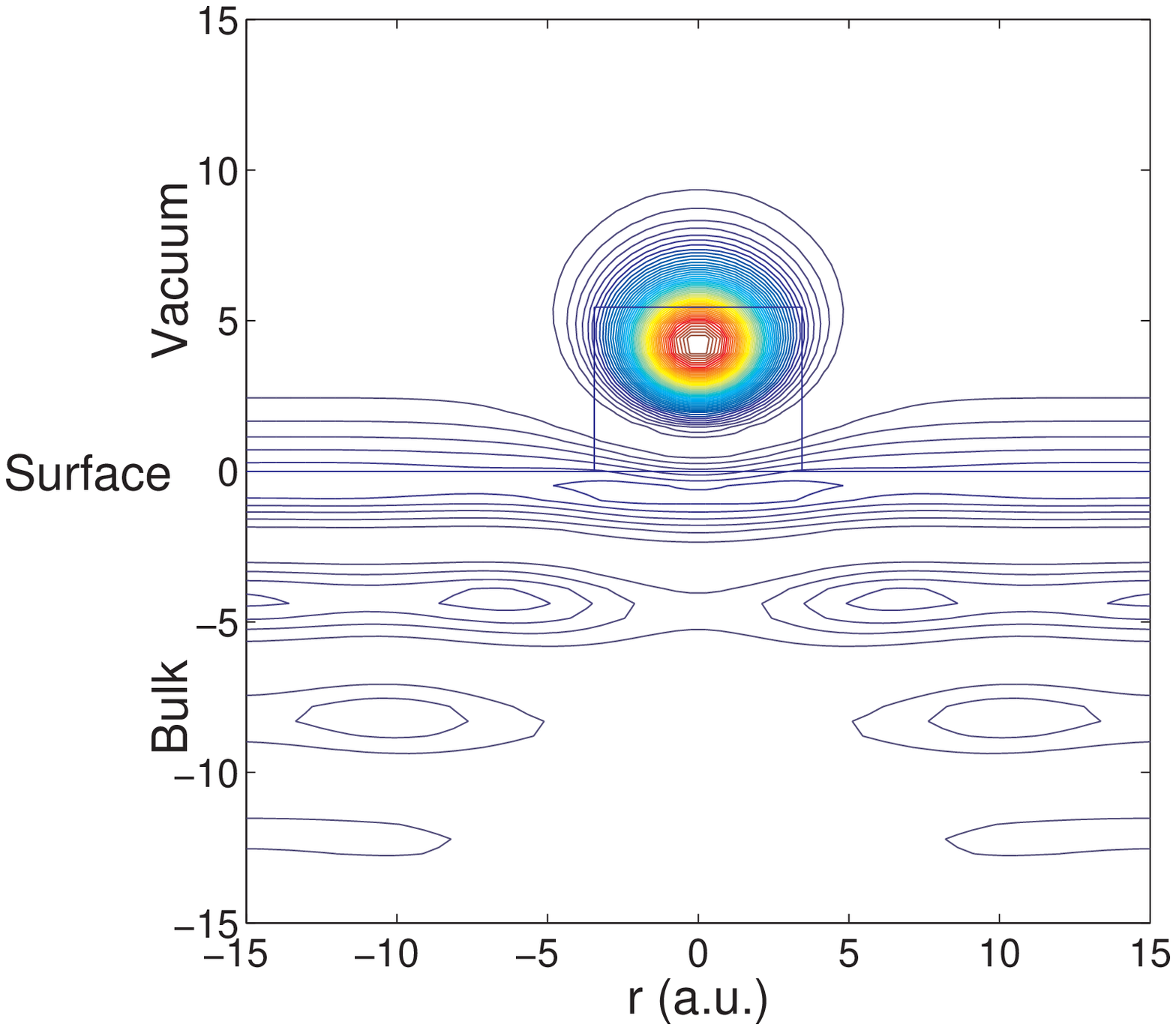}&
\includegraphics[		
  height=0.25\paperwidth,
  keepaspectratio]{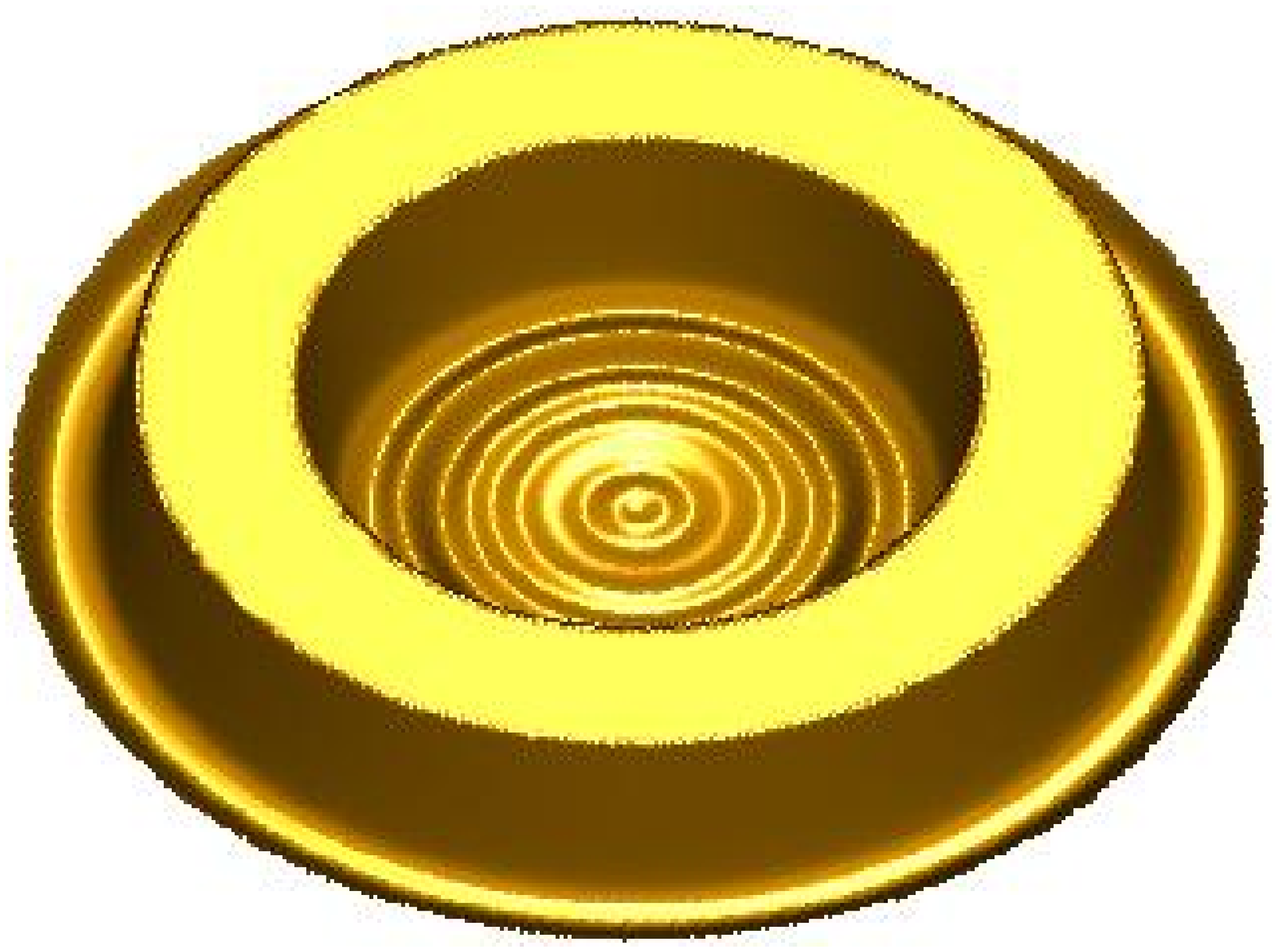}\tabularnewline
\end{tabular}

\caption{\label{cap:haksufig}Two example calculations with \cyl. Left:
modulus of the first surface state wave function localized at the
single Pb adatom deposited on the Cu(111) surface. Jellium profile 
of the adsorbate is pictured with the rectangular box. Right: 3D picture
of the oscillations of the surface electron density inside a small,
circular quantum corral.}
\end{figure}

So far we have studied e.g. the behaviour of surface state wave functions
when a single Pb-adatom is deposited on the surface. On the plain surface
the state is delocalized and has a constant amplitude along
the surface, where as with Pb-adatom, it is localised in
the Pb adsorbate (Fig.\ref{cap:haksufig}). In a more complex case
we have examined surface state electrons inside a circular ring of
adatoms deposited on the surface. These kind of quantum corrals have
been extensively studied experimentally and theoretically within the
scattering theory, where oscillations on the surface LDOS have been
observed \cite{heller}. Our calculations represent a somewhat different
viewpoint as the LDOS is calculated starting from the Kohn-Sham orbitals
instead of the scattering Green function.

				 
 \subsection{Positron calculations with \doppler  \label{dopplersec}}

\begin{figure}
\begin{minipage}[b]{0.49\textwidth}
\begin{center}
a)
\includegraphics[width=0.6\textwidth]{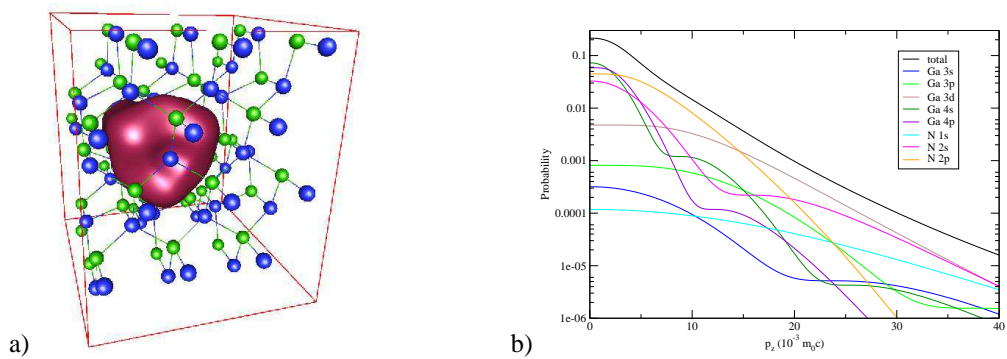}
\end{center}
\end{minipage}
\hfill
\begin{minipage}[b]{0.49\textwidth}
\begin{center}
b)
\includegraphics[width=0.85\textwidth]{spagetti}
\end{center}
\end{minipage}
\caption{a) An isosurface of the positron wave function localized at a Ga
vacancy in GaN. The Ga atoms are shown by green and the N atoms by
blue spheres. b) The total and decomposed Doppler spectrum for the Ga
vacancy in GaN calculated using the \doppler\ program. The
contributions of the individual orbitals of the Ga and N atoms are
shown by colored lines.}\label{VGa}
\end{figure}

The use of positron annihilation in defect studies is based on
the trapping of positrons from a delocalized bulk state to a
localized state at the defect (see Fig.~\ref{VGa}a). The trapping is due
to the reduced nuclear repulsion at the open-volume defects. Because
the electronic structure seen by the positron at the defect differs
from that in the  perfect bulk crystal the annihilation
characteristics change. The
positron lifetime increases because the average electron density
decreases. For the same reason the momentum distribution of annihilating
electron-positron pairs becomes more peaked at low momenta. However,
the positron density may sample the different atomic species of 
a compound material with different relative probabilities in the
bulk and at a defect. The defect may be surrounded by impurity
atoms. In these cases the high-momentum region of the distribution,
which is mainly due to annihilation with core electrons, reflects
the chemical structure of the defect. The changes
in the bond structure between the atoms neighboring the defect may
also affect the low-momentum part of the distribution. In order to
understand these changes and fully benefit from them in defect
identification, theoretical calculations with high predictive power
are indispensable.

The description of the electron-positron system can be formulated
as a two-component density-functional theory~\cite{boronski}. In the
measurements there is only one positron in the solid sample at a
time. Therefore, the density-functional scheme has to be
properly purified from positron self-interaction effects. Comparisons
between theoretical two-component DFT results and experimental results have
shown that  the following scheme is adequate. First, the electron
density $n(\mathbf{r})$
of the system is solved without the effect of the positron. This can be
done using different (all-electron) electronic structure calculation
methods. A surprisingly good approximation for the positron lifetime
and core-electron momentum calculations is to simply superimpose free atom 
charges. Then the potential $V_+(\mathbf{r})$ felt by positron is
constructed as a sum of the Coulomb potential $\phi(\mathbf{r})$ due
to electrons and ions and
the so-called (electron-positron) correlation potential
$V_{\mathrm{corr}}(\mathbf{r})$ which is treated in a local density
approximation (LDA). In the zero-positron-density limit it is a
function of the electron density $n_{-}(\mathbf{r})$ only, i.e.
\begin{equation}
V_+({\bf r}) = \phi({\bf r})+V_{\mathrm{corr}}(n_-({\bf r})),\label{pospot}
\end{equation}
The ensuing single-particle Schr\"odinger equation can be solved
using similar techniques as used for the electron states. For example, we
use the three-dimensional real-space Schr\"odinger equation solver
of the {\tt MIKA} package. When using self-consistent electronic charge
density we use also the Poisson equation solver of the {\tt MIKA} package to
calculate in Eq.~(\ref{pospot}) the Coulomb potential due to valence electrons.

The scheme described above is for a delocalized positron the exact
zero-positron-density limit of the two-component DFT. However, the
approximation made in the case of a localized positron can be
justified by considering the positron and its screening cloud as a
neutral quasiparticle which does not affect the average electron density.

When the electron density $n(\mathbf{r})$ 
and the positron density $n_+(\mathbf{r})= \vert \psi^+(\mathbf{r})
\vert^2$ are known the positron annihilation rate is calculated within
the LDA (in the zero-positron-density limit) as an overlap integral
\begin{equation}
\lambda = \frac{1}{\tau}=\pi r_e^2c\int \mathrm{d}{\bf r}\,n_+({\bf r})n_-({\bf r})
\gamma(n_-({\bf r})),
\label{lambda}
\end{equation}
where $r_{e}$ is the classical electron radius, $c$ the speed of
light, and $\gamma$ the enhancement factor taking into account
the pile-up of electron density at the positron (a correlation effect).
The inverse of the annihilation rate is the positron lifetime $\tau$.

We calculate the momentum distribution of the annihilating
electron-positron pairs using the so-called state-dependent
enhancement scheme~\cite{Alatalo96} as
\begin{equation}
\rho ({\bf p}) = 
\pi r_e^2c\sum_j \gamma_{j}\bigg\vert\int \mathrm{d}{\bf r}\,e^{-{\rm i}{\bf p}\cdot {\bf r}
}
\psi^{+}(\mathbf{r})\psi_{j}(\mathbf{r})\bigg\vert^2,
\label{momdis}
\end{equation}
where the state-dependent enhancement factor is written as
$\gamma_{j}=\lambda_{j}/\lambda_{j}^{\mathrm{IPM}}$. $\lambda_{j}$ is
the annihilation rate of the state $j$ within the LDA,
\begin{equation}
\lambda_{j}=\pi r_{e}^{2}c\int\mathrm{d}\mathbf{r}\,n_{+}(\mathbf{r})n_{j}(\mathbf{r})\gamma(n_{-}(\mathbf{r})),
\end{equation}
and $\lambda_{j}^{\mathrm{IPM}}$ is the annihilation rate of the state
$j$ within the independent-particle model (IPM, $\gamma\equiv
1$). Above, $n_{j}(\mathbf{r})=|\psi_{j}(\mathbf{r})|^{2}$ is the
electron density of the state $j$. One can directly compare
computational results with experimentally measured Doppler
broadening of the 511~keV annihilation line or with experimentally
measured angular correlation of the annihilation gammas.
We have found that the use of the commonly employed position-dependent
enhancement factor [enhancement taken into account with the factor
$\sqrt{\gamma(n_{-}(\mathbf{r}))}$ inside the Fourier transform in
Eq.~(\ref{momdis})] leads to unphysical results when one compares the
ratio of two Doppler spectra with the experiment~\cite{Makkonen05b}.

The \doppler\ program uses the
atomic superposition method. One can either use the LDA
parametrizations (enhancement factor and correlation potential) by
Boro\'nski and Nieminen~\cite{boronski} or the gradient-corrected
scheme by Barbiellini \textit{et
al}~\cite{Barbiellini95,Barbiellini96}. However, the atomic
superposition method cannot be used for the low-momentum part due to
valence electrons. For that purpose self-consistent all-electron
valence wavefunctions have to be constructed. For example, we have
used the projector augmented-wave (PAW) method implemented in the
plane-wave code Vienna \textit{Ab initio} Simulation Package (VASP)
\cite{kresse96PRB,kresseCMS96,Kresse99,Makkonen05,Makkonen05b}. Recent
applications of the real-space solvers of the {\tt MIKA} package to positron
studies include, for example, a study of vacancy-impurity complexes in
highly Sb-doped Si~\cite{Rummukainen05} in which computational Doppler
spectra were used to identify experimentally detected unknown
vacancy-type defects and and a study of the effects of positron
localization on the Doppler broadening of the annihilation line in the
case of Al~\cite{Calloni05}.

Figure~\ref{VGa}a shows an example of a positron state calculated for
a Ga vacancy in GaN. The corresponding Doppler spectrum calculated
with the \doppler\ program is shown in Fig.~\ref{VGa}b. The figure shows
also the decomposed momentum distributions corresponding to annihilations with
electrons on different orbitals. Their weights are calculated with Eq.~(\ref{momdis}).

The \doppler\ program also enables one to calculate the forces on
ions due to a localized positron within the atomic superposition
approximation. These forces can be used together with electron-ion and ion-ion
forces calculated with standard methods in order to find the relaxed
ionic configuration of a vacancy-type defect at which there is a
localized positron~\cite{Makkonen05b}.

\subsection{All-electron finite-element calculations} \label{elmersec}

\begin{figure}
\centerline{\includegraphics[width=0.2\textwidth]{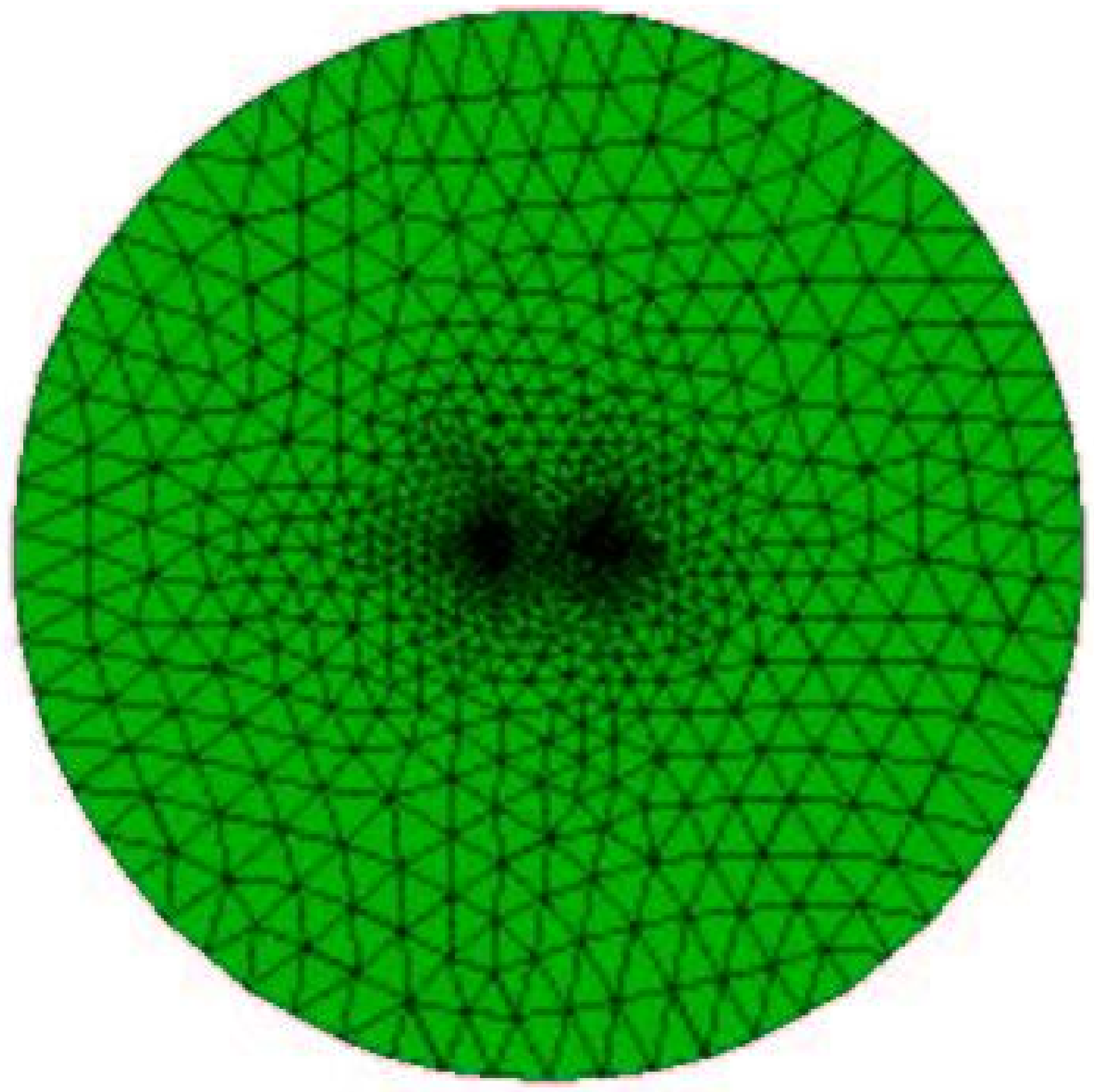}
            \hspace{0.7cm}
            \includegraphics[width=0.2\textwidth]{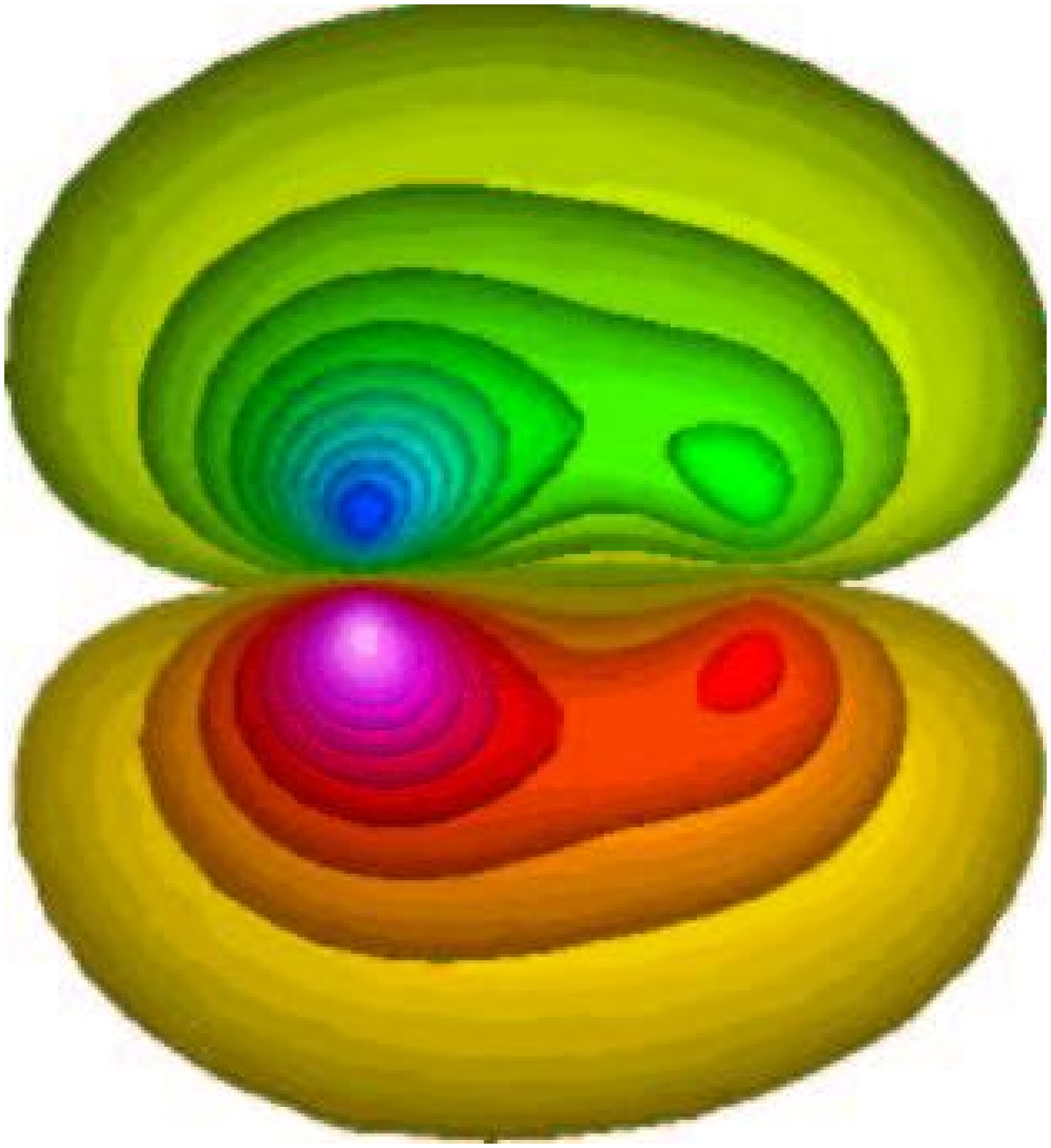}
            \hspace{0.7cm}
            \includegraphics[width=0.2\textwidth]{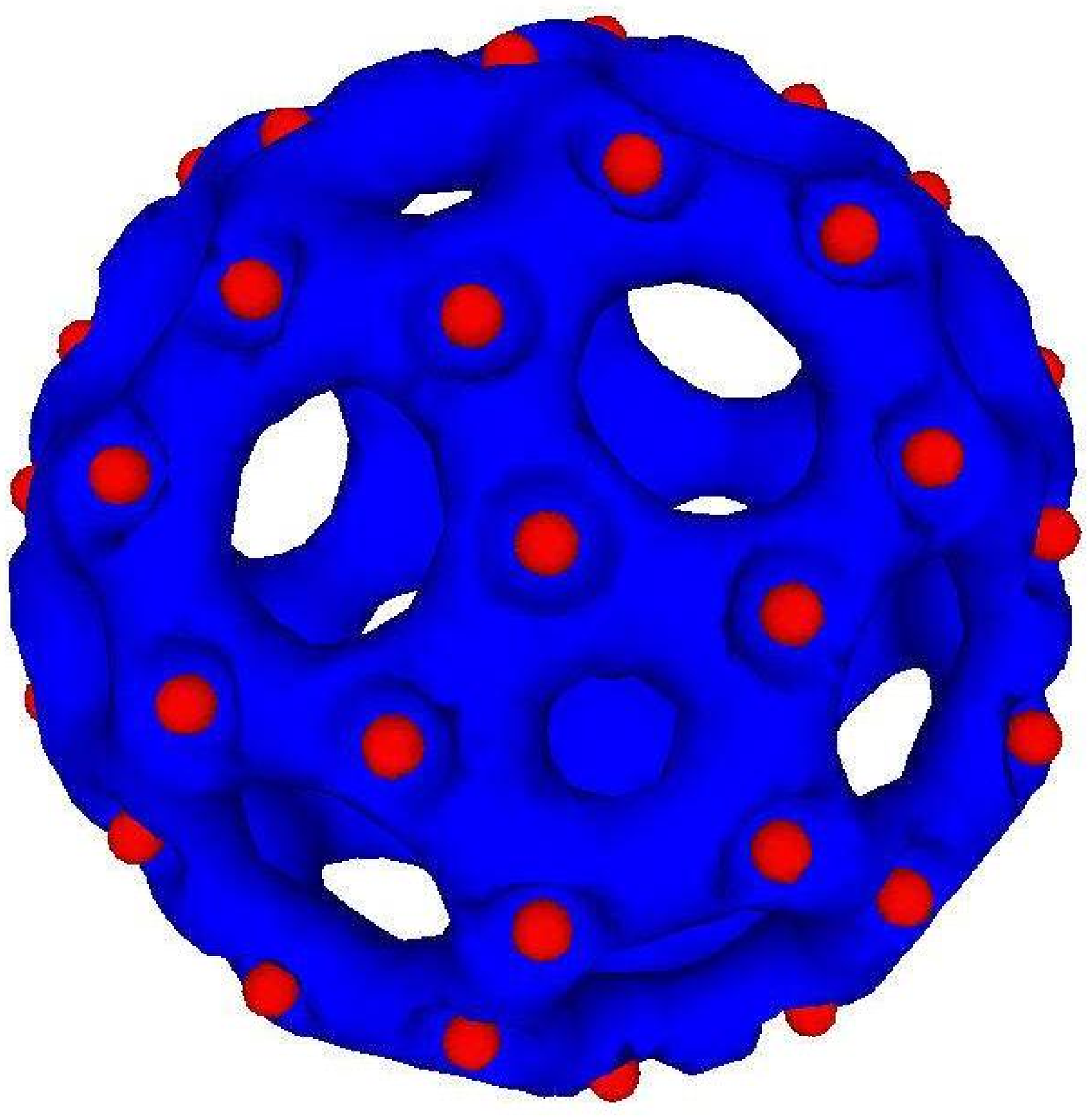}}
\centerline{\vspace{0.5cm}}
\centerline{\includegraphics[width=0.2\textwidth]{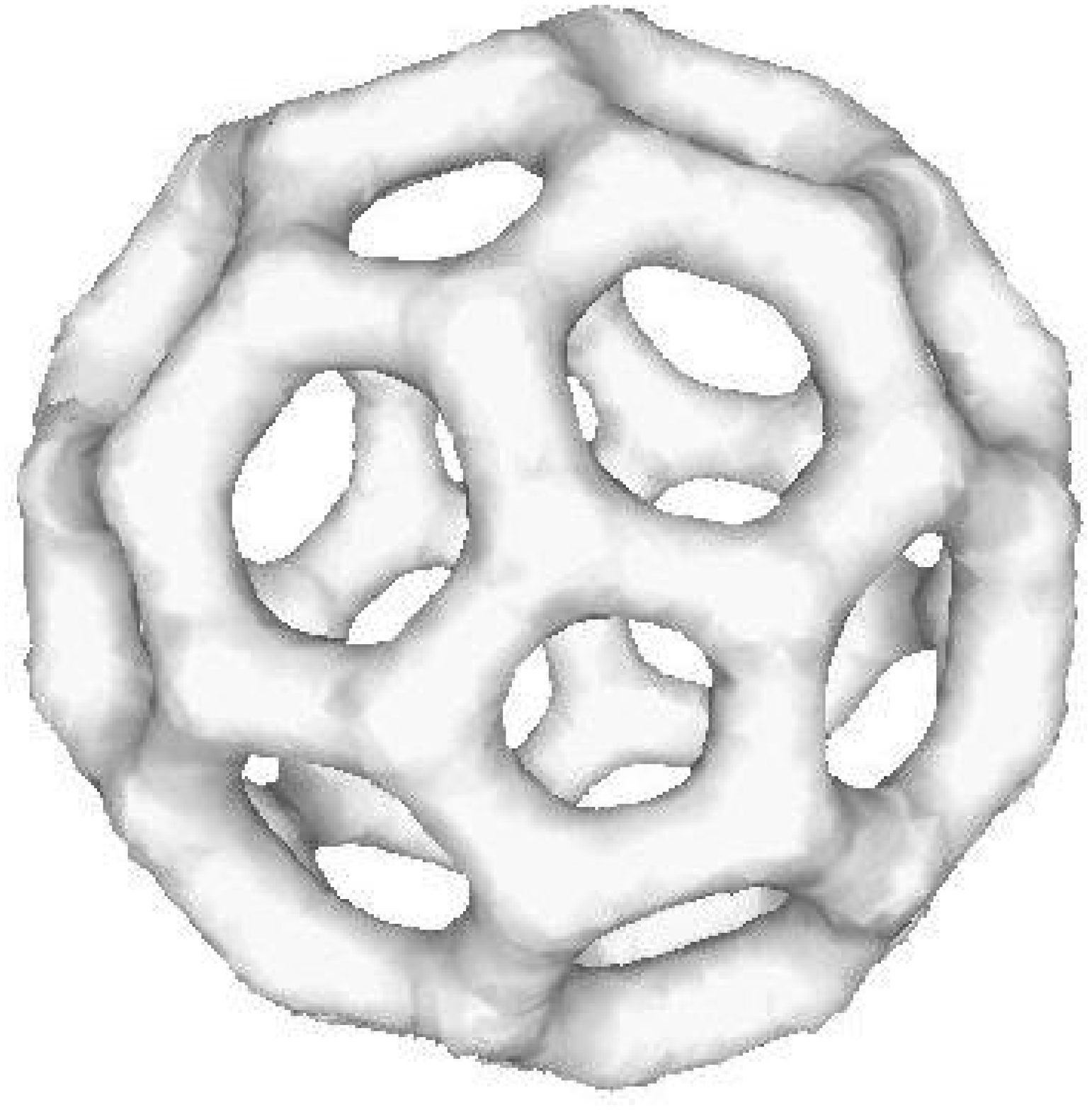}
            \hspace{0.7cm}
            \includegraphics[width=0.2\textwidth]{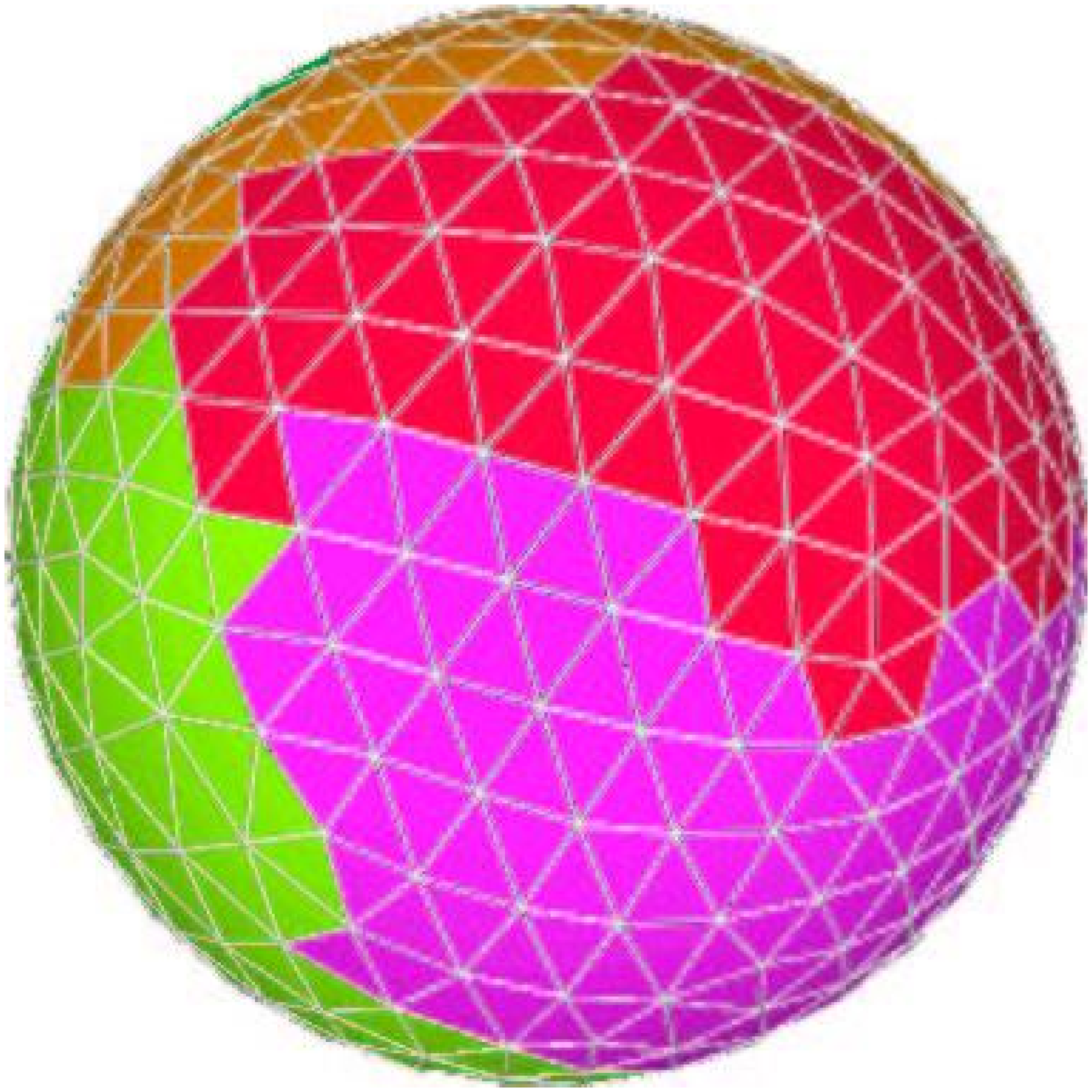}
            \hspace{0.7cm}
            \includegraphics[width=0.2\textwidth]{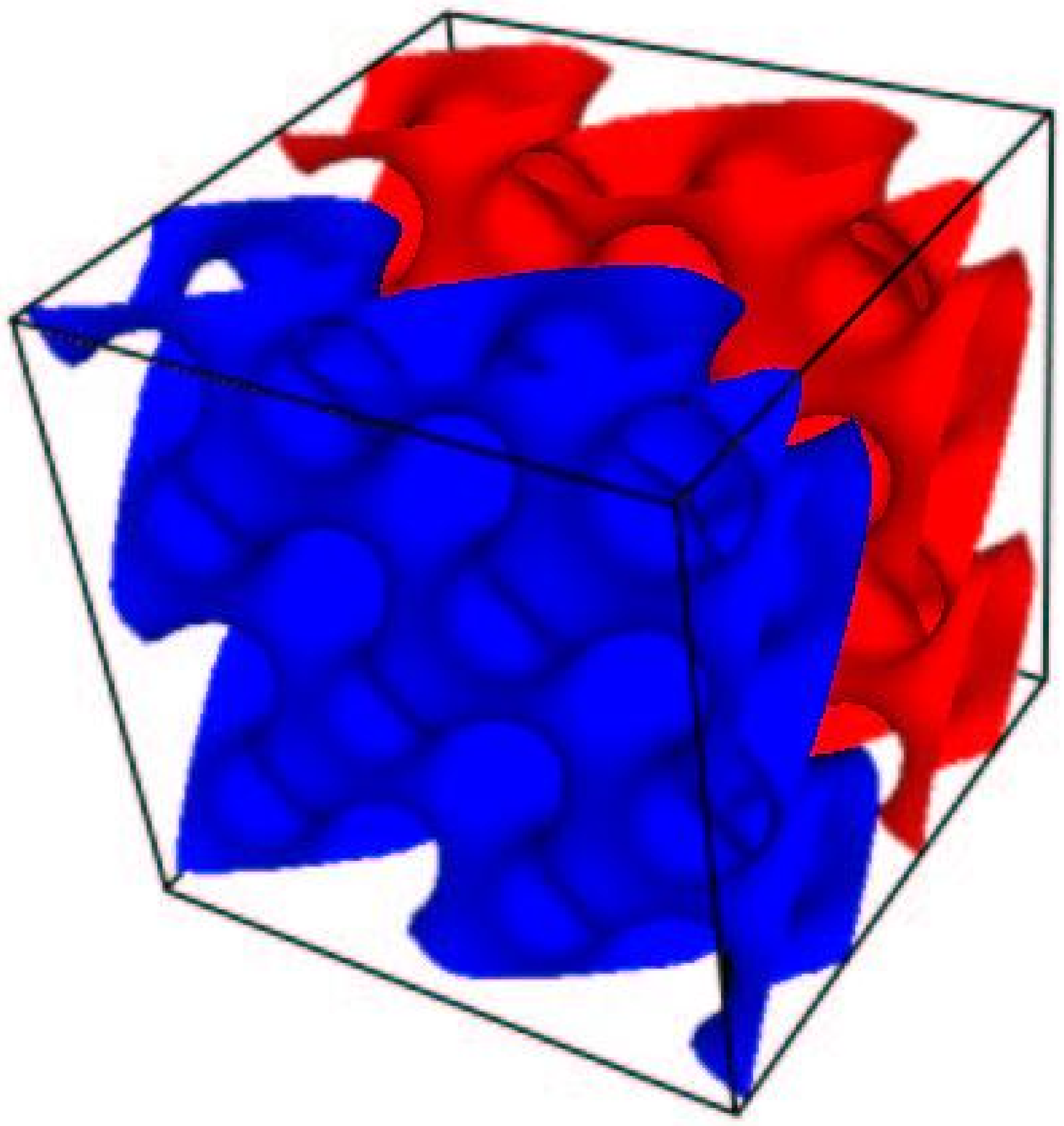}}

\caption{\label{elmerfig} Some electronic structure calculations using
{\bf Elmer}. All-electron calculations for the molecules CO and	%
C$_{60}$ were performed. Implementation of periodic boundary
conditions was tested  in the case of bulk silicon. See also the text.}
\end{figure}

Although the main emphasis of the MIKA-project has been thus far on the support
and development of the MIKA-package, which is based on finite-differences, it has been
interesting to perform some simple test calculations using the general purpose 
finite-element package {\bf Elmer} (Sec.~\ref{sec:elmer}). Some of these calculations
have been already reported in Ref.~\cite{tuomas03thesis}.
 

Fig.~\ref{elmerfig} illustrates
three test cases. 
As the first self-consistent calculation within the local-density
approximation of the density-functional theory using {\bf Elmer}, an
all-electron calculation for the carbon monoxide molecule was performed.
The top left panel of Fig.~\ref{elmerfig} illustrates the  main features
of the selected
finite-element mesh\footnote{To be precise, the mesh in the picture is a two-dimensional
mesh and not a cutplane through the three-dimensional mesh, as we were not able to draw
such a cutplane}. Since the divergent ($1/r$) potential has to be
represented on the mesh, a very fine mesh is used in the immediate
neighbourhood of the singularity. The mesh was generated using 
{\bf Netgen} \cite{netgen}, consisted of 120 000 quadratic elements
and 170 000 node points. Each node point corresponds to one basis function.
To solve the Kohn-Sham eigenvalue problem, the {\bf Arpack} package 
was utilized, with incomplete LU-factorization \cite{ilu86} as
the preconditioner within the computationally expensive inversion 
part of the shift and invert step of the Lanczos method
 as discussed in Sec.~\ref{sec:lanczos} and Sec.~\ref{sec:iterative}.
The initial guess for the effective potential
was the sum of the bare nuclear potentials.  
We used an {\it ad hoc} linear mixing scheme, where the effective 
potential at each iteration was a linear combination of the input
and output potential, with an exponentially decreasing  weight for 
the input potential, resulting in fast convergence once the
decay parameters were  properly adjusted. In the middle panel of the upper row of
Fig.~\ref{elmerfig} a contour plot of a single-particle wave function
provided by the Kohn-Sham scheme is shown.  Also larger molecules can
be treated using this  all-electron scheme within
{\bf Elmer}. The top right panel of Fig.~\ref{elmerfig} shows a
selected orbital from an all-electron calculation of the
C$_{60}$-molecule. The lower left panel illustrates the electron
density of C$_{60}$, obtained from a parallel calculation
involving eight processors and $4 \times 10^5$ degrees of freedom.
In this calculation the preconditioner used in the CG-method of
the inversion step was a multigrid method.
The lower middle panel of
Fig.~\ref{elmerfig} illustrates the partitioning
of the mesh that was used -- domains with different colours were mapped
to different processors with the help of the  
{\bf Metis} program \cite{metis}. 
Periodic boundary conditions were implemented
within {\bf Elmer}, and in order to compare the computational
efficiency of the RQMG method and the Lanczos method, the Schr\"odinger
equation was solved (\ie\ a non-self-consistent calculation was	
done) in a periodic potential corresponding to bulk silicon using both
methods. A uniform grid consisting of 32$^3$ points was used in the
64-atom supercell.  The computational efficiencies of the two methods
were found to be similar. 
 The lower right panel of Fig.~\ref{elmerfig} illustrates
a selected eigenfunction from this periodic test case.

The calculation on carbon monoxide reported above, originally performed in 2002,
was revisited in 2005, taking advantage of the novel mesh generator 
{\bf GiD} \cite{gid}. This time only 28 000 second order 
elements (39 000 basis functions) were required. The accuracy of the computation
was checked by evaluating the dipole moment of the molecule, which is a well
known basis-set sensitive quantity. The result $-0.235 D$ was obtained, whereas
the basis-set limit for this quantity is $-0.226 D$ \cite{he00MP}. 

There is plenty of room for improvement
in the computational efficiency of our {\bf Elmer}-based solver, 
and the next steps required
seem to be straightforward.
Implementing norm-conserving pseudopotentials or PAW,
gathering experience on making meshes with 
higher-order $p$-elements, possibly with the order variable in space 
(these are now available in {\bf Elmer}) 
as well as paying more attention on the choice of the 
eigenproblem solver and mixing scheme
will improve the efficiency considerably. 

\subsection{Ballistic transport calculations using 
the Green's function
method \label{greensec}} 

In this section we present calculations of finite element
implementation for the transport properties of the nanostructures. 

In the DFT the electron density of a system is constructed from
eigenfunctions of the single-particle Hamiltonian. For
isolated systems, such as molecules or quantum dots, the whole
system can be included in the calculation volume. However, this is 
not possible in general and some approximations have to be done. In the
case of bulk materials the systems are treated as infinite by using
periodic boundary conditions. In transport problems the
nanostructure modeled is connected to two or more leads so that
the system size is infinite but does not have periodicity. A
solution to this problem is to calculate the electron density from 
the single-particle Green's function with the open boundary conditions
\cite{gre_datta}. This makes finite-size effects small.
The Green's function method has also other useful properties in transport
calculations. For example, it is possible to apply a bias voltage across 
the system and calculate the current through it.

Green's functions in transport calculations are heavy to calculate 
numerically which has delayed  their first-principles implementations 
until the recent years.  The computational methods have to be chosen 
carefully.  Typically one uses a special basis set, such as atomic
orbitals \cite{transiesta1,transiesta2}, tailored to describe the
equilibrium electronic properties of the system in question.  
Because a systematic error control is not straightforward with atomic 
orbitals, the development of other types of solvers is well motivated.
We have implemented the Green's function transport formalism using 
the FEM with $p$-elements up to the fourth order polynomials
\cite{oma3D}. The solver has one, two, and three-dimensional versions
so that different types of nanostructures can be modeled.  The other
implementations beyond the atomic-orbitals ones employ an
$O(N)$ optimized basis
\cite{bernholc}, a wavelet basis \cite{wavelet}, full-potential
linearized augmented plane-waves \cite{embedded}, maximally localized
Wannier functions \cite{wannier}, a finite-difference method
\cite{differenssi}, and a 
 finite-element method fith linear basis functions \cite{fem-green}.
\begin{figure}[htb!]
\begin{center}
\epsfig{file=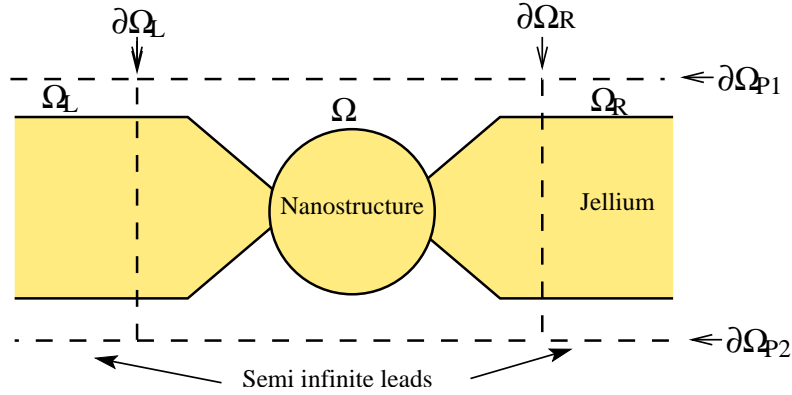,width=0.7\textwidth}
\end{center}
\caption{\label{nanorakenne2} Schematic sketch of modelling
a nanostructure. }
\end{figure}

A brief introduction to the Green's function method is based on the 
schematic sketch in Fig.~\ref{nanorakenne2}. The calculation region 
is divided into the three parts $\Omega_R$, $\Omega$, and $\Omega_L$. 
The nanostructure is located in $\Omega$ and it is connected to the two
leads $\Omega_R$ and $\Omega_L$ using the open boundary conditions on
$\partial \Omega_{L/R}$. This means that electrons can travel
through the boundaries $\partial \Omega_{L/R}$ without
reflection. The system is also open in the sense that during the
iterations towards the self-consistent solution the number of electrons 
in $\Omega$ is not a constant. On the boundaries 
$\partial \Omega_{P1/P2/P3/P4}$ Dirichlet's or  periodic boundary 
conditions are applied.

In the Green's function method the electron density is calculated from
the retarded Green's function $G^r$. This requires first the solving of the
equation
\begin{equation}\label{greenR}
\Big{(}\omega+ \frac{1}{2}\nabla^2-V_{\rm eff}({\bf r}) \Big{)} G^r({\bf
r},{\bf r}';\omega) = \delta({\bf r}-{\bf r}'),
\end{equation}
where $\omega$ is the electron energy and $V_{\rm eff}$ is the
effective potential in the system. The separation between the two
solutions $G^r$ and $G^a$ is made using the boundary conditions. In
practice, most of the calculation time is spent in solving the
corresponding equation. The calculation of the electron density requires
integration over the electron energy so that $G^r$ has
to be calculated at several energy values. However, when the
integration path is moved to the complex plane, the numerical
integration does not require many points. Moreover, the integration 
is easy to parallelize.

In transport calculations we are interested in the electron
current. In the Green's function formulation the current is calculated
using the Landauer type of formulation
\begin{equation}
I = \frac{1}{\pi} \int_{-\infty}^{\infty} T(\omega) \left(
f_{L}(\omega) - f_R(\omega) \right) d\omega.
\end{equation}
Note that the above tunneling probability $T(\omega)$ is not the real
probability in the sense that it can be larger than unity because it
includes the summation over the different conducting channels. A
channel can contribute to the conductance by one conductance quantum
at maximum. Similarly to the electron density, $T$ is also calculated 
using the $G^r$ function.

In order to use the FEM we first write the equations of the Green's
function method in the so-called variational form.  This is how also the
numerical form for the open boundary conditions can be derived. The
result is analogous to the truncated matrix derivation used, for example,
in the case of the finite-difference method \cite{datta}.  First we take a
nicely-behaving arbitrary function $v({\bf r})$ and multiply both
sides of Eq.~(\ref{greenR}) by it. Then we integrate over the
calculation domain $\Omega$. We use the properties of the open
boundary system and make some manipulations
\cite{fem_avoin_reuna}, which are shown in detail in Ref.
\cite{oma_johto}. After this the equation takes the form
\begin{equation}
  \label{eq:variational_formulation}
  \begin{aligned}
    \int_\Omega &\Big{\{} - \nabla v({\bf r}) \cdot \frac{1}{2} \nabla
    G^r ({\bf r,r'};\omega) + v({\bf r}) \big{[} \omega-V_{\rm eff}({\bf r})
    \big{]} G^r({\bf r,r'} ; \omega) \Big{\}} \,d{\bf r}\\ &- \langle
    \hat{\Sigma}_L G^r,v \rangle - \langle \hat{\Sigma}_R G^r,v
    \rangle = \, v({\bf r'}),
%
  \end{aligned}
\end{equation}
where the so-called self-energy operators of the leads have the form
\begin{equation}
  \label{eq:dton}
  \begin{aligned}
   &\langle \hat{\Sigma}_{L/R} G^r,v \rangle = \\ &\int_{\partial
  \Omega_{L/R}} \int_{\partial \Omega_{L/R}} \frac{1}{4} G^r({\bf
  r}_{L/R}', {\bf r'} ; \omega) \frac{\partial^2 g_e ({\bf r}_{L'/R'},
  {\bf r}_{L/R} ; \omega)} {\partial {\bf n}_{L/R} \partial {\bf
  n}_{L'/R'}}\, v ({\bf r}_{L/R}) \, d{\bf r}_{L'/R'} d {\bf r}_{L/R}.
  \end{aligned}
\end{equation}
The open boundary conditions of the system are included in the
self-energy operators. They are the surface integrals over the open
boundaries $\partial \Omega_{L/R}$. The function $g_e ({\bf
r}_{L'/R'},{\bf r}_{L/R};\omega)$ is the Green's function of the
isolated lead $\Omega_{L/R}$ with the zero boundary condition on the
boundary $\partial \Omega_{L/R}$ \cite{fem_avoin_reuna}. In this
equation $g_e ({\bf r}_{L'/R'}, {\bf r}_{L/R};\omega)$ is
differentiated with respect to both arguments in the direction of the
normal vector ${\bf n}_{L/R}$ on the surface $\partial \Omega_{L/R}$.

The solution for $G^r$ can be calculated from
\ref{eq:variational_formulation} by using the finite-element
approximation $G^r( {\bf r,r'};\omega) \approx \sum_{i,j=1}^N
g_{ij}(\omega) \, \phi_i({\bf r}) \, \phi_j({\bf r'})$ and choosing
$v({\bf r}) = \phi_p({\bf r})$.

The calculation of the $G^r$ function requires inverting of large
matrixes. We use a direct sparse solver for this purpose
\cite{hsl}. A direct solver routine requires more computer memory than
iterative methods. However, when one has to solve a lot of linear 
equations with the same coefficient matrix but different left hand sides 
direct routines are more efficient than iterative ones.

We have used our FEM implementation for modeling of two-dimensional
quantum point contacts \cite{oma07}, Na atomic wires \cite{oma3D}, and
thin HfO$_2$ layers \cite{oma3D}.

\begin{figure}[htb!]
\begin{center}
\epsfig{file=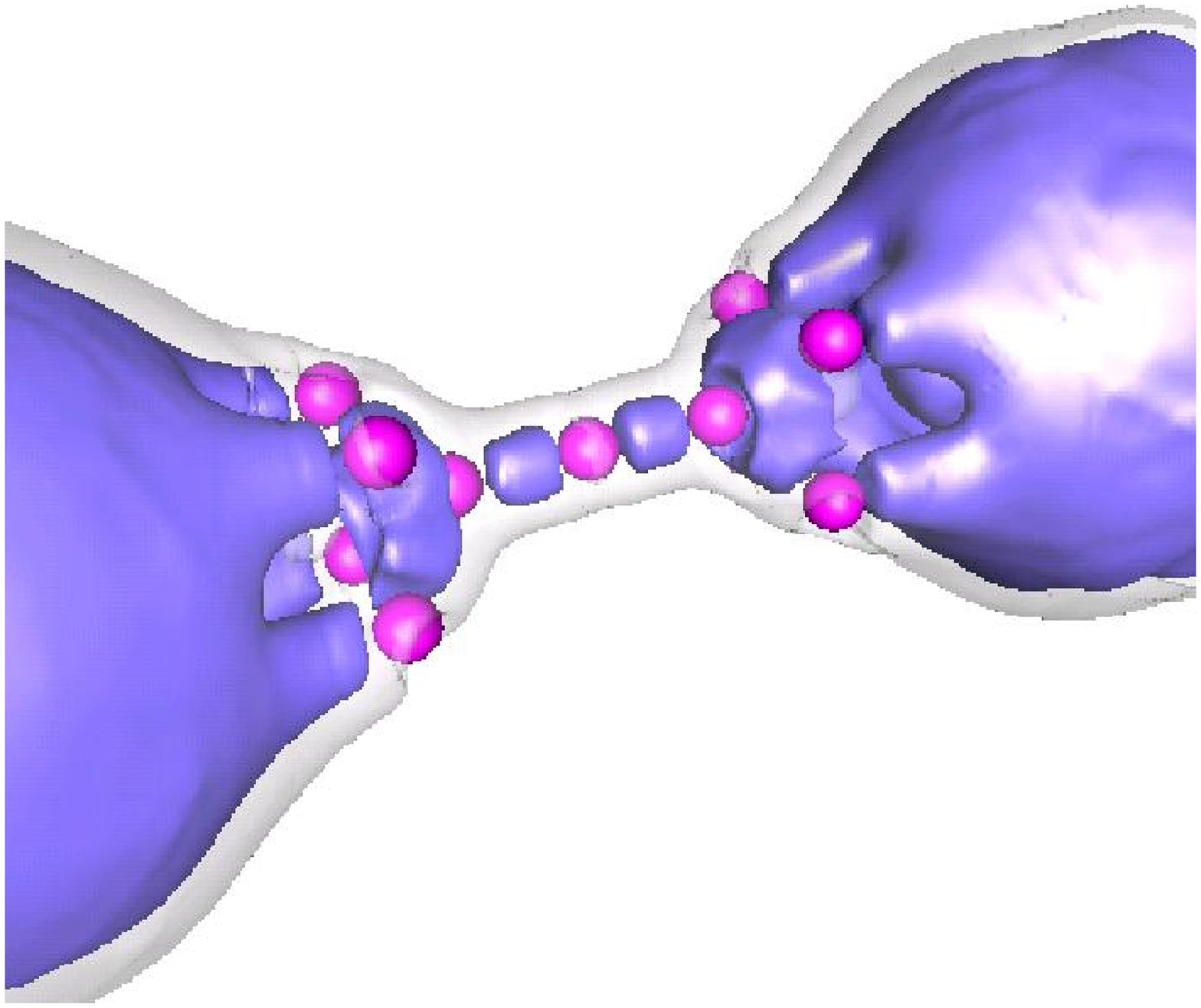,width=0.48\textwidth}
\epsfig{file=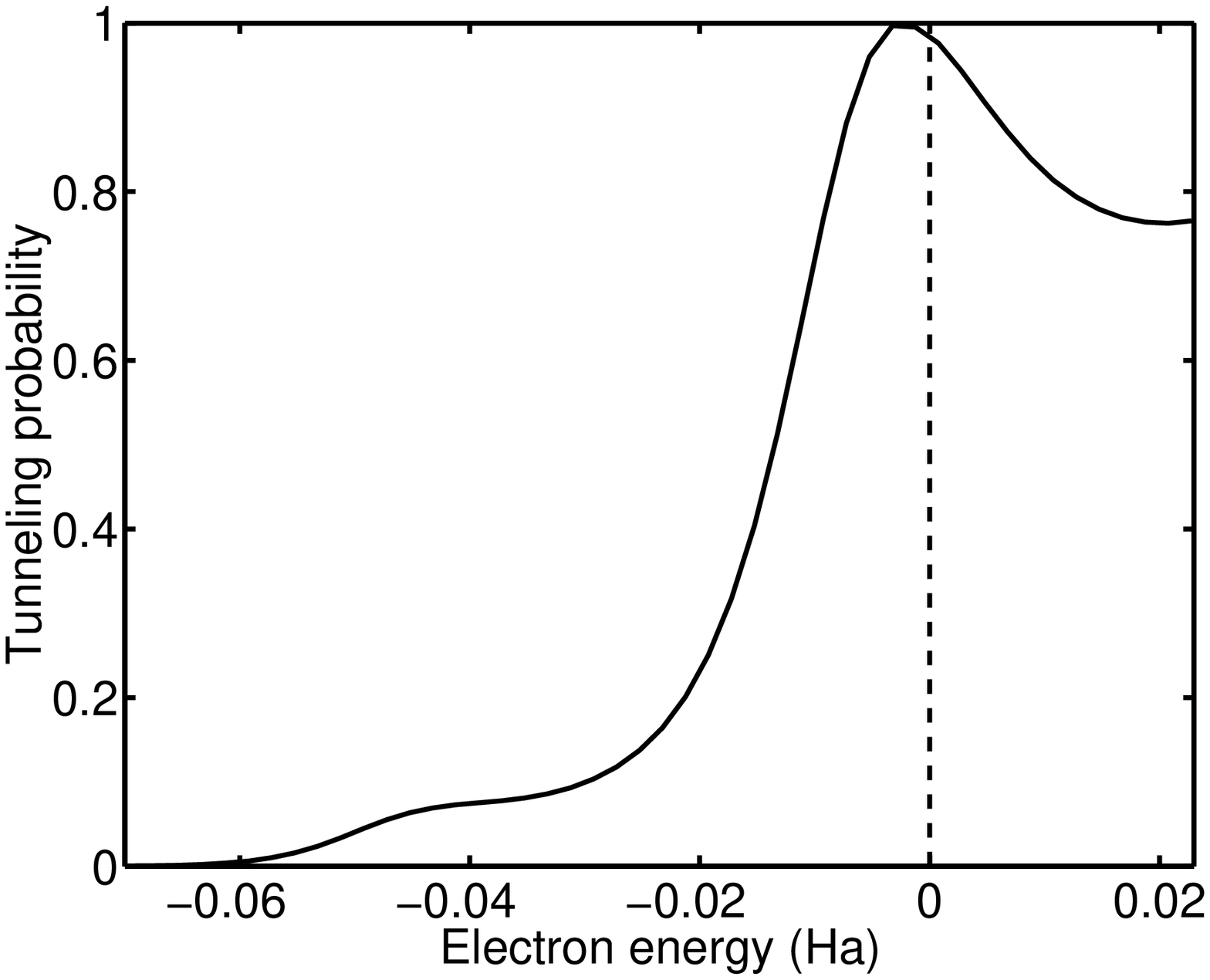,width=0.48\textwidth}
\end{center}
\caption{\label{NaKetju} Example of a nanostructure transport problem,
i.e., a chain of three Na atoms. The left-hand side shows constant electron
density surfaces against the backbone of the Na pseudoatoms (red spheres).
The bulk electrodes are described using the jellium background charge.
The right-hand side gives the tunneling probability as a function of the
electron energy. The energy zero (dashed line) corresponds to the Fermi
energy. }			%
\end{figure}




\subsection{Solution of atomic orbitals using interpolating wavelets} \label{sec:waveletcalc}

\begin{figure}[\figopt]
  \centering
  \includegraphics[width=0.49\textwidth]{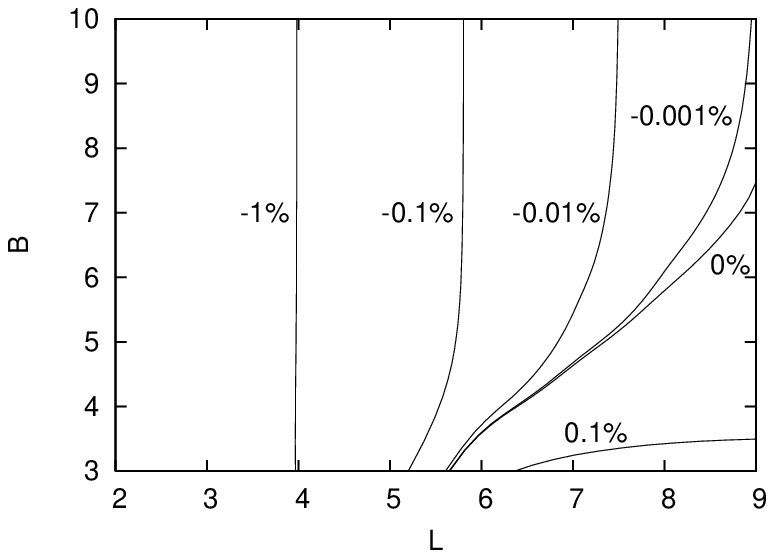}
   \includegraphics[width=0.49\textwidth]{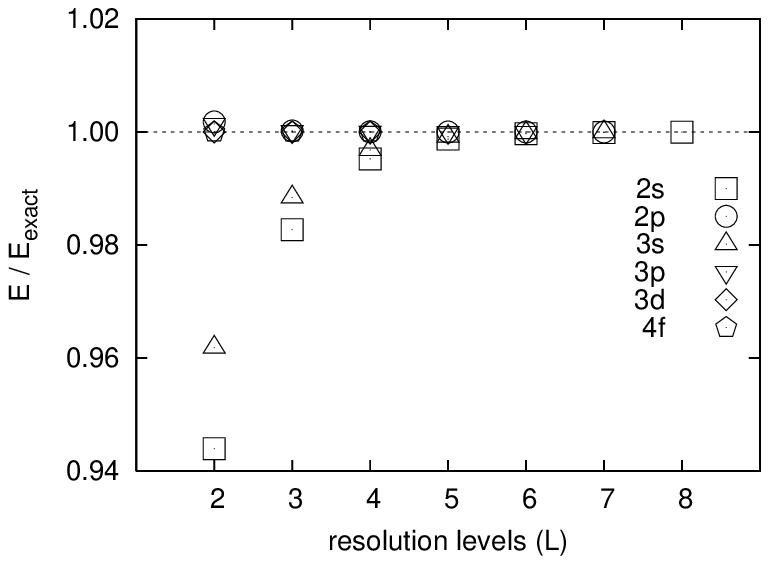}
  \caption{(a) Relative error of the hydrogen 1s orbital eigenvalue,
    unit of length \(u=1.0\).
    (b) Energy eigenvalues of 2s, 2p, 3s, 3p, 3d, and 4f orbitals of
     hydrogen with 
     \(B = 15, 20, 30, 30, 30,\;\textrm{and}\;45\), respectively. Ratio
     of the computed and exact values is shown. 
Reprinted with permission from T. H\"oyn\"al\"anmaa et al., Phys. Rev. E 70, 066701 (2005).
Copyright (2005) by the American Physical Society. }
   \label{fig:wavelet-hydr}
\end{figure}


Wavelets have been used for solving partial differential equations and
the electronic structure, in particular, only recently
\cite{ty97,ca93,g98,wc96}.  Most authors have chosen compactly
supported orthonormal wavelets. Daubechies wavelets
\cite{fd98,wc96}, Meyer wavelets \cite{ym96}, and Mexican hat wavelets
\cite{ca93} have been used.
Three of the authors of the present paper performed
electronic structure calculations
for hydrogen and some many electron atoms with interpolating wavelets
\cite{hrr04}.
Goedecker and Ivanov used interpolating wavelets to solve the Poisson
equation \cite{gi98}.

Orthonormal wavelet families provide useful properties like recursive
refinement relations and they lead to fast discrete wavelet transform
for multiresolution analysis.  Wavelet methods are closely connected
to point-grid based methods that also generalize to higher than one
dimensions \cite{vps95,a99}.  Lippert \etal\  \cite{lae98} have used
interpolating wavelets in point-grid based methods.

Representation of operators in orthonormal wavelet bases has also
been studied \cite{b92,bk97,bcr91} and we have introduced an algorithm to
compute the standard operator form of an arbitrary operator from its
nonstandard operator form.  Nonstandard operator form decouples
different resolution levels, which turns out to be an important aspect
for numerical approaches.

\begin{figure}[\figopt]
  \centering
  \begin{center}
    \includegraphics[width=0.49\textwidth]{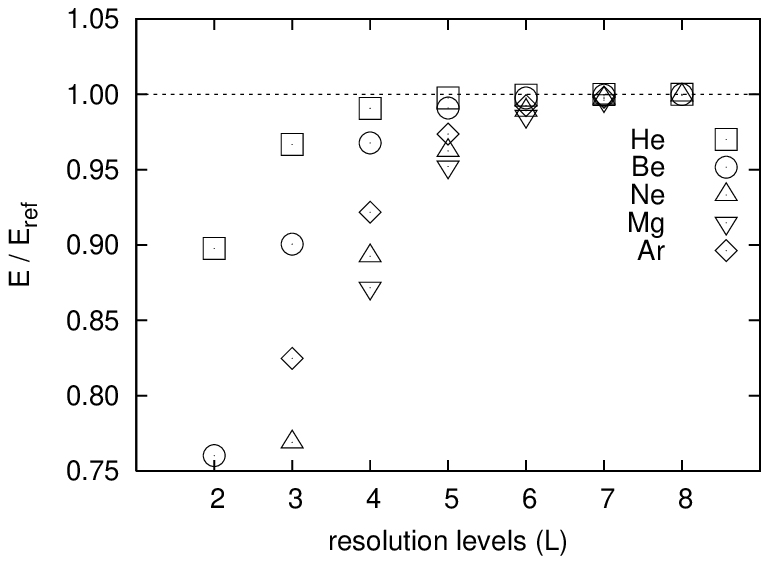}
    \includegraphics[width=0.49\textwidth]{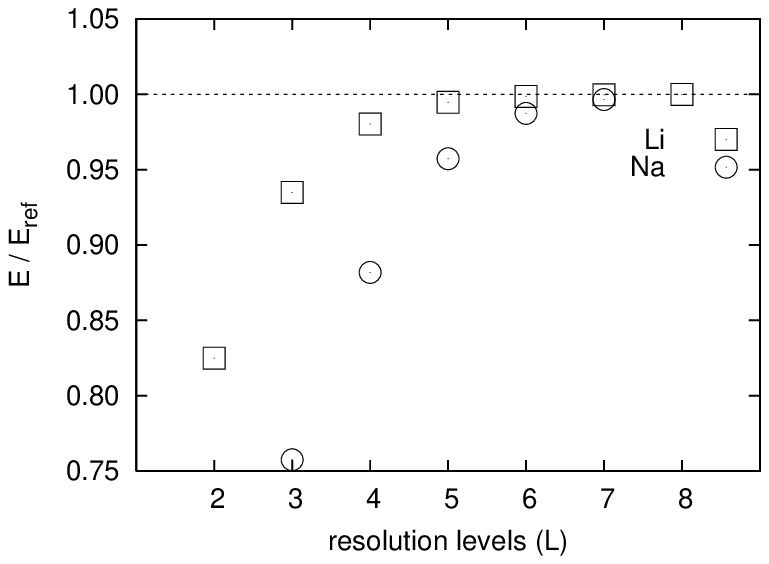}
    \caption{(a) Computed total energies for helium, beryllium, neon,
      magnesium, and argon. Ratio of computed and accurate \cite{f77}
      values is shown.
      (b) Computed total energies for lithium and sodium. Ratio of
      computed and accurate \cite{f77} values is shown.
Reprinted with permission from T. H\"oyn\"al\"anmaa et al., Phys. Rev. E 70, 066701 (2005).
Copyright (2005) by the American Physical Society. }
    \label{fig:wavelet-hf}
  \end{center}
\end{figure}


Due to spherical symmetry of atoms the three %
dimensional 
single-particle equation
\(\hamiltonianop \psi = \varepsilon \psi \) of the hydrogen-like atoms is
separated to the one dimensional radial part and the two
dimensional angular part.  Exact solutions to the latter are spherical
harmonics \( Y_{\ell m_\ell} \) where \( \ell \) and \( m_\ell \) are
the orbital quantum number (angular momentum quantum number) and
orbital magnetic quantum numbers, respectively.
Writing \( R_{n \ell} \) for the radial wavefunction the hydrogen-like
orbitals are \( \psi_{n \ell m_\ell}(r,\theta,\phi) = R_{n
\ell}(r) \, Y_{\ell m_\ell}(\theta, \phi) \), where \(n\) is the principal quantum number.
The radial Schr\"{o}dinger equation for a hydrogen-like atom is
\begin{equation}
  \label{eq:schrodinger-hydrogen}
  \left[ -\frac{1}{2} \frac{d^2}{dr^2} 
  - \frac{Z}{r} + \frac{\ell(\ell+1)}{2r^2} \right] P_{n \ell}(r) =
  \varepsilon_{n \ell} P_{n \ell}(r) ,
\end{equation}
where \(Z\) is the atomic number, \( \varepsilon_{n \ell} \) are the
orbital energies, and \( P_{n \ell}(r) = r R_{n \ell}(r) \) are the radial
wavefunctions multiplied by \(r\). Functions \(P_{n \ell}(r)\) are
called wavefunctions, too. In general, the radial wavefunction
\(P_{n\ell}\) has \(n-\ell-1\) nodes \cite{c81}.

The Schr\"odinger equation \(\hamiltonianop \Psi = E \Psi\) of a
many-electron atom leads to Hartree-Fock equations in the central field
approximation and with Slater determinant wavefunctions \cite{af97,c81,s97}.
The Hartree-Fock equations for many-electron atoms are presented
e.g. in our previous article \cite{hrr04}.
We omit the nondiagonal Lagrange multipliers and HF equations can be
written as matrix eigenvalue equations
\cite{l83}
  \begin{equation}
    \label{eq:hartree-fock-op}
    \mx{F}_{\orbital{i}} \mx{c}_{\orbital{i}} =
    \varepsilon_{\orbital{i}} \mx{c}_{\orbital{i}} .
  \end{equation}


The basis sets used for the computations were formed so that there are
S type basis functions \linebreak
\(\varphi^{k_{\smin}}_{-2B},\ldots,\varphi^{k_{\smin}}_{2B}\) in
resolution level \(k_{\smin}\) and
and D type basis functions \(\psi^k_{-B},\ldots,\psi^k_{B-1}\) for \linebreak
\(k=k_{\smin},\ldots,k_{\smin}+L-2\). Here \(L\) is the number of resolution
levels and \(B\) is a parameter describing the number of basis
functions in each resolution level.  

Relative error of the hydrogen 1s orbital eigenvalue is plotted in
figure \ref{fig:wavelet-hydr} (a). Here the computation parameters \(L\) and
\(B\) are varied. In figure \ref{fig:wavelet-hydr} (b) the energy
eigenvalues for several excited states of hydrogen are plotted. Here
the number of resolution levels \(L\) is varied. Results for HF
computations for closed-shell atoms are presented in figure
\ref{fig:wavelet-hf} (a) and for open-shell atoms in figure
\ref{fig:wavelet-hf} (b).


To summarize, we have demonstrated that interpolating wavelets can be successfully
used to solve the atomic orbitals and the electronic structure of
atoms.  We were able to systematically increase the accuracy of the
calculations by choosing the number of resolution levels and the
number of basis functions at each level.  This kind of flexibility is
an important benefit of wavelets.  The numerical results were found to
converge to the exact ones as the number of resolution levels
increases.  However, with a large number of resolution levels the
needed computation capacity increases considerably.
A noticeable feature in our development for the Hartree-Fock formalism
is that all relevent operators can be evaluated analytically.


\section{Future developments} \label{sec:future}
In the following we list of our future plans with regard to the
further development of software based on each of the three
discretization methods. This is done in order to shed light on the
motivation behind the work reported in Sec.~5 and to stimulate
discussions and collaborations.

\subsection{Finite differences \label{jatkoprojsec}}

The emphasis on the MIKA-project will be in the development of the
finite difference method within the {\bf GridPaw}-code.  The main
scientific objective in the near future is the implementation of
time-dependent density functional theory (TDDFT) both in the linear
response limit and in the real-time domain. These formulations can be
used, for example, for the calculation of optical absorption spectra
and band gaps of semiconductors and insulators. The real-time
formulation is needed for non-linear phenomena such as the electron
dynamics under strong femtosecond laser pulses.  Recently, a useful
survey of time-propagation methods has been publihed by Castro et
al. \cite{Castro04}.

The time dependent DFT problems are often solved using operator
splitting methods. This means that the Laplacian part is transferred
to Fourier side with FFT and propagated there, while the potential is
exponentiated directly in real space.  Mostly second or fourth order
splitting schemes are used. This is in fact completely analogous to
the diffusion algorithm of Sec.~\ref{diffusionsec}, which can be seen
as propagation in imaginary time.

Recently, the so called exponential integrators have turned out to be
very effective, especially in parabolic problems \cite{timoc}. These
integrators use the exponential of the linear approximation. The
development to hyperbolic systems, as TDDFT, is under active study.
An interesting option for equation $x'=f(x)$ is, for example, the
following exponential explicit midpoint rule \cite{timoa}
\begin{equation}
  x^{k+1}=x^k+e^{h H}(x^{k-1}-x^k)+2h\,\phi(h H)\,f(x^k)\ ,
\end{equation}
where $H$ is an approximation of $f'(x^k)$ and $\phi(z)=(e^z-1)/z$.
This produces a time reversible propagator, with good norm and energy
preservation properties. Computations of $e^{h H}v$ will be done using
Krylov subspace techniques. Efficient preconditioning for these, is
under active study.  Also variants of the use of Magnus expansions for
the exponential seem to be promising \cite{timob}.

The search for new, better exchange correlation functionals is an
ever-growing quest within DFT. One problem in the standard LDA and GGA
functionals is that part of the self-interaction contribution of the
Hartree- and Coulomb-potentials remain in the total energy
functional. One result of this remaining self-interaction contribution
is that in finite systems the potential decays exponentially instead
of the correct $1/r$ behaviour. In contrast to the LDA, the
Hartree-Fock scheme is self-interaction free and has the correct
asymptotic behaviour. One approach to improve from the LDA is the
so-called optimized potential method (OPM) where a multiplicative
potential is constructed from a Hartree-Fock-like exchange energy
functional
\cite{Sahni82,Langreth83}. This potential, which depends explicitly on the
Kohn-Sham orbitals, has to solved from a complicated integral
equation.  We are going to implement this exact-exchange scheme by
using the KLI-approximation \cite{KLI90} for the OPM integral
equation.

\subsection{Finite element methods} \label{fepropaganda}

Even though finite differences methods are currently the most well
established in the context of the electronic structure calculations,
it is clear that FE-methods could provide certain advantages. Thus,
further assessment of the feasibility of FE methods in electronic
structure calculations is one of the objectives of the current
MIKA-project

In this section we briefly summarize what we already have and what is
still missing. All-electron ground-state calculations for molecules
can be performed today with {\bf Elmer}, as described in
Sec.~\ref{elmersec}.  With the commercial mesh generator {\bf GiD}
significant reduction in the number of second order basis functions
for a high quality computation of the carbon monoxide molecule was
observed (Sec.~\ref{elmersec}) in comparison with the 2002 version of
{\bf Netgen}. Interestingly, both of these discretizations yielded
smaller matrix problems than a state-of-the-art finite-difference
pseudopotential approach (see Sec.~\ref{quantco}), and also
computations of larger molecules (C$_{60}$) were shown to be
feasible. Periodic boundary conditions have been implemented.  Further
improvements of efficiency 
utilization of $hp$-elements, pseudopotentials (or PAW) and
fine-tuning of the eigenproblem solver and mixing scheme.  On the
other hand, if we aim at an efficient method for real-time TDDFT
calculations, the performance (in terms of cpu-time) of the
eigenproblem solver and mixing scheme becomes less critical, whereas
the reduction of the size of the Hamiltonian matrix becomes even more
important.

An alternative route to a general purpose FE-package may be provided
by the code described in Sec.~\ref{greensec}.  This program already
includes an implementation of norm-conserving pseudopotentials.  This
implementation can possibly be adapted to {\bf Elmer}, or an
eigenproblem solver can be added into the transport code.

The question of the feasibity of molecular dynamics with an
unstructured nonuniform FE-mesh is often raised.  The FE-community
provides the following suggestion, based on experience in other
fields: each atom is attached to the mesh, and the mesh is considered
to be made of "rubber". As the atom moves, the mesh moves along, but
only within a finite range, and towards the boundary of the range the
mesh moves less. Therefore, the basis functions are indeed functions
of atomic coordinates and Pulay forces occur.  If the atoms move over
large distances, they may "tear the mesh apart" locally, by making
certain corners of tetrahedra too sharp.  In this case local remeshing
is applied. In this context, it should be also pointed out that
successful molecular dynamics simulations using structured nonuniform
finite-element meshes (in so called adaptive coordinates) have been
performed already by Tsuchida \cite{tsuchida04JCP}.  As a final remark
we add, that the possibility to use nonuniform unstructured meshes
within FEM is a very useful, but optional feature. It is also possible
to work with uniform meshes, and indeed implementation of molecular
dynamics is easier in that case. Even a uniform high-order FE-mesh is
expected to be more efficient than a uniform FD-grid -- at least in
the simplest case where no interpolation tricks are applied on the
FD-side -- in terms of the required number of basis functions (or grid
points) for a given accuracy.

\subsection{Wavelet package} \label{waveletfuture}

The wavelet based approximation scheme is still quite recent
computational tool in electronic structure calculations even though
it has shown to be very successful in other applications in
science and engineering. Its use so far has been mostly restricted to
cases which reduce to one-dimensional problems and
whose solutions
by other methods are well known in the literature. 
However, a promising recent development should be
noticed in this context \cite{arias03PRL}.

In near future we will continue the investigation to which
extent the wavelet method can speed up the computation of the
atomic orbitals. Especially we will study two and
three-dimensional problems. We are currently investigating the use of
interpolating tensor product wavelets for these calculations.
Special kind of nonseparable wavelets may be better for this purpose.
They allow achieving the full benefit of the fast evaluation of the discrete
wavelet transform. Also the methods will be compared with the
more traditional solution methods.

\section{Discussion}


In this section we compare the different methods of discretization
from various points of view. The discussion is biased towards
comparing the FE and FD methods to each other, the wavelet approach is
not equally well represented here due to the lack of experience on
wavelets of the authors of the present section.

In Sec.~\ref{similar} we discuss the similarities and differences of
the discretization methods, emphasizing aspects of local refinements,
ease of implementation and linear algebra. In Sec.~\ref{quantco} a
quantitative comparison between the sizes of the resulting matrix
problems in a pseudopotential FD method and an all-electron
FE-calculation for CO is made.  Also some speculation based on
FE-results from a transport calculation with pseudopotentials is
presented.  In Sec.~\ref{gendg} we present an awkward transition (in
latin: pons asini) from the double grid method of Ono and Hirose
\cite{ono99PRL} to the state-of-the art finite-element method.

\subsection{Similarities and differences between the methods of discretization} \label{similar}

The three methods for the discretization of the eigenvalue equation
considered in this paper have significant similarities but also some
differences. The most notable difference is that in the finite-element
and the wavelet method a basis set is present whereas the
finite-difference method relies on representing functions by their
values at grid points in space. This difference has two main
consequences.
\begin{enumerate}
\item The use of a basis set gives more freedom for constructions with varying spatial resolution. It is also easy to take into account details in geometry and general forms of boundary conditions. On the other hand, one must be able to generate non-uniform meshes conforming to details in geometry in three dimensions.
\item The basis set nature of the finite-element and the wavelet methods allows the use of variational arguments in the development of the method. This decreases the regularity requirements imposed on the solution and leads to a simpler error analysis.
\end{enumerate}

Let us discuss the question of varying spatial resolution a bit more
closely. In the finite-difference context the main approach is the
method of adaptive coordinates
\cite{waghmareCPC01,modine97PRB,gygi95PRB} 
This method involves a mapping from a regular grid in
three-dimensional parameter-space to a curvilinear coordinate system
in real-space. This approach obviously has some limitations with
regard to the ratio of smallest and largest local grid-spacing, and at
least cannot be regarded as a promising approach for all-electron
calculations of heavy elements.  Special care has to be exercised in
order to obtain symmetric finite-difference matrix representation of
the Laplacian in adaptive coordinates
\cite{modine97PRB,castro-private}.  Nonsymmetric matrices would render
standard iterative methods from linear algebra, such as the conjugate
gradient method, or Lanczos method for matrix exponentiation required
in time-propagation schemes \cite{Castro04} unreliable.  On the
finite-element side, an implementation of adaptive coordinates has
been introduced as well
\cite{tsuchidaJPSJ98}, but the pure FE-approach involves instead
general unstructured meshes consisting of elements of various, often
tetrahedral, shapes (see Sec.~\ref{sec:fem}).  This approach allows
greater spatial variations in the element size, granting accurate
all-electron calculations with problem sizes much smaller than the
corresponding FD-grids in pseudopotential calculations (see
Sec.~\ref{quantco}).  The resulting matrices ($H,S,A$) in
finite-element methods are symmetric by construction.
For the sake of completeness, we mention that 
another approach, namely that of composite grids \cite{beck00RMP, bylaska95SIAM},  
has been introduced in the context of finite-differences for
varying spatial resolution. In this method, local patches of finer
grids are applied. In the wavelet approach, local refinements
are  applied in a similar way \cite{a99}.

From an algorithmical point of view the finite-difference psedopotential
method with uniform grid and trapetsoid rule (Eq.~\ref{bbfdpseudo}) seems
to be the simplest one from the first outlook. The resulting matrices
are well-structured with a regular hierarchy. On the other hand, in
the basis set methods each element of the mesh can be processed using
the same algorithm as long as all the elements are affine images of
some basic reference element. This is especially easy in the wavelet
method where the basis functions are obtained via a simple scaling
methodology. When implementing a finite-element code, it is possible 
and important to reuse large parts of the general purpose finite-element
codes which are freely available to the general public, such
as those of Refs.~\cite{elmerweb,netgen}. When basic routines are recycled,
the implementation of an FE-approach becomes very simple as well.

From the viewpoint of linear algebra all the methods have rather
similar characteristics\footnote{Note however, that whereas the FE-approach
leads to symmetric matrices for the discretized Hamiltonian $H$, Laplacian $A$ 
and overlap matrix $S$, the FD-method tends to make $H$ and/or $A$ nonsymmetric
in the case where adaptive coordinates or generalized FD-schemes of 
\cite{briggs95PRB,collatz60,heiskanen98hodie} are applied, respectively.}
They all result in sparse matrices whose condition number is governed
by the underlying differential operator. They all allow the use of
multilevel and domain decomposition methods with effective
preconditioners. The major differences in favor of FD appear only for
very large systems with several hundred million degrees of freedom
when the well-structured nature of the finite-difference grid makes it
possible to apply the matrix on a vector in an iterative scheme
without explicitly storing it at any stage of the computation. In our
FE-examples presented below, where second and fourth order tetrahedral
meshes were considered, the storage of the necessary matrices (not
counting pseudotentials as dense blocks) required memory equivalent to
the storage of the order of one hundred eigenfunctions. This remains
the case for larger systems as well, thus the relative extra storage
becomes increasingly insignificant as larger systems are addressed, in
the typical applications where a set of all occupied (and possibly
some unoccupied) states need to be calculated.

\subsection{Quantitative comparisons} \label{quantco}

It is difficult to make objective comparisons between two different
discretization methods based on two different implementations as
computer programs. There are very many parameters that affect the
efficiency of the programs, and in the end, the user is only
interested in the cpu-time required for a calculation with given
accuracy.  An objective comparison of the efficiencies of the
discretization step, however, has to focus on the size of the matrix
problem\footnote{The size of a sparse $N\times N$ matrix $A$ naturally
involves, besides the number of rows $N$, also the average number $M$
of nonzero elements per row in $A$. When $A$ involves pseudopotential
operators, it becomes also relevant whether these are stored as dense
blocks, as is necessary in the context of direct methods
(Sec.~\ref{sec:direct}) or if only the projector vectors are stored,
as is common in FD-methods, and the pseudopotential part of the
Hamiltonian is evaluated in a "matrix free" manner. } produced, rather
than on the cpu-time spent by the entire calculation (unless exactly
the same implementation is used for the solution of the resulting
matrix problem). The cpu-time spent at the solution step then depends,
apart from the size of the matrix problem, also on the implementation
details and on the algorithms chosen for the eigenproblem, linear
systems of equations, and mixing.

The dipole-moment of the carbon monoxide molecule\footnote{Within the
local density approximation, its basis-set limit is $-0.226 D$
\cite{he00MP}.} is a well known basis-set sensitive quantity. Even
with the high-order FD-method and the straightforward trapetsoid rule
of Eq.~\ref{bbfdpseudo}, a rather small grid spacing of 0.2 a.u. is
required for a converged result (with a tolerance of $0.01 D$) when
norm conserving Trollier-Martins (TM) psedopotentials are
used\footnote{The required grid-spacing is dictated by the specific
choice of the pseudopotential, in this example $r_s =r_p =1.3a.u.$ was
used for both C and O.  Admittedly, it may be possible to find a
pseudopotential which allows a slightly larger grid spacing for a
converged result. Especially so, if the KB-pseudopotential is replaced
by the PAW formalism.} \cite{kronik_private_communication}.  When the
computational volume is a sphere of radius 10 a.u., this results in
520 000 grid points. A smaller sphere results in loss of accuracy for
the value of the dipole moment. Interestingly, the number of degrees
of freedom in this calculation is greater than the number of the 39
000 quadratic basis functions in the all-electron FE-calculation
within a sphere of radius 20 a.u., reported in
Sec.~\ref{elmersec}. There is no essential difference in the average
number of nonzero matrix elements per row in the two
calculations. This number was 42 in the FD case\footnote{37 matrix
elements per row for the kinetic energy operator with N=6 in
Eq.~\ref{fdkinetic} plus two additional dense blocks for the
pseudopotentials. The dense blocks were not stored in the actual
implementation. } and 55 in the FE case.

In Sec~\ref{greensec} and Ref.~\cite{oma3D} a thin HfO$_2$ layer was
modelled with Green's functions, TM pseudopotentials and fourth order
$p$-elements.  The unit cell, periodic in two dimensions, consisted of
85 atoms, with dimensions of $27\times21\times21 a_0^3$.  28 000 basis
functions were considered sufficient for converged results\footnote{
The matrix for the Laplacian had 75 nonzero elements per row.  In the
Hamiltonian matrix, the pseudopotentials were stored as dense blocks
as necessitated by the direct method of Sec.~\ref{sec:direct}, and
thus 893 matrix elements per row were needed. In other applications,
it may be more efficient to just store the projector functions for the
pseudopotentials, which is the standard method also in
FD-applications.}.  This was a transport calculation where Green's
functions were evaluated, and we have no experience on the equivalent
calculations using FD-methods. However, extrapolating from experience
with Kohn-Sham-equations, and also in light of the previous example it
is plausible to conclude that the grid spacing of 0.75 a.u., that
would result in 28 000 grid points in this case, can hardly give
reliable results.

\subsection{Generalizing the double grid method}  \label{gendg}

At the end of Sec.~\ref{sec:fd} we recognized, that the double-grid
method of Ono and Hirose \cite{ono99PRL} in fact amounts to computing
the matrix elements of the pseudopotential operator in a special basis
set arising from the polynomial interpolation used. We repeat here the
formula for the matrix elements of the pseudopotential operator
\begin{equation}
V^{\rm nl}_{ijk,i'j'k'} =  \sum_l c_l \tilde{x}_{lijk}\tilde{x}_{li'j'k'} = \int \phi_{ijk} (\rr)
 \hat{V}^{\rm nl} \phi_{i'j'k'}(\rr) d \rr,
\end{equation}
where the required integrals are approximated by a trapetsoid rule on
a fine uniform grid. For the sake of consistency, it is tempting to
ask if not the same idea could be applied to the entire hamiltonian
$H$, i.e., replace it with a discrete version
\begin{equation}
H_{ijk,i'j'k} = \int \phi_{ijk} (\rr) \hat{H} \phi_{i'j'k'}(\rr) d \rr.
\label{hmatrixel}
\end{equation}
Indeed, this can be done, but then we must keep in mind that the basis 
set of ${\phi_{ijk}}$'s is not orthonormal, but their overlap matrix
\begin{equation}
S_{ijk,i'j'k} = \int \phi_{ijk} (\rr)  \phi_{i'j'k'}(\rr) d \rr.
\label{smatrixel}
\end{equation}
also enters the formalism\footnote{This reasoning also reveals, that
the original implementation of the double grid method, where no $S$
matrix appears, may be improvable also if an orthogonal set of
$\phi_{ijk}$'s can be found.}, and we end up with the matrix equation
$Hc = \varepsilon Sc$ for the unknown coefficients in the expansion
\begin{equation}
\psi(\rr) = \sum_{ijk} c_{ijk} \phi_{ijk}(\rr).
\end{equation}
The procedure outlined above can be recognized as a variational method
of Sec.~\ref{sec:galerkin} with the basis set of $\phi_{ijk}$'s spanning
both the trial and test spaces. Thus the
double grid method itself is somewhere between the straightforward 
FD-scheme of Eqs~(\ref{fdkinetic},\ref{bbfdpseudo}) and the
finite-element method. Some similarities with the present
considerations can be seen in the approach based on the ``Galerkin
transfer'', outlined in Refs.~\cite{torsti04PSIK,torsti05nano}, where
a discretization on a grid was obtained from an FD-discretization on
an auxiliary finer grid, taking advantage of the prolongator and
restrictor operarors from the multigrid formalism. The present
approach can be seen as a limiting case, where the fine grid is
replaced by continuous space.

We add two more comments.  i) When numerically evaluating integrals
such as Eqs. (\ref{tildex}, \ref{hmatrixel}, \ref{smatrixel}) the
trapetsoid rule with uniform sampling is not the most efficient
method\footnote{The question of numerical integration always appears
when a basis set is introduced in DFT calculations. For example,
within the package Amsterdam Density Functional (ADF), Slater-type
basis funcions are used. The numerical integration routines
\cite{tevelde90thesis,tevelde92JCP} play a central role in making ADF
one of the most accurate codes in the market.}.  The integrands have
support only within an orthorhombic volume, and very efficient
numerical cubature formulas within such standard volumes are well
known \cite{cools03JC}. ii) Although the thought experiment above gave
a continuous path from the double-grid method to the FE-method with a
basis consisting of the functions $\phi_{ijk}$, we do not suggest
extending this basis set to very high order polynomials as it may
result in matrices with high condition number. 

Other bases of low order polynomial product form are e.g. the
"blip"-function basis (in the special case of $k=0$) of Hernandez and
Gillan \cite{hernandez97PRB} as well as the polynomial product basis
of Tsuchida and Tsukada \cite{tsuchida98JPSJ}.  The nonseparable cubic
"serendipity" element \cite{pask04MSMSE} gives rise to a useful basis
set which is no longer of product form, but still is associated with a
simple uniform grid structure.  To complete our awkward transition
from the double grid method to the state-of the art FE-methods
\cite{braess,daBible,szabobook}, we repeat here that the tetrahedral
$p$-element basis \cite{p_elementit,oma3D} discussed in Sec.~\ref{psection}
associated with a good mesh generator such as the rapidly developing
open source {\bf Netgen}-package \cite{netgen} grants us, in
principle, with complete flexibility in locally selecting the
polynomial order $p$ as well as the element size $h$.

\section{Summary} \label{sec:conclusions}

In this paper we have discussed our efforts to develop systematic
real-space methods as alternatives to the standard plane-wave and
atom-centered basis function methods. We aim at a general purpose
infrastructure for electronic structure calculations based on the
Kohn-Sham density-functional theory \cite{hohenberg64PR,kohn65PR} and
its time-dependent generalization \cite{runge84PRL}.  Our first line
of work is the MIKA-project. In its early stages, this project has
been rather tightly bound to the finite-difference method, the
Rayleigh-quotient multigrid method, and the {\bf MIKA} program
package.  In the near future the main development effort is directed
to the implementation of software tools for time-dependent
density-functional theory within the {\bf GridPaw}-code
\cite{mortensenPRB}. It is based on a different implementation of the
FD method. Its attractive feature is a successful FD-implementation of
the Projector Augmented Wave method, originally developed in the
plane-wave context, which allows the use of relatively coarse uniform
FD-grids for the accurate description of a wide range of elements.

The second theme of this paper has been the promotion of the
applications of FE methods within electronic structure calculations.
This work extends the scope of the MIKA project.  Nowadays, there are
general-purpose open-source FE-packages available
\cite{netgen,elmerweb}, that are extremely well suited as starting
points for the development of solvers for the Kohn-Sham
equations. Thus the argument that implementation of FE-methods for
electronic structure problems is more difficult than that of the
FD-methods has to be reconsidered.  We have compared a very
preliminary  all-electron example calculation for the carbon monoxide
molecule\footnote{We have also performed an all-electron calculation
with multimesh preconditioning for the C$_{60}$ molecule,
demonstrating that our scheme is not limited to small systems.} with a
state-of-the-art pseudopotential FD-calculation, in terms of the size
of the matrix problem and seen that the spatially adaptive
FE-discretization yields a matrix problem which is more than an order
of magnitude smaller.  Further reductions in the size of the problem
are expected when pseudopotentials or the PAW method is implemented,
and/or full advantage is taken of $hp$-FEM methods, where spatial variations
in the order $p$  of the piecewise polynomial basis functions are allowed
simultaneously with variations in the element size $h$.

Reductions in the CPU-time required by the finite-element ground-state
computations may also be available via a careful selection of the
eigenproblem solver and mixing scheme. However, as the ultimate goal
is to utilize also the FE-approach in real-time propagation within
TDDFT, the optimal performance of the ground state solver is of
secondary importance (as long as it delivers highly accurate results
-- this may be critical for the stability of the time propagation --
and as it is reliable in the sense that not too much labour of the
users needs to be invested in obtaining a convergent ground state
calculation) since the time propagation will always consume the
dominant part of the cpu-time. The reduction of the size of the
Hamiltonian matrix is directly relevant also for the performance of
time propagation. From a somewhat more philosophical point of view, we
have contemplated on the now popular double grid method due to Ono and
Hirose \cite{ono99PRL}, and seen it as the first step in a gradual
transformation from the FD-method to the FE-method.

The third line of work discussed in this paper is the development of a
program package based on the wavelet approach. This effort is still at
a very preliminary stage (thus far only spherically symmetric,
computationally one-dimensional results have been reported), and it
has been included here mainly with the dual aim of increasing the
level of wavelet awareness among the authors of the present paper and
of stimulating discussions with the various wavelet specialists. Based
on the current results, it is obvious that our wavelet package has not
yet reached a level of maturity comparable to that of the FD and FE
packages. Thus we will base our TDDFT developments on either one, or
both, of these two methods.

\begin{acknowledgement}		
T. T.  would like to thank the organizers (E. Artacho, E. Hernandez
and T. Beck) of the wonderful CECAM-workshop ``State of the art
developments and perspectives of real-space electronic structure
techniques in condended matter and molecular physics'', that took place in Lyon
in June 2005.  The discussions in the workshop gave
the necessary motivation to assemble this paper. The authors are
grateful to the advisory board of the MIKA-project for encouragement
and discussions during the process of writing this paper. Members of
the advisory board that are not also authors of this paper are K. W.
Jacobsen (CAMP and CSC), R. M.  Nieminen (COMP) and H. H\"akkinen
(ACQD).  E. Krotscheck and M. Aichinger are acknowledged for their
collaboration related to the response iteration scheme
(see. preamble of Sec.~\ref{sec:solve}, Sec.~\ref{rsdotsec} and Sec.~\ref{axialsec}).  
We are grateful to L. Kronik,
M. Johansson and D. Sundholm  for useful discussions
on the dipole moment of carbon monoxide (see Sec.~\ref{quantco}).  
Thanks to J. J.  Mortensen and
A. Foster for comments on the manuscript during the
process of its preparation. T. T., J. E. and T. H. acknowledge
financial support from the Finnish technology agency TEKES.
The work of E. R. was partially  supported by the Austrian Science Fund
FWF under project P15083-N08. T. T. is grateful for financial support
from the Magnus Ehrnrooth foundation.
\end{acknowledgement}


\bibliography{paula,biblio,positron,qd} 

\begin{thebibliography}{100}

\bibitem{kohn65PR}
W.~Kohn and L.~J. Sham.
\newblock Phys. Rev. {\bf 140}, A1133 (1965).

\bibitem{hohenberg64PR}
P.~Hohenberg and W.~Kohn.
\newblock Phys. Rev. {\bf 136}, B864 (1964).

\bibitem{runge84PRL}
E.~Runge and E.~K.~U. Gross.
\newblock Phys. Rev. Lett. {\bf 52}, 997 (1984).

\bibitem{waghmareCPC01}
U.~V. Waghmare, H.~Kim, I.~J. Park, N.~Modine, P.~Maragakis, and E.~Kaxiras.
\newblock Comp.\ Phys.\ Comm. {\bf 137}, 341 (2001).

\bibitem{castro-private}
A.~Castro.
\newblock Private communication  (2005).

\bibitem{tsuchidaJPSJ98}
E.~Tsuchida and M.~Tsukada.
\newblock J. Phys. Soc. Jap. {\bf 67}, 3844 (1998).

\bibitem{gygi92EL}
F.~Gygi.
\newblock Europhys. Lett. {\bf 19}, 617 (1992).

\bibitem{gygi93PRB}
F.~Gygi.
\newblock Phys. Rev.~B {\bf 48}, 11692 (1993).

\bibitem{blochl94PRB}
P.~Bl{\"o}chl.
\newblock Phys. Rev.~B {\bf 50}, 17953 (1994).

\bibitem{blochl03}
P.~Bl{\"o}chl, C.~J. F{\"o}rst, and J.~Schimpl.
\newblock Bull. Mater. Sci {\bf 26}, 33 (2003).

\bibitem{hernandez97PRB}
E.~Hern{\'a}ndez and M.~J. Gillan.
\newblock Phys. Rev.~B {\bf 55}, 13485 (1997).

\bibitem{skylaris02PRB}
C.-K. Skylaris, A.~A. Mostofi, P.~D. Haynes, O.~Dieguez, and M.~C. Payne.
\newblock Phys. Rev.~B {\bf 66}, 035119 (2002).

\bibitem{fattebert00PRB}
J.-L. Fattebert and J.~Bernholc.
\newblock Phys. Rev.~B {\bf 62}, 1713 (2000).

\bibitem{netgen}
Three-dimensional mesh generator {\bf Netgen}, see
  http://www.hpfem.jku.at/netgen/index.html.

\bibitem{elmerweb}
Elmer -- Finite element software for multiphysical problems, see
  http://www.csc.fi/elmer.

\bibitem{ono99PRL}
T.~Ono and K.~Hirose.
\newblock Phys. Rev. Lett. {\bf 82}, 5016 (1999).

\bibitem{siesta}
J.~M. Soler, E.~Artacho, J.~D. Gale, A.~García, J.~Junquera, P.~Ordejón, and
  D.~Sánchez-Portal.
\newblock J. Phys. : Condens. Matter {\bf 14}, 2745 (2002).

\bibitem{car85PRL}
R.~Car and M.~Parrinello.
\newblock Phys. Rev. Lett. {\bf 55}, 2471 (1985).

\bibitem{barnett93PRB}
R.~N. Barnett and U.~Landman.
\newblock Phys. Rev.~B {\bf 48}, 2081 (1993).

\bibitem{tsuchida04JCP}
E.~Tsuchida.
\newblock J.~Chem. Phys. {\bf 121}, 4740 (2004).

\bibitem{schmid04JCC}
R.~Schmid.
\newblock J. Comput. Chem. {\bf 25}, 799 (2004).

\bibitem{kleinmanPRL82}
L.~Kleinman and D.~Bylander.
\newblock Phys. Rev. Lett. {\bf 48}, 1425 (1982).

\bibitem{liu03PRB}
Y.~Liu, D.~A. Yarne, and M.~E. Tuckerman.
\newblock Phys. Rev.~B {\bf 68}, 125110 (2003).

\bibitem{baye02PRE}
D.~Baye, M.~Hesse, and M.~Vincke.
\newblock Phys. Rev.~E {\bf 65}, 026701 (2002).

\bibitem{kobus96CPC}
J.~Kobus, L.~Laaksonen, and D.~Sundholm.
\newblock Comp.\ Phys.\ Comm. {\bf 98}, 346 (1996).

\bibitem{becke86PRA}
A.~D. Becke.
\newblock Phys. Rev. A {\bf 33}, 2786 (1986).

\bibitem{becke88JCP}
A.~D. Becke.
\newblock J.~Chem. Phys. {\bf 88}, 2547 (1988).

\bibitem{springer98PRB}
M.~Springer.
\newblock Phys. Rev.~B {\bf 58}, 1939 (1998).

\bibitem{briggs95PRB}
E.~L. Briggs, D.~J. Sullivan, and J.~Bernholc.
\newblock Phys. Rev.~B {\bf 52}, R5471 (1995).

\bibitem{collatz60}
L.~Collatz.
\newblock {\em The numerical treatment of differential equations\/}
  (Springer-Verlag, Berlin, 1960).

\bibitem{heiskanen98hodie}
M.~Heiskanen.
\newblock unpublished  (1998).

\bibitem{chelikowsky94PRB}
J.~R. Chelikowsky, N.~Troullier, K.~Wu, and Y.~Saad.
\newblock Phys. Rev.~B {\bf 50}, 11355 (1994).

\bibitem{nogueiraAPDFT}
F.~Nogueira, A.~Castro, and M.~Marques.
\newblock In {\em A Primer in Density Functional Theory\/}, edited by
  C.~Fiolhais, F.~Nogueira, and M.~Marques (Springer-Verlag, Heidelberg, 2003),
  pp. 218--256.

\bibitem{braess}
D.~Braess.
\newblock {\em Finite Elements\/} (Cambridge University Press, Cambridge,
  1997).

\bibitem{daBible}
P.~G. Ciarlet and J.~L. Lions (editors).
\newblock {\em Finite Element Methods\/}, volume~II of {\em Handbook of
  Numerical Analysis\/} (North-Holland, Amsterdam, 1991).

\bibitem{szabobook}
B.~Szabo and I.~Babuska.
\newblock {\em Finite Element Analysis\/} (John Wiley \& Sons, Inc., New York,
  1991).

\bibitem{pask04MSMSE}
{J. E. Pask and P. A. Sterne}.
\newblock {Modelling Simul. Mater. Sci. Eng.} {\bf 13}, R71 (2005).

\bibitem{easymesh}
Two-dimensional mesh generator {\bf Easymesh}, \\ see
  http://www-dinma.univ.trieste.it/nirftc/research/easymesh/Default.htm.

\bibitem{p_elementit}
M.~Ainsworth and J.~Coyle.
\newblock Int. J. Numer. Meth. Eng. {\bf 58}, 2103 (2003).

\bibitem{stevenson2003}
R.~Stevenson.
\newblock SIAM J. Numer. Anal. {\bf 41}, 1074 (2003).

\bibitem{dahlke2002}
S.~Dahlke, W.~Dahmen, and K.~Urban.
\newblock SIAM J. Numer. Anal. {\bf 40}, 1230 (2002).

\bibitem{dahmen1997}
W.~Dahmen.
\newblock Acta Numerica {\bf 6}, 55 (1997).

\bibitem{lindemann05thesis}
M.~Lindemann.
\newblock {\em Approximation properties of non-separable wavelet bases with
  isotropic scaling matrices\/}.
\newblock {Ph.D.} dissertation, Universit{\"a}t Bremen (2005).

\bibitem{g98}
{S. Goedecker}.
\newblock {\em {Wavelets and their application for the solution of partial
  differential equations in physics}\/} ({Presses Polytechniques et
  Universitaires Romandes}, 1998).

\bibitem{chui-li1996}
C.~K. Chui and C.~Li.
\newblock SIAM J. Math. Anal. {\bf 27}, 865 (1996).

\bibitem{donoho1992}
D.~L. Donoho.
\newblock {\em Interpolating wavelet transform\/}.
\newblock Technical report, Dept. of Statistics Stanford Univ. (1992).

\bibitem{d92}
{I. Daubechies}.
\newblock {\em {Ten Lectures on Wavelets}\/}, volume~61 of {\em {CBMS-NSF
  regional conference series in applied mathematics}\/} ({SIAM}, 1992).

\bibitem{golub89MC}
G.~H. Golub and C.~F.~V. Loan.
\newblock {\em Matrix Computations\/} (The Johns Hopkins University Press,
  London, 1989), second edition.

\bibitem{dongarraCRS}
J.~Dongarra.
\newblock In {\em Templates for the solution of Algrebraic Eigenvalue Problems:
  A Practical Guide\/}, edited by Z.~Bai, J.~Demmel, J.~Dongarra, A.~Ruhe, and
  H.~van~der Vorst (SIAM, Philadelphia, 2000), p. 372.

\bibitem{lapack}
{\bf Lapack}, Linear Algebra PACKage, see http://www.netlib.org/lapack.

\bibitem{sparskit}
A basic tool-kit for sparse matrix computations {\bf Sparskit}, \\ see
  http://www-users.cs.umn.edu/~saad/software/SPARSKIT/sparskit.html.

\bibitem{brandt77MOC}
A.~Brandt.
\newblock Math. Comput. {\bf 31}, 333 (1977).

\bibitem{Brandt83SIAM}
A.~Brandt, S.~F. McCormick, and J.~W. Ruge.
\newblock SIAM J. Sci. Comput. (USA) {\bf 4}, 244 (1983).

\bibitem{wang00JCP}
J.~Wang and T.~L. Beck.
\newblock J. Chem. Phys. {\bf 112}, 9223 (2000).

\bibitem{wijesekara03JTCC}
N.~Wijesekera, G.~Feng, and T.~L. Beck.
\newblock J. Theor. Comput. Chem. {\bf 2}, 553 (2003).

\bibitem{costinerPRE95b}
S.~Costiner and S.~Ta'asan.
\newblock Phys. Rev.~E {\bf 52}, 1181 (1995).

\bibitem{briggs96PRB}
E.~L. Briggs, D.~J. Sullivan, and J.~Bernholc.
\newblock Phys. Rev.~B {\bf 54}, 14362 (1996).

\bibitem{payne92RMP}
M.~C. Payne, M.~P. Teter, D.~C. Allan, T.~A. Arias, and J.~D. Joannopoulos.
\newblock Rev. Mod. Phys. {\bf 64}, 1045 (1992).

\bibitem{kresse96PRB}
G.~Kresse and J.~Furthm{\"u}ller.
\newblock Phys. Rev.~B {\bf 54}, 11169 (1996).

\bibitem{kresseCMS96}
G.~Kresse and J.~Furthm{\"u}ller.
\newblock Comp. Mat. Sci. {\bf 6}, 15 (1996).

\bibitem{bowler00CPL}
D.~R. Bowler and M.~J. Gillan.
\newblock Chem. Phys. Lett. {\bf 325}, 473 (2000).

\bibitem{newton}
J.Auer and E.~Krotscheck.
\newblock Comp. Phys. Comm. {\bf 118}, 139 (1999).

\bibitem{Collective}
J.~Auer and E.~Krotscheck.
\newblock Comp. Phys. Comm. {\bf 151}, 265 (2003).

\bibitem{torsti04PSIK}
T.~Torsti, V.~Lindberg, I.~Makkonen, E.~Ogando, E.~R{\"a}s{\"a}nen,
  H.~Saarikoski, M.~J. Puska, and R.~M. Nieminen.
\newblock $\Psi_k$ Newsletter {\bf 65}, 105 (2004).

\bibitem{torsti05nano}
T.~Torsti, V.~Lindberg, I.~Makkonen, E.~Ogando, E.~R{\"a}s{\"a}nen,
  H.~Saarikoski, M.~J. Puska, and R.~M. Nieminen.
\newblock Handbook of Theoretical and Computational nanotechnology  (2006).

\bibitem{Castro04}
A.~Castro, M.~A.~L. Marques, and A.~Rubio.
\newblock J. Chem. Phys. {\bf 121}, 3425 (2004).

\bibitem{vorstbook}
H.~A. van~der Vorst.
\newblock {\em Iterative Krylov Methods for large linear systems\/} (Cambridge
  University Press, 2003).

\bibitem{ilu86}
H.~C. Elman.
\newblock Math. Comput. {\bf 47}, 191 (1986).

\bibitem{briggs00MTSE}
W.~L. Briggs, V.~E. Henson, and S.~F. McCormick.
\newblock {\em A Multigrid Tutorial, Second Edition\/} (SIAM, 2000).

\bibitem{hackbush85MGMA}
W.~Hackbush.
\newblock {\em Multi-Grid Methods and Applications\/} (Springer-Verlag, Berlin,
  1985).

\bibitem{wesseling92AITMM}
P.~Wesseling.
\newblock {\em An Introduction to Multigrid Methods\/} (John Wiley \& Sons,
  Inc., New York, 1992).

\bibitem{duff}
I.~S. Duff.
\newblock Comp. Phys. Comm. {\bf 97}, 45 (1996).

\bibitem{duff-reid}
J.~K. Reid and I.~S. Duff.
\newblock ACM Trans. on Math. Software {\bf 9}, 302 (1983).

\bibitem{hsl}
The Harwell Subroutine Library, see http://www.cse.clrc.ac.uk/nag/hsl/.

\bibitem{umfpack}
T.~A. Davis.
\newblock ACM Trans. Math. Software {\bf 30}, 353 (2004).

\bibitem{watson}
A.~Gupta.
\newblock IBM Research Report, RC {\bf 21886}, 98462 (2000).

\bibitem{saad92NMLEP}
Y.~Saad.
\newblock {\em Numerical Methods for Large Eigenvalue Problems\/} (Manchester
  University Press, Manchester, 1992).

\bibitem{wang94CMS}
L.-W. Wang and A.~Zunger.
\newblock Comp. Mat. Sci {\bf 2}, 326 (1994).

\bibitem{modine97PRB}
N.~A. Modine, G.~Zumbach, and E.~Kaxiras.
\newblock Phys. Rev.~B {\bf 55}, 10289 (1997).

\bibitem{tuomas03thesis}
T.~Torsti.
\newblock {\em Real-Space Electronic Structure Calculations for Nanoscale
  Systems\/}.
\newblock {Ph.D.} dissertation, Helsinki University of Technology (2003).
\newblock Http://lib.hut.fi/Diss/2003/isbn9512264706/.

\bibitem{stathopoulos00CSE}
{A. Stathopoulos and S. {\"O}g\"ut and Y. Saad and J. Chelikowsky and H. Kim}.
\newblock Comput. Sci. Eng. {\bf 2}, 19 (2000).

\bibitem{knyazevSIAM}
A.~Knyazev.
\newblock Siam. J. Sci. Comput. {\bf 23}, 517 (2001).

\bibitem{knyazevETNA}
A.~Knyazev and K.~Neymeyr.
\newblock El. Trans. Numer. Anal. {\bf 15}, 38 (2003).

\bibitem{casida95RADFM}
M.~E. Casida.
\newblock In {\em Recent Advances in Density Functional Methods, Part I\/},
  edited by D.~P. Chong (Singapore, 1995), p. 155.

\bibitem{jamorski96JCP}
C.~Jamorski, M.~E. Casida, and D.~R. Salahub.
\newblock J. Chem. Phys. {\bf 104}, 5134 (1996).

\bibitem{davidsonJCP75}
E.~R. Davidson.
\newblock J. Comp. Phys. {\bf 17}, 87 (1975).

\bibitem{davidsonCP93}
E.~R. Davidson.
\newblock Computers in Physics {\bf 7}, 519 (1993).

\bibitem{gisbergenCPC99}
S.~J.~A. Gisbergen, J.~G. Snijders, and E.~J. Baerends.
\newblock Comp. Phys. Comm. {\bf 118}, 119 (1999).

\bibitem{stratmanJChP98}
R.~E. Stratman, G.~E. Scuseria, and M.~J. Frisch.
\newblock J. Chem. Phys. {\bf 109}, 8218 (1998).

\bibitem{burdickCPC03}
W.~R. Burdick, Y.~Saad, L.~Kronik, I.~Vasiliev, M.~Jain, and J.~R. Chelikowsky.
\newblock Comp. Phys. Comm. {\bf 156}, 22 (2003).

\bibitem{scalapack}
The {\bf ScaLapack} project, see http://www.netlib.org/scalapack.

\bibitem{arpack}
{\bf Arpack} -- Arnoldi package, see http://www.caam.rice.edu/software/ARPACK/.

\bibitem{Heiskanen}
M.~Heiskanen, T.~Torsti, M.~Puska, and R.~Nieminen.
\newblock Phys.\ Rev.\ B {\bf 63}, 245106 (2001).

\bibitem{torsti03IJQC}
T.~Torsti, M.~Heiskanen, M.~Puska, and R.~Nieminen.
\newblock Int. J. Quantum Chem. {\bf 91}, 171 (2003).

\bibitem{mandel89JCC}
J.~Mandel and S.~McCormick.
\newblock J. Comput. Phys. {\bf 80}, 442 (1989).

\bibitem{wood85JPA}
D.~M. Wood and A.~Zunger.
\newblock J. Phys. A {\bf 18}, 1343 (1985).

\bibitem{mortensenPRB}
J.~J. Mortensen, L.~B. Hansen, and K.~W. Jacobsen.
\newblock Phys. Rev.~B {\bf 71}, 035109 (2005).

\bibitem{pulay80CPL}
P.~Pulay.
\newblock Chem. Phys. Lett. {\bf 73}, 393 (1980).

\bibitem{pulay82JCC}
P.~Pulay.
\newblock J. Comp. Chem. {\bf 3}, 556 (1982).

\bibitem{imstep}
J.~Auer, E.~Krotscheck, and S.~A. Chin.
\newblock J. Chem. Phys. {\bf 115}, 6841 (2001).

\bibitem{michael}
M.~Aichinger and E.~Krotscheck.
\newblock Comp. Mater. Sci. {\bf 34}, 188 (2005).

\bibitem{magalg}
M.~Aichinger, S.~A. Chin, and E.~Krotscheck.
\newblock Comp. Phys. Comm. {\bf 171}, 197 (2005).

\bibitem{Paulan_vaikkari}
P.~Havu.
\newblock {\em Modeling of Electronic Transport in Nanostructures\/}.
\newblock {Ph.D.} dissertation, Helsinki University of Technology (2005).
\newblock Http://lib.tkk.fi/Diss/2005/isbn9512278596/.

\bibitem{quantumdots}
For a review, see, e.g., L. P. Kouwenhoven, D. G. Austing, and S. Tarucha, Rep.
  Prog. Phys. {\bf 64}, 701 (2001); S. M. Reimann and M. Manninen, Rev. Mod.
  Phys. {\bf 74}, 1283 (2002).

\bibitem{lsda}
H.~Saarikoski, E.~R\"as\"anen, S.~Siljam\"aki, A.~Harju, M.~J. Puska, and R.~M.
  Nieminen.
\newblock Phys. Rev. B {\bf 67}, 205327 (2003).

\bibitem{statistics1}
M.~Aichinger and E.~R\"as\"anen.
\newblock Phys. Rev. B {\bf 71}, 165302 (2005).

\bibitem{statistics2}
E.~R\"as\"anen and M.~Aichinger.
\newblock Phys. Rev. B {\bf 72}, 045352 (2005).

\bibitem{sivan}
U.~Sivan, R.~Berkovits, Y.~Aloni, O.~Prus, A.~Auerbach, and G.~Ben-Yoseph.
\newblock Phys. Rev. Lett. {\bf 77}, 1123 (1996).

\bibitem{patel}
S.~R. Patel, S.~M. Cronenwett, D.~R. Stewart, A.~G. Huibers, C.~M. Marcus,
  C.~I. Duruoz, J.~S. Harris, K.~Campman, and A.~C. Gossard.
\newblock Phys. Rev. Lett. {\bf 80}, 4522 (1998).

\bibitem{hirose2}
K.~Hirose, F.~Zhou, and N.~S. Wingreen.
\newblock Phys. Rev. B {\bf 63}, 75301 (2001).

\bibitem{jiang}
H.~Jiang, D.~Ullmo, W.~Yang, and H.~U. Baranger.
\newblock Phys. Rev. B {\bf 69}, 235326 (2004).

\bibitem{henricluster}
H.~Saarikoski, A.~Harju, M.~J. Puska, and R.~M. Nieminen.
\newblock Phys. Rev. Lett. {\bf 93}, 116802 (2004).

\bibitem{henristability}
H.~Saarikoski, S.~M. Reimann, E.~R{\"a}s{\"a}nen, A.~Harju, and M.~J. Puska.
\newblock Phys. Rev. B {\bf 71}, 035421 (2005).

\bibitem{henrienergy}
H.~Saarikoski and A.~Harju.
\newblock Phys. Rev. Lett. {\bf 94}, 246803 (2005).

\bibitem{esagiant}
E.~R{\"a}s{\"a}nen, H.~Saarikoski, A.~Harju, M.~J. Puska, Y.~Yu, and S.~M.
  Reimann.
\newblock To be published (cond-mat/0509660).

\bibitem{henrimagnet}
A.~Harju, H.~Saarikoski, and E.~R{\"a}s{\"a}nen.
\newblock To be published.

\bibitem{chakraborty}
T.~Chakraborty and P.~Pietil{\"a}inen.
\newblock {\em The Quantum Hall Effects: Fractional and Integral\/}
  (Springer-Verlag, Berlin, 1995).

\bibitem{oosterkamp}
T.~H. Oosterkamp, J.~W. Janssen, L.~P. Kouwenhoven, D.~G. Austing, T.~Honda,
  and S.~Tarucha.
\newblock Phys. Rev. Lett. {\bf 82}, 2931 (1999).

\bibitem{markkumusta}
M.~P. Schwarz, D.~Grundler, C.~Heyn, D.~Heitmann, D.~Reuter, and A.~Wieck.
\newblock Phys. Rev. B {\bf 68}, 245315 (2003).

\bibitem{manninen}
M.~Manninen, S.~M. Reimann, M.~Koskinen, Y.~Yu, and M.~Toreblad.
\newblock Phys. Rev. Lett {\bf 94}, 106405 (2005).

\bibitem{chulkov-ps}
E.~Chulkov, V.~M. Silkin, and P.~M. Echenique.
\newblock Surf.Sci. {\bf 437}, 330 (1999).

\bibitem{chulk-selfe-im}
I.~Sarria, J.~Osama, E.~Chulkov, J.~Pitarke, and P.~Echenique.
\newblock Phys. Rev. B {\bf 60}, 11795 (1999).

\bibitem{edu-slabs}
E.~Ogando, N.~Zabala, E.~V. Chulkov, and M.~J. Puska.
\newblock Phys. Rev. B {\bf 71}, 205401 (2005).

\bibitem{heller}
G.~A. Fiete and E.~J. Heller.
\newblock Rev. Mod. Phys. {\bf 75}, 993 (2003).

\bibitem{boronski}
E.~Boro\'nski and R.~M. Nieminen.
\newblock Phys. Rev. B {\bf 34}, 3820 (1986).

\bibitem{Alatalo96}
M.~Alatalo, B.~Barbiellini, M.~Hakala, H.~Kauppinen, T.~Korhonen, M.~J. Puska,
  K.~Saarinen, P.~Hautoj\"arvi, and R.~M. Nieminen.
\newblock Phys. Rev. B {\bf 54}, 2397 (1996).

\bibitem{Makkonen05b}
I.~Makkonen, M.~Hakala, and M.~J. Puska.
\newblock Phys. Rev. B {\bf 73}, 035103 (2006).
\newblock (cond-mat/0509025).

\bibitem{Barbiellini95}
B.~Barbiellini, M.~J. Puska, T.~Torsti, and R.~M. Nieminen.
\newblock Phys. Rev. B {\bf 51}, 7341 (1995).

\bibitem{Barbiellini96}
B.~Barbiellini, M.~J. Puska, T.~Korhonen, A.~Harju, T.~Torsti, and R.~M.
  Nieminen.
\newblock Phys.\ Rev.\ B {\bf 53}, 16201 (1996).

\bibitem{Kresse99}
G.~Kresse and D.~Joubert.
\newblock Phys.\ Rev.\ B {\bf 59}, 1758 (1999).

\bibitem{Makkonen05}
I.~Makkonen, M.~Hakala, and M.~J. Puska.
\newblock J.\ Phys.\ Chem.\ Solids {\bf 66}, 1128 (2005).

\bibitem{Rummukainen05}
M.~Rummukainen, I.~Makkonen, V.~Ranki, M.~J. Puska, K.~Saarinen, and H.-J.~L.
  Gossmann.
\newblock Phys.\ Rev.\ Lett. {\bf 94}, 165501 (2005).

\bibitem{Calloni05}
A.~Calloni, A.~Dupasquier, R.~Ferragut, P.~Folegati, M.~M. Iglesias,
  I.~Makkonen, and M.~J. Puska.
\newblock Phys.\ Rev.\ B {\bf 72}, 054112 (2005).

\bibitem{metis}
G.~Karypis and V.~Kumar.
\newblock SIAM Journal on Scientific Computing {\bf 20}, 359 (1998).

\bibitem{gid}
The personal pre- and postrocessor {\bf GiD}, see
  http://gid.cimne.upc.es/index.html.

\bibitem{he00MP}
Y.~He, J.~Gr{\"a}fenstein, E.~Kraka, and D.~Cremer.
\newblock Molecular Physics {\bf 98}, 1639 (2000).

\bibitem{gre_datta}
S.~Datta.
\newblock {\em Electronic transport in mesoscopic systems\/} (Cambridge
  University Press, Cambridge, 1995).

\bibitem{transiesta1}
J.~Taylor, H.~Guo, and J.~Wang.
\newblock Phys. Rev. B {\bf 63}, 245407 (2001).

\bibitem{transiesta2}
M.~Brandbyge, J.~Mozos, P.~Ordej\'on, J.~L. Taylor, and K.~Stokbro.
\newblock Phys. Rev. B {\bf 65}, 165401 (2002).

\bibitem{oma3D}
P.~Havu, V.~Havu, M.~J. Puska, M.~H. Hakala, A.~S. Foster, and R.~M. Nieminen.
\newblock J. Chem. Phys. (in print); http://arxiv.org/abs/physics/0506159
  (2006).

\bibitem{bernholc}
M.~B. Nardelli, J.-L. Fattebert, and J.~Bernholc.
\newblock Phys. Rev. B {\bf 64}, 245423 (2001).

\bibitem{wavelet}
K.~S. Thygesen, M.~V. Bollinger, and K.~W. Jacobsen.
\newblock Phys. Rev. B {\bf 67}, 115404 (2003).

\bibitem{embedded}
D.~Wortmann, H.~Ishida, and S.~Bluegel.
\newblock Phys. Rev. B {\bf 66}, 075113 (2002).

\bibitem{wannier}
A.~Calzolari, N.~Marzari, I.~Souza, and M.~B. Nardelli.
\newblock Phys. Rev. B {\bf 69}, 035108 (2004).

\bibitem{differenssi}
P.~A. Khomyakov and G.~Brocks.
\newblock Phys. Rev. B {\bf 70}, 195402 (2004).

\bibitem{fem-green}
E.~Polizzi and A.~N. Ben.
\newblock J. Comput. Phys. {\bf 202}, 150 (2005).

\bibitem{datta}
S.~Datta and W.~Tian.
\newblock Phys. Rev. B {\bf 55}, R1914 (1997).

\bibitem{fem_avoin_reuna}
T.~J.~R. Hughes.
\newblock Comput. Methods Appl. Mech Engrg. {\bf 127}, 387 (1995).

\bibitem{oma_johto}
P.~Havu, T.~Torsti, M.~J. Puska, and R.~M. Nieminen.
\newblock Phys. Rev. B {\bf 66}, 075401 (2002).

\bibitem{oma07}
P.~Havu, M.~J. Puska, R.~M. Nieminen, and V.~Havu.
\newblock Phys. Rev. B {\bf 70}, 233308 (2004).

\bibitem{ty97}
{C. J. Tymczak and Xiao-Qian Wang}.
\newblock {Phys. Rev. Lett.} {\bf 78}, 3654 (1997).

\bibitem{ca93}
{K. Cho, T. A. Arias, J. D. Joannopoulos, and P. K. Lam}.
\newblock {Phys. Rev. Lett.} {\bf 71}, 1808 (1993).

\bibitem{wc96}
{Siqing Wei and M. Y. Chou}.
\newblock {Phys. Rev. Lett.} {\bf 76}, 2650 (1996).

\bibitem{fd98}
{P. Fischer and M. Defranceschi}.
\newblock {SIAM J. Numer. Anal.} {\bf 35}, 1 (1998).

\bibitem{ym96}
{K. Yamaguchi and T. Mukoyama}.
\newblock {J. Phys. B} {\bf 29}, 4059 (1996).

\bibitem{hrr04}
{T. H\"oyn\"al\"anmaa, T. T. Rantala, and K. Ruotsalainen}.
\newblock Phys. Rev. E {\bf 70}, 066701 (2004).

\bibitem{gi98}
{S. Goedecker and O. V. Ivanov}.
\newblock {Solid State Commun.} {\bf 105}, 665 (1998).

\bibitem{vps95}
{O. V. Vasilyev, S. Paolucci, and M. Sen}.
\newblock {J. Comput. Phys.} {\bf 120}, 33 (1995).

\bibitem{a99}
{T. A. Arias}.
\newblock {Rev. Mod. Phys.} {\bf 71}, 267 (1999).

\bibitem{lae98}
{R. A. Lippert, T. A. Arias, and A. Edelman}.
\newblock {J. Comput. Phys.} {\bf 140}, 278 (1998).

\bibitem{b92}
{G. Beylkin}.
\newblock {SIAM J. Numer. Anal.} {\bf 6}, 1716 (1992).

\bibitem{bk97}
{G. Beylkin and J. M. Keiser}.
\newblock {J. Comput. Phys.} {\bf 132}, 233 (1997).

\bibitem{bcr91}
{G. Beylkin, R. Coifman, and V. Rokhlin}.
\newblock {Commun. Pure Appl. Math.} {\bf XLIV}, 141 (1991).

\bibitem{f77}
{Ch. Froese Fischer}.
\newblock {\em {The Hartree-Fock Method for Atoms -- A Numerical Approach}\/}
  ({John Wiley \& Sons}, 1977).

\bibitem{c81}
{R. D. Cowan}.
\newblock {\em {The theory of atomic structure and spectra}\/} ({University of
  California Press}, 1981).

\bibitem{af97}
{P. W. Atkins and R. S. Friedman}.
\newblock {\em {Molecular Quantum Mechanics}\/} ({Oxford University Press},
  1997).

\bibitem{s97}
{V. Schmidt}.
\newblock {\em {Electron Spectrometry of Atoms using Synchrotron Radiation}\/}
  ({Cambridge University Press}, 1997).

\bibitem{l83}
{I. N. Levine}.
\newblock {\em {Quantum chemistry}\/} ({Allyn and Bacon}, 1983).

\bibitem{timoc}
M.~Hochbruck and A.~Ostermann.
\newblock Appl. Numer. Math. {\bf 53} (2005).

\bibitem{timoa}
M.~Hochbruck, C.~Lubich, and H.~Selhofer.
\newblock SIAM J. Sci. Comput. {\bf 19} (1998).

\bibitem{timob}
M.~Hochbruck and C.~Lubich.
\newblock SIAM J. Numer. Anal. {\bf 41} (2003).

\bibitem{Sahni82}
V.~S.~J. Gruenebaum and J.~P. Perdew.
\newblock Phys. Rev. B {\bf 26}, 4371 (1982).

\bibitem{Langreth83}
D.~C. Langreth and M.~J. Mehl.
\newblock Phys. Rev. B {\bf 28}, 1809 (1983).

\bibitem{KLI90}
J.~B. Krieger, Y.~Li, and G.~J. Iafrate{}.
\newblock Phys. Lett. A {\bf 146}, 256 (1990).

\bibitem{arias03PRL}
I.~Daykov, T.~Engeness, and T.~Arias.
\newblock Phys. Rev. Lett. {\bf 90}, 216402 (2003).

\bibitem{gygi95PRB}
F.~Gygi and G.~Galli.
\newblock Phys. Rev.~B {\bf 52}, R2229 (1995).

\bibitem{beck00RMP}
T.~L. Beck.
\newblock Rev. Mod. Phys. {\bf 72}, 1041 (2000).

\bibitem{bylaska95SIAM}
E.~J. Bylaska, S.~R. Kohn, S.~B. Baden, A.~Edelman, R.~Kawai, M.~E.~G. Ong, and
  J.~H. Weare.
\newblock In {\em Proceedings of the 7th SIAM Conference on Parallel Processing
  for Scientific Computing\/}, edited by D.~H. Bailey{\it \ et al.} (1995), p.
  219.

\bibitem{kronik_private_communication}
L.~Kronik.
\newblock {\it private communication}.

\bibitem{tevelde90thesis}
G.~te~Velde.
\newblock {\em Numerical Integration and other methodological aspects of
  bandstructure calculations\/}.
\newblock {Ph.D.} dissertation, Vrije Universiteit te Amsterdam (1990).

\bibitem{tevelde92JCP}
G.~te~Velde and E.~J. Baerends.
\newblock J. Comput. Phys. {\bf 99}, 84 (1992).

\bibitem{cools03JC}
J.~Cools.
\newblock J. Complexity {\bf 19}, 445 (2003).

\bibitem{tsuchida98JPSJ}
E.~Tsuchida and M.~Tsukada.
\newblock J. Phys. Soc. Japan {\bf 67}, 3844 (1998).

\end{thebibliography}

\end{document}